\documentclass[sigplan,screen]{acmart}
\usepackage{datetime2}
\DTMsavedate{date}{2021-02-23}

\setcopyright{rightsretained}
\acmPrice{15.00}
\acmDOI{}
\acmYear{2021}
\copyrightyear{2021}
\acmISBN{}
\acmConference[]{}{}{}

\startPage{1}

\makeatletter
\@ifundefined{lhs2tex.lhs2tex.sty.read}%
  {\@namedef{lhs2tex.lhs2tex.sty.read}{}%
   \newcommand\SkipToFmtEnd{}%
   \newcommand\EndFmtInput{}%
   \long\def\SkipToFmtEnd#1\EndFmtInput{}%
  }\SkipToFmtEnd

\newcommand\ReadOnlyOnce[1]{\@ifundefined{#1}{\@namedef{#1}{}}\SkipToFmtEnd}
\usepackage{amstext}
\usepackage{amssymb}
\usepackage{stmaryrd}
\DeclareFontFamily{OT1}{cmtex}{}
\DeclareFontShape{OT1}{cmtex}{m}{n}
  {<5><6><7><8>cmtex8
   <9>cmtex9
   <10><10.95><12><14.4><17.28><20.74><24.88>cmtex10}{}
\DeclareFontShape{OT1}{cmtex}{m}{it}
  {<-> ssub * cmtt/m/it}{}

\DeclareFontShape{OT1}{cmtt}{bx}{n}
  {<5><6><7><8>cmtt8
   <9>cmbtt9
   <10><10.95><12><14.4><17.28><20.74><24.88>cmbtt10}{}
\DeclareFontShape{OT1}{cmtex}{bx}{n}
  {<-> ssub * cmtt/bx/n}{}

\newcommand{\anonymous}{\kern0.06em \vbox{\hrule\@width.5em}}
\newcommand{\plus}{\mathbin{+\!\!\!+}}

\renewcommand{\leq}{\leqslant}
\renewcommand{\geq}{\geqslant}
\usepackage{polytable}

\@ifundefined{mathindent}%
  {\newdimen\mathindent\mathindent\leftmargini}%
  {}%

\def\resethooks{%
  \global\let\SaveRestoreHook\empty
  \global\let\ColumnHook\empty}
\newcommand*{\savecolumns}[1][default]%
  {\g@addto@macro\SaveRestoreHook{\savecolumns[#1]}}
\newcommand*{\restorecolumns}[1][default]%
  {\g@addto@macro\SaveRestoreHook{\restorecolumns[#1]}}
\newcommand*{\aligncolumn}[2]%
  {\g@addto@macro\ColumnHook{\column{#1}{#2}}}

\resethooks

\newcommand{\onelinecommentchars}{\quad-{}- }
\newcommand{\commentbeginchars}{\enskip\{-}
\newcommand{\commentendchars}{-\}\enskip}

\newcommand{\visiblecomments}{%
  \let\onelinecomment=\onelinecommentchars
  \let\commentbegin=\commentbeginchars
  \let\commentend=\commentendchars}

\newcommand{\invisiblecomments}{%
  \let\onelinecomment=\empty
  \let\commentbegin=\empty
  \let\commentend=\empty}

\visiblecomments

\newlength{\blanklineskip}
\newcommand{\hsindent}[1]{\quad}
\let\hspre\empty
\let\hspost\empty

\EndFmtInput
\makeatother
\ReadOnlyOnce{polycode.fmt}%
\makeatletter

\newcommand{\hsnewpar}[1]%
  {{\parskip=0pt\parindent=0pt\par\vskip #1\noindent}}

\newcommand{\hscodestyle}{}

\newcommand{\sethscode}[1]%
  {\expandafter\let\expandafter\hscode\csname #1\endcsname
   \expandafter\let\expandafter\endhscode\csname end#1\endcsname}

  {\par\noindent
   \advance\leftskip\mathindent
   \hscodestyle
   \let\\=\@normalcr
   \let\hspre\(\let\hspost\)%
   \pboxed}%
  {\endpboxed\)%
   \par\noindent
   \ignorespacesafterend}

  {\hsnewpar\abovedisplayskip
   \advance\leftskip\mathindent
   \hscodestyle
   \let\hspre\(\let\hspost\)%
   \pboxed}%
  {\endpboxed%
   \hsnewpar\belowdisplayskip
   \ignorespacesafterend}

  {\hsnewpar\abovedisplayskip
   \advance\leftskip\mathindent
   \hscodestyle
   \let\\=\@normalcr
   \(\pboxed}%
  {\endpboxed\)%
   \hsnewpar\belowdisplayskip
   \ignorespacesafterend}

\newcommand{\plainhs}{\sethscode{plainhscode}}

\plainhs

  {\hsnewpar\abovedisplayskip
   \advance\leftskip\mathindent
   \hscodestyle
   \let\\=\@normalcr
   \(\parray}%
  {\endparray\)%
   \hsnewpar\belowdisplayskip
   \ignorespacesafterend}

  {\parray}{\endparray}

  {\(\parray}{\endparray\)}

\def\codeframewidth{\arrayrulewidth}
\RequirePackage{calc}

  {\parskip=\abovedisplayskip\par\noindent
   \hscodestyle
   \arrayrulewidth=\codeframewidth
   \tabular{@{}|p{\linewidth-2\arraycolsep-2\arrayrulewidth-2pt}|@{}}%
   \hline\framedhslinecorrect\\{-1.5ex}%
   \let\endoflinesave=\\
   \let\\=\@normalcr
   \(\pboxed}%
  {\endpboxed\)%
   \framedhslinecorrect\endoflinesave{.5ex}\hline
   \endtabular
   \parskip=\belowdisplayskip\par\noindent
   \ignorespacesafterend}

\newcommand{\framedhslinecorrect}[2]%
  {#1[#2]}

  {\(\def\column##1##2{}%
   \let\>\undefined\let\<\undefined\let\\\undefined
   \newcommand\>[1][]{}\newcommand\<[1][]{}\newcommand\\[1][]{}%
   \def\fromto##1##2##3{##3}%
   }{\) }%

  {\let\orighscode=\hscode
   \let\origendhscode=\endhscode
   \def\endhscode{\def\hscode{\endgroup\def\@currenvir{hscode}\\}\begingroup}
   \orighscode\def\hscode{\endgroup\def\@currenvir{hscode}}}%
  {\origendhscode
   \global\let\hscode=\orighscode
   \global\let\endhscode=\origendhscode}%

\makeatother
\EndFmtInput
\usepackage{booktabs}
\usepackage{subfigure}
\usepackage{wrapfig}

\usepackage{amsmath}
\usepackage{xcolor}
\usepackage{booktabs}

\usepackage{graphicx}
\usepackage{tikz}
\usepackage{pgfplots}
\usetikzlibrary{positioning}
\usetikzlibrary{calc}
\usetikzlibrary{plotmarks}

\usepackage{cleveref}

\definecolor{C1}{RGB}{0,153,204}
\definecolor{C2}{RGB}{89,0,179}

\newcounter{commentctr}[section]
\newenvironment{myhs}{\vspace{0.10cm}\par\noindent\begin{minipage}{\textwidth}\small}{\end{minipage}\vspace{0.10cm}}

\def\dotminus{\mathbin{\ooalign{\hss\raise1ex\hbox{.}\hss\cr
  \mathsurround=0pt$-$}}}

\definecolor{hsblack}{RGB}{45,32,3}
\definecolor{hsgold1}{RGB}{179,169,149}
\definecolor{hsgold2}{RGB}{177,149,90}
\definecolor{hsgold3}{RGB}{190,106,13}
\definecolor{hsblue1}{RGB}{173,176,182}
\definecolor{hsblue2}{RGB}{113,142,205}
\definecolor{hsblue3}{RGB}{0,33,132}
\definecolor{hsblue4}{RGB}{97,108,132}
\definecolor{hsblue5}{RGB}{34,50,68}
\definecolor{hsred2}{RGB}{191,121,103}
\definecolor{hsred3}{RGB}{171,72,46}

\newcommand*{\mathcolor}{}
\def\mathcolor#1#{\mathcoloraux{#1}}
\newcommand*{\mathcoloraux}[3]{%
  \protect\leavevmode
  \begingroup
    \color#1{#2}#3%
  \endgroup
}
\newcommand{\HSKeyword}[1]{\mathcolor{hsgold3}{\textbf{#1}}}
\newcommand{\HSNumeral}[1]{\mathcolor{hsred3}{#1}}

\newcommand{\HSString}[1]{\mathcolor{hsred2}{#1}}
\newcommand{\HSSpecial}[1]{\mathcolor{hsblue4}{\ensuremath{#1}}}
\newcommand{\HSSym}[1]{\mathcolor{hsblue4}{\ensuremath{#1}}}
\newcommand{\HSCon}[1]{\mathcolor{hsblue3}{#1}}
\newcommand{\HSVar}[1]{\mathcolor{hsblue5}{\mathit{\ensuremath{#1}}}}
\newcommand{\HSComment}[1]{\mathcolor{hsgold2}{\textit{#1}}}

\newcommand{\HT}[1]{\ifdefined\HSCon\HSCon{#1}\else#1\fi}
\newcommand{\HS}[1]{\ifdefined\HSSym\HSSym{#1}\else#1\fi}
\newcommand{\HV}[1]{\ifdefined\HSVar\HSVar{#1}\else#1\fi}

\begin{document}
\title{Formal Verification of Authenticated, Append-Only Skip Lists in Agda: Extended Version}
\titlenote{This is an extended version of our paper~\cite{Miraldo2021} published in the 10th ACM SIGPLAN International Conference on Certified Programs and Proofs (CPP '21). This version (\DTMusedate{date}) presents a stronger version of the \ensuremath{\HSVar{evocr}} property than originally presented, and provides a link to our development in open source~\cite{aaosl-agda}.}
%
%

\author{Victor Cacciari Miraldo}
\affiliation{
  \institution{Dfinity Foundation}
  \country{The Netherlands}
}
\authornote{Work performed while an intern at Oracle Labs.}
\email{victor.miraldo@dfinity.org}

\author{Harold Carr}
\affiliation{
  \institution{Oracle Labs}
  \country{United States}
}
\email{harold.carr@oracle.com}

\author{Mark Moir}
\affiliation{
  \institution{Oracle Labs}
  \country{New Zealand}
}
\email{mark.moir@oracle.com}

\author{Lisandra Silva}
\affiliation{
  \institution{Inesc Tec}
  \country{Portugal}
}
\email{lisandra.m.silva@inesctec.pt}
\authornotemark[2]

\author{Guy L. Steele Jr.}
\affiliation{
  \institution{Oracle Labs}
  \country{United States}
}
\email{guy.steele@oracle.com}

\begin{abstract}

Authenticated Append-Only Skiplists (AAOSLs) enable maintenance and
querying of an authenticated log (such as a blockchain) without
requiring any single party to store or verify the entire log, or to
trust another party regarding its contents.  AAOSLs
can help to enable efficient dynamic participation (e.g., in consensus) and
reduce storage overhead. 

  In this paper, we formalize an AAOSL originally described by Maniatis and Baker,
and prove its key correctness properties.  Our model and proofs are machine
checked in Agda.
Our proofs apply to a generalization of the original construction and
provide confidence that instances of this
generalization can be used in practice.  Our formalization effort has also
yielded some simplifications and optimizations.

\end{abstract}

\ccsdesc[500]{Software and its engineering~Formal methods}
\ccsdesc[300]{Theory of computation~Data structures design and analysis}
\ccsdesc[300]{Security and privacy~Distributed systems security}


\bibliographystyle{ACM-Reference-Format}

\maketitle              
%


\newcommand{\guydash}{-{\hskip-0.3em}-}


\newcommand{\calcauth}{auth}
\newcommand{\authType}{Auth}
\newcommand{\mkauthType}{mkauth}
\newcommand{\mkadvType}{mkadv}


\newcommand{\hopsrc}{\hbox{\it hop\guydash{}src}\;}
\newcommand{\hoptgt}{\hbox{\it hop\guydash{}tgt}\;}


\newcommand{\todo}[1]{\textit{\textbf{TODO:} #1}}

\section{Introduction}

\emph{Decentralized} technologies---such as
blockchains~\cite{nakamoto}---that enable reliable coordination among
parties that do not trust each other are of increasing interest and
importance.  It is common to maintain a self-authenticating \emph{log}
using \emph{hash chaining}~\cite{Spreitzer1997}: each entry (e.g., block or
transaction) includes a cryptographic hash based on its own contents
as well as on the previous entry.  This technique ensures that a dishonest
participant can ``rewrite history'' without detection
only if it can find a collision for the cryptographic hash function
used; it is a standard assumption that this is infeasible for a
computationally bounded adversary.

Participating in consensus to add more entries to the log typically
requires knowledge of the current ``state''\negthinspace, which is a
function of all previous transactions.  However, downloading and
verifying the entire history and directly computing the state is too
slow and expensive for many purposes, such as bringing a new
participant online quickly.

One option is to provide a state and
a quorum of signed digests for that state
to a new participant $p$. We could further
enable $p$ to download only parts of the state that it needs by using more
structured state types~\cite{Miraldo2018}.  Although this enables $p$
to participate in consensus for appending \emph{new} blocks to the
chain, receiving and verifying the
state at index $j$ does \emph{not} enable $p$ to verify claims about transactions
at indexes $i < j$ without significant work.  This is because
hash-chaining typically used in blockchains is \emph{linear}:
to confirm a claim that a particular transaction is at index $i$,
$p$ must fetch and verify \emph{all} transactions between $i$ and
$j$.

By instead using an \emph{Authenticated Append-Only Skip List}~\cite{Baker2003}
(AAOSL) to represent the log, an existing
participant (the \emph{prover}) can provide to a new participant (the
\emph{verifier}) an \emph{advancement proof} from some previous log
entry (perhaps the initial---or \emph{genesis}---entry) to the new
root digest that the verifier has obtained from a quorum of
participants.  The verifier can then receive a \emph{membership proof}
containing a claim about a previous log entry, and
confirm that the claim is consistent with the log on which the
previously verified advancement proof was based.  The new participant
can also construct advancement proofs and/or membership proofs
pertaining to new log entries;
thus it can assist another new participant to
bootstrap in future.  Together, these ideas enable
dynamic participation without requiring any participant to
possess all historical information.

To provide this functionality efficiently,
the hash stored in each log entry is made to directly depend not only on the
previous entry but also on selected entries further back in the log.  The
``hops'' back to these previous entries are arranged so that
advancement and membership proofs are logarithmic in the length of
the log.  As a result, these ideas enable dynamic participation that
can be sustained over a long period of time, because the work required
for a new participant to join grows only logarithmically with the
length of the log.

These advantages come at the cost of using mechanisms that are significantly more
complicated than linear hash-chaining.  Therefore, a formally verified
model is paramount before using these techniques in practice.

  Our primary contribution is a formal model of a class of AAOSLs
and proofs of important properties about them in the Agda language~\cite{Norell2008}.
Moreover, we prove that the original AAOSL~\cite{Baker2003}
is an instance of this class and, consequently, enjoys the same properties.
Finally, our formal verification work enabled us to simplify
some aspects of the original AAOSL, such as the encoding
of advancement proofs and the definition of authenticators.

  We provide some background in \Cref{sec:background} and then present
a specification of the original AAOSL~\cite{Baker2003}
in \Cref{sec:skiplog}.  In \Cref{sec:verifying}, we present our
Agda model of a class of AAOSLs, proofs of key properties about them,
and our proof that the original AAOSL is an
instance of the class.  We discuss related and
future work in \Cref{sec:relwork}, and conclude in \Cref{sec:conc}.

\section{Background}
\label{sec:background}

  To illustrate traditional linear hash chaining using Haskell,
we define the following datatype and type synonyms.

\vspace{-1em}
\begin{minipage}[t]{.24\textwidth}
\begin{myhs}
\begin{hscode}\SaveRestoreHook
\column{B}{@{}>{\hspre}l<{\hspost}@{}}%
\column{19}{@{}>{\hspre}l<{\hspost}@{}}%
\column{E}{@{}>{\hspre}l<{\hspost}@{}}%
\>[B]{}\HSKeyword{data}\;\HT{\hbox{\it \authType}}\;\HSVar{a}{}\<[19]%
\>[19]{}\HSSym{\mathrel{=}}\HT{\hbox{\it \authType}}\;\HSVar{a}\;\HSCon{Digest}{}\<[E]%
\ColumnHook
\end{hscode}\resethooks
\end{myhs}
\end{minipage}\begin{minipage}[t]{.24\textwidth}
\begin{myhs}
\begin{hscode}\SaveRestoreHook
\column{B}{@{}>{\hspre}l<{\hspost}@{}}%
\column{14}{@{}>{\hspre}l<{\hspost}@{}}%
\column{E}{@{}>{\hspre}l<{\hspost}@{}}%
\>[B]{}\HSKeyword{type}\;\HSCon{Log}\;\HSVar{a}{}\<[14]%
\>[14]{}\HSSym{\mathrel{=}}\HSSpecial{\HSSym{[\mskip1.5mu} }\HT{\hbox{\it \authType}}\;\HSVar{a}\HSSpecial{\HSSym{\mskip1.5mu]}}{}\<[E]%
\ColumnHook
\end{hscode}\resethooks
\end{myhs}
\end{minipage}

\ensuremath{\HT{\hbox{\it \authType}}\;\HSVar{a}}---the type of authenticated values of some type \ensuremath{\HSVar{a}}---comprises
a value of type \ensuremath{\HSVar{a}} and its \ensuremath{\HSCon{Digest}}.  We assume (via a type constraint \ensuremath{\HSCon{Hashable}\;\HSVar{a}}) that
there is a function \ensuremath{\HSVar{hash}} that receives an \ensuremath{\HSVar{a}} and returns a fixed-length bytestring,
which we call a \emph{digest} (or \emph{hash}).

The ``genesis'' log is represented as \ensuremath{\HSSpecial{\HSSym{[\mskip1.5mu} }\HT{\hbox{\it \authType}}\;\HV{a_\star}\;\HV{h_\star}\HSSpecial{\HSSym{\mskip1.5mu]}}},
where \ensuremath{\HV{a_\star}} is agreed in advance and is used only to compute
\ensuremath{\HV{h_\star}\HSSym{\mathrel{=}}\HSVar{hash}\;\HV{a_\star}}; we sometimes refer to index 0 as \ensuremath{\HV{\star}}.  The \ensuremath{\HSVar{append}} function
illustrates the hash chaining technique.

\begin{myhs}
\begin{hscode}\SaveRestoreHook
\column{B}{@{}>{\hspre}l<{\hspost}@{}}%
\column{3}{@{}>{\hspre}c<{\hspost}@{}}%
\column{3E}{@{}l@{}}%
\column{6}{@{}>{\hspre}l<{\hspost}@{}}%
\column{E}{@{}>{\hspre}l<{\hspost}@{}}%
\>[B]{}\HSVar{append}\HSSym{::}\HSSpecial{(}\HSCon{Hashable}\;\HSVar{a}\HSSpecial{)}\HSSym{\Rightarrow} \HSVar{a}\HSSym{\to} \HSCon{Log}\;\HSVar{a}\HSSym{\to} \HSCon{Log}\;\HSVar{a}{}\<[E]%
\\
\>[B]{}\HSVar{append}\;\HSVar{x}\;\HSSpecial{\HSSym{[\mskip1.5mu} }\HSSpecial{\HSSym{\mskip1.5mu]}}\HSSym{\mathrel{=}}\HSVar{error}\;\HSString{``Log~not~initialized\char34 }{}\<[E]%
\\
\>[B]{}\HSVar{append}\;\HSVar{x}\;\HSSpecial{(}\HT{\hbox{\it \authType}}\;\HSVar{y}\;\HSVar{dy}\HSCon{\mathbin{:}}\HSVar{l}\HSSpecial{)}{}\<[E]%
\\
\>[B]{}\hsindent{3}{}\<[3]%
\>[3]{}\HSSym{\mathrel{=}}{}\<[3E]%
\>[6]{}\HT{\hbox{\it \authType}}\;\HSVar{x}\;\HSSpecial{(}\HSVar{hash}\;\HSSpecial{(}\HSVar{hash}\;\HSVar{x}\HSSym{\plus} \HSVar{dy}\HSSpecial{)}\HSSpecial{)}\HSCon{\mathbin{:}}\HT{\hbox{\it \authType}}\;\HSVar{y}\;\HSVar{dy}\HSCon{\mathbin{:}}\HSVar{l}{}\<[E]%
\ColumnHook
\end{hscode}\resethooks
\end{myhs}

An initialized log comprises a head \ensuremath{\HT{\hbox{\it \authType}}\;\HSVar{y}\;\HSVar{dy}} and a tail \ensuremath{\HSVar{l}}.  When
appending \ensuremath{\HSVar{x}}, the new
log is constructed by hashing \ensuremath{\HSVar{hash}\;\HSVar{x}} concatenated with the latest hash in the
log, \ensuremath{\HSVar{dy}}.  We chose to present \ensuremath{\HSVar{append}} as adding an element to
the \emph{front} of the list for conciseness. An implementation that
adds elements to the \emph{tail} of a list would use the same hash chaining:
the hash for the element at position $i+1$ depends exclusively on the hash
of the element at position $i$. Whether we interpret position $0$ to
be the leftmost or the rightmost element of the list is irrelevant for
the remainder of our formalization.

With a traditional linear log, a client that knows the $j^{th}$ log entry
can verify the $i^{th}$ entry for $i < j$ only by recomputing each
digest for entries $i+1$ to $j$ to confirm that the recomputed digest
for $j$ matches. This is inefficient and requires possession of all
entries between $i$ and $j$.

An AAOSL~\cite{Baker2003} stores a (partial) log linearly as usual,
but enables verifying earlier contents in the log, or verifying that a
recent segment of the log builds on a known previous log, by verifying
only a logarithmic amount of data.

\section{The SkipLog}
\label{sec:skiplog}

\begin{figure}
\centering
\subfigure[A \emph{skiplog} with entries indexed 1 through 12]{
\includegraphics[scale=0.35]{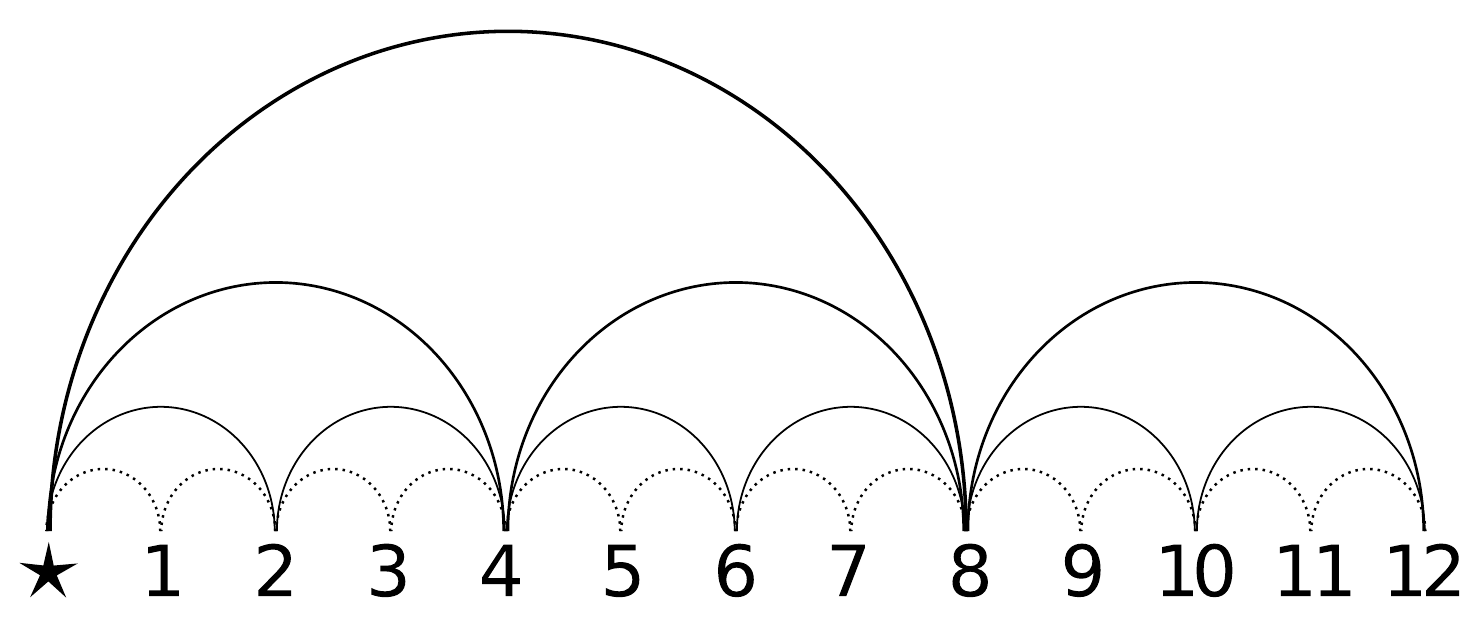}
\label{fig:skiplog-example}}
\quad\quad
\subfigure[%
The direct dependencies of $h_8$ and $h_{12}$, needed to verify a proof of advancement from $h_7$ to $h_{12}$ assuming $h_7$ is already known and trusted.
]{
\includegraphics[scale=0.35]{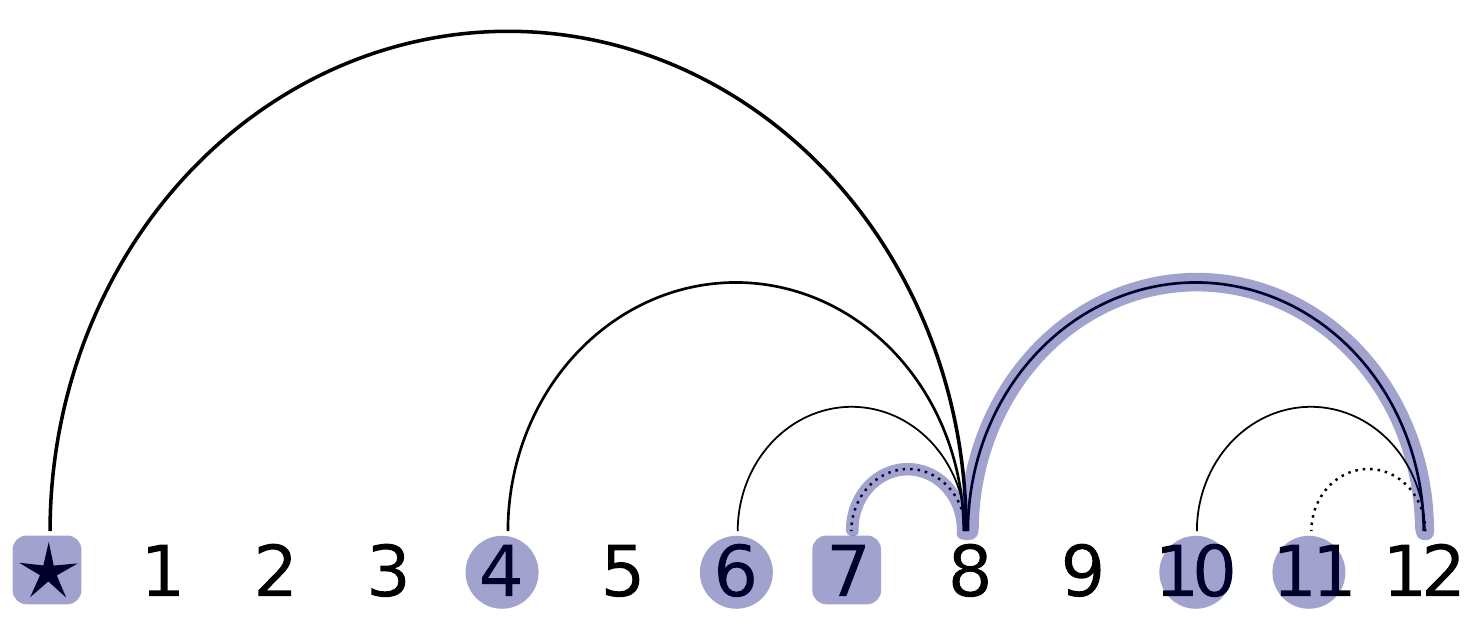}
\label{fig:skiplog-adv-proof}}
\caption{SkipLog illustrations}
\label{fig:bigfig}
\end{figure}

  In traditional skip lists~\cite{Pugh1990}, for each inserted index, a ``height'' is probabilistically
chosen, and for each level up to that height, a ``skip pointer'' to the previous
index with a height of at least that level is stored.

A \emph{skiplog} is an append-only variant, enabling
the use of \emph{deterministic} skip pointers per power of two, as seen in
\Cref{fig:skiplog-example}.  These skip pointers define a
\emph{dependency relation} (or \emph{hop relation}) between indexes,
and thus a \emph{dependency graph}.

When traversing the list
from index 12 to index 4, the
shortest path is to take the skip pointer at level 3, from 12 to 8,
then the one at level 3, from 8 to 4. A skip pointer at level $l$
skips $2^{l-1}$ entries back. The \emph{maximum skip level} for an index $i > 0$ is defined
as $k+1$ where $k$ is the largest integer such that $2^k$ divides $i$.
We use the type synonyms \ensuremath{\HSCon{Index}} and \ensuremath{\HSCon{Level}} (both \ensuremath{\HSCon{Int}}s) to
make the role of these values evident.

\begin{myhs}
\begin{hscode}\SaveRestoreHook
\column{B}{@{}>{\hspre}l<{\hspost}@{}}%
\column{3}{@{}>{\hspre}l<{\hspost}@{}}%
\column{11}{@{}>{\hspre}l<{\hspost}@{}}%
\column{E}{@{}>{\hspre}l<{\hspost}@{}}%
\>[B]{}\HSVar{maxLvl}\HSSym{::}\HSCon{Index}\HSSym{\to} \HSCon{Level}{}\<[E]%
\\
\>[B]{}\HSVar{maxLvl}\;\HSNumeral{0}{}\<[11]%
\>[11]{}\HSSym{\mathrel{=}}\HSNumeral{0}{}\<[E]%
\\
\>[B]{}\HSVar{maxLvl}\;\HSVar{j}{}\<[11]%
\>[11]{}\HSSym{\mathrel{=}}\HSNumeral{1}\HSSym{+}\HSVar{go}\;\HSVar{j}{}\<[E]%
\\
\>[B]{}\hsindent{3}{}\<[3]%
\>[3]{}\HSKeyword{where}\;\HSVar{go}\;\HSVar{j}\HSSym{\mathrel{=}}\HSKeyword{if}\;\HSVar{even}\;\HSVar{j}\;\HSKeyword{then}\;\HSNumeral{1}\HSSym{+}\HSVar{go}\;\HSSpecial{(}\HSVar{j}\HSSym{\mathbin{`div`}}\HSNumeral{2}\HSSpecial{)}\;\HSKeyword{else}\;\HSNumeral{0}{}\<[E]%
\ColumnHook
\end{hscode}\resethooks
\end{myhs}

The maximum skip level of a given index is
exactly the number of other indexes it depends on.  Therefore, the maximum
skip level of index $0$ is defined as 0, indicating that there is no
skip pointer from 0. In our skiplogs, the hash of each
entry is made to depend on all the entries one skip pointer away from it.

  Appending an entry is similar to the description
in \Cref{sec:background}. This time, however, we change how
the hash of the appended entry is computed, ensuring that
it depends directly on every log entry one skip pointer away from it.
Thus, the hash for log index 8 depends on entries for 7, 6, 4, and $\star$.

\begin{myhs}
\begin{hscode}\SaveRestoreHook
\column{B}{@{}>{\hspre}l<{\hspost}@{}}%
\column{E}{@{}>{\hspre}l<{\hspost}@{}}%
\>[B]{}\HSVar{append}\HSSym{::}\HSSpecial{(}\HSCon{Hashable}\;\HSVar{a}\HSSpecial{)}\HSSym{\Rightarrow} \HSVar{a}\HSSym{\to} \HSCon{Log}\;\HSVar{a}\HSSym{\to} \HSCon{Log}\;\HSVar{a}{}\<[E]%
\\
\>[B]{}\HSVar{append}\;\HSVar{x}\;\HSVar{log}\HSSym{\mathrel{=}}\HV{\hbox{\it \mkauthType}}\;\HSVar{x}\;\HSVar{log}\HSCon{\mathbin{:}}\HSVar{log}{}\<[E]%
\ColumnHook
\end{hscode}\resethooks
\end{myhs}

  The \ensuremath{\HV{\hbox{\it \calcauth}}\;\HSVar{x}\;\HSVar{log}} call computes the \emph{authenticator}~\cite{Baker2003}
of \ensuremath{\HSVar{x}} at the new position in the log.
The \emph{authenticator} for position \ensuremath{\HSVar{j}} is a hash computed using the
hash of the data to be stored at \ensuremath{\HSVar{j}} and the hashes of the entries one
skip pointer back from \ensuremath{\HSVar{j}}, namely entries at indexes $j - 2^{l-1}$
for \ensuremath{\HSVar{l}\HSSym{\mathrel{=}}\HSNumeral{1}\HSSpecial{,}\HSNumeral{2}\HSSpecial{,}\HS{\dots}\HSSpecial{,}\HSVar{maxLvl}\;\HSVar{j}}.

\begin{myhs}
\begin{hscode}\SaveRestoreHook
\column{B}{@{}>{\hspre}l<{\hspost}@{}}%
\column{3}{@{}>{\hspre}l<{\hspost}@{}}%
\column{8}{@{}>{\hspre}l<{\hspost}@{}}%
\column{14}{@{}>{\hspre}l<{\hspost}@{}}%
\column{30}{@{}>{\hspre}l<{\hspost}@{}}%
\column{E}{@{}>{\hspre}l<{\hspost}@{}}%
\>[B]{}\HV{\hbox{\it \mkauthType}}\HSSym{::}\HSSpecial{(}\HSCon{Hashable}\;\HSVar{a}\HSSpecial{)}\HSSym{\Rightarrow} \HSVar{a}\HSSym{\to} \HSCon{Log}\;\HSVar{a}\HSSym{\to} \HT{\hbox{\it \authType}}\;\HSVar{a}{}\<[E]%
\\
\>[B]{}\HV{\hbox{\it \mkauthType}}\;\HSVar{a}\;\HSVar{log}\HSSym{\mathrel{=}}{}\<[E]%
\\
\>[B]{}\hsindent{3}{}\<[3]%
\>[3]{}\HSKeyword{let}\;{}\<[8]%
\>[8]{}\HSVar{j}{}\<[14]%
\>[14]{}\HSSym{\mathrel{=}}\HSVar{length}\;\HSVar{log}{}\<[E]%
\\
\>[8]{}\HSVar{deps}{}\<[14]%
\>[14]{}\HSSym{\mathrel{=}}\HSSpecial{\HSSym{[\mskip1.5mu} }\HSVar{lookupHash}\;{}\<[30]%
\>[30]{}\HSVar{log}\;\HSSpecial{(}\HSVar{j}\HSSym{-}\HSNumeral{2}^{\HSVar{l}\HSSym{-}\HSNumeral{1}}\HSSpecial{)}\HSSym{\mid} \HSVar{l}\HSSym{\leftarrow} \HSSpecial{\HSSym{[\mskip1.5mu} }\HSNumeral{1}\HSSym{\mathinner{\ldotp\ldotp}}\HSVar{maxLvl}\;\HSVar{j}\HSSpecial{\HSSym{\mskip1.5mu]}}\HSSpecial{\HSSym{\mskip1.5mu]}}{}\<[E]%
\\
\>[B]{}\hsindent{3}{}\<[3]%
\>[3]{}\HSKeyword{in}\;\HT{\hbox{\it \authType}}\;\HSVar{a}\;\HSSpecial{(}\HV{\hbox{\it \calcauth}}\;\HSVar{j}\;\HSSpecial{(}\HSVar{hash}\;\HSVar{a}\HSSpecial{)}\;\HSVar{deps}\HSSpecial{)}{}\<[E]%
\ColumnHook
\end{hscode}\resethooks
\end{myhs}

  The \ensuremath{\HSVar{lookupHash}\;\HSVar{log}\;\HSVar{i}} function returns the authenticator for the entry at position \ensuremath{\HSVar{i}}
and the \ensuremath{\HV{\hbox{\it \calcauth}}} auxiliary function assembles the necessary data into a
single hash. This hash is determined by hashing the concatenation of
the \emph{partial authenticators}~\cite{Baker2003} of each dependency
in \ensuremath{\HSVar{deps}}.  In the original presentation, \emph{partial authenticators} are computed by hashing
the current index, level, data, and the authenticator of the
dependency, illustrated by the following pseudo-Haskell:

\begin{myhs}
\begin{hscode}\SaveRestoreHook
\column{B}{@{}>{\hspre}l<{\hspost}@{}}%
\column{3}{@{}>{\hspre}l<{\hspost}@{}}%
\column{6}{@{}>{\hspre}l<{\hspost}@{}}%
\column{22}{@{}>{\hspre}l<{\hspost}@{}}%
\column{43}{@{}>{\hspre}l<{\hspost}@{}}%
\column{E}{@{}>{\hspre}l<{\hspost}@{}}%
\>[B]{}\HV{\hbox{\it \calcauth}}\HSSym{::}\HSCon{Index}\HSSym{\to} \HSCon{Digest}\HSSym{\to} \HSSpecial{\HSSym{[\mskip1.5mu} }\HSCon{Digest}\HSSpecial{\HSSym{\mskip1.5mu]}}\HSSym{\to} \HSCon{Digest}{}\<[E]%
\\
\>[B]{}\HV{\hbox{\it \calcauth}}\;\HSVar{j}\;\HSVar{datumDig}\;\HSVar{lvlDigs}\HSSym{\mathrel{=}}\HSVar{hash}\;\HSSpecial{(}\HSVar{concat}{}\<[E]%
\\
\>[B]{}\hsindent{3}{}\<[3]%
\>[3]{}\HSSpecial{\HSSym{[\mskip1.5mu} }{}\<[6]%
\>[6]{}\HSVar{hash}\;\HSSpecial{(}\HSVar{encode}\;\HSVar{j}{}\<[22]%
\>[22]{}\HSSym{\plus} \HSVar{encode}\;\HSNumeral{1}{}\<[43]%
\>[43]{}\HSSym{\plus} \HSVar{datumDig}\HSSym{\plus} \HSVar{lookup}\;\HSNumeral{0}\;\HSVar{lvlDigs}\HSSpecial{)}{}\<[E]%
\\
\>[B]{}\hsindent{3}{}\<[3]%
\>[3]{}\HSSpecial{,}{}\<[6]%
\>[6]{}\HSVar{hash}\;\HSSpecial{(}\HSVar{encode}\;\HSVar{j}{}\<[22]%
\>[22]{}\HSSym{\plus} \HSVar{encode}\;\HSNumeral{2}{}\<[43]%
\>[43]{}\HSSym{\plus} \HSVar{datumDig}\HSSym{\plus} \HSVar{lookup}\;\HSNumeral{1}\;\HSVar{lvlDigs}\HSSpecial{)}{}\<[E]%
\\
\>[B]{}\hsindent{3}{}\<[3]%
\>[3]{}\HSSpecial{,}{}\<[6]%
\>[6]{}\HS{\dots}{}\<[E]%
\\
\>[B]{}\hsindent{3}{}\<[3]%
\>[3]{}\HSSpecial{,}{}\<[6]%
\>[6]{}\HSVar{hash}\;\HSSpecial{(}\HSVar{encode}\;\HSVar{j}{}\<[22]%
\>[22]{}\HSSym{\plus} \HSVar{encode}\;\HSVar{l}{}\<[43]%
\>[43]{}\HSSym{\plus} \HSVar{datumDig}\HSSym{\plus} \HSVar{lookup}\;\HSSpecial{(}\HSVar{l}\HSSym{-}\HSNumeral{1}\HSSpecial{)}\;\HSVar{lvlDigs}\HSSpecial{)}{}\<[E]%
\\
\>[B]{}\hsindent{3}{}\<[3]%
\>[3]{}\HSSpecial{\HSSym{\mskip1.5mu]}}\HSSpecial{)}\;\HSKeyword{where}\;\HSVar{l}\HSSym{\mathrel{=}}\HSVar{maxLvl}\;\HSVar{j}{}\<[E]%
\ColumnHook
\end{hscode}\resethooks
\end{myhs}

  Translating the pseudo-Haskell above to working Haskell yields:

\begin{myhs}
\begin{hscode}\SaveRestoreHook
\column{B}{@{}>{\hspre}l<{\hspost}@{}}%
\column{3}{@{}>{\hspre}l<{\hspost}@{}}%
\column{42}{@{}>{\hspre}l<{\hspost}@{}}%
\column{E}{@{}>{\hspre}l<{\hspost}@{}}%
\>[B]{}\HV{\hbox{\it \calcauth}}\HSSym{::}\HSCon{Index}\HSSym{\to} \HSCon{Digest}\HSSym{\to} \HSSpecial{\HSSym{[\mskip1.5mu} }\HSCon{Digest}\HSSpecial{\HSSym{\mskip1.5mu]}}\HSSym{\to} \HSCon{Digest}{}\<[E]%
\\
\>[B]{}\HV{\hbox{\it \calcauth}}\;\HSVar{j}\;\HSVar{datumDig}\;\HSVar{lvlDigs}\HSSym{\mathrel{=}}{}\<[E]%
\\
\>[B]{}\hsindent{3}{}\<[3]%
\>[3]{}\HSKeyword{let}\;\HSVar{pauth}\;\HSVar{lvl}\;\HSVar{lvldig}\HSSym{\mathrel{=}}\HSVar{hash}\;\HSSpecial{(}\HSVar{encode}\;\HSVar{j}{}\<[42]%
\>[42]{}\HSSym{\plus} \HSVar{encode}\;\HSVar{lvl}{}\<[E]%
\\
\>[42]{}\HSSym{\plus} \HSVar{datumDig}\HSSym{\plus} \HSVar{lvldig}\HSSpecial{)}{}\<[E]%
\\
\>[B]{}\hsindent{3}{}\<[3]%
\>[3]{}\HSKeyword{in}\;\HSVar{hash}\;\HSSpecial{(}\HSVar{concat}\;\HSSpecial{(}\HV{\mathit{zipWith}}\;\HSVar{pauth}\;\HSSpecial{\HSSym{[\mskip1.5mu} }\HSNumeral{1}\HSSym{\mathinner{\ldotp\ldotp}}\HSSpecial{\HSSym{\mskip1.5mu]}}\;\HSVar{lvlDigs}\HSSpecial{)}\HSSpecial{)}{}\<[E]%
\ColumnHook
\end{hscode}\resethooks
\end{myhs}

  The \ensuremath{\HSVar{encode}} function encodes data in a \ensuremath{\HSCon{ByteString}}, facilitating
concatenation and hashing with the rest of the data.  As in
\Cref{sec:background}, the authenticator for the last position in the
log depends (directly or indirectly) on the authenticators of every
prior position and thus provides a ``cumulative'' hash of the whole
structure.

We started our work using the original \ensuremath{\HV{\hbox{\it \calcauth}}} definition~\cite{Baker2003} shown above.
However, our formalization effort
revealed that any definition of \ensuremath{\HV{\hbox{\it \calcauth}}} that is injective in its second and third
parameters suffices. That is, any \ensuremath{\HV{\hbox{\it \calcauth}}} such that \ensuremath{\HV{\hbox{\it \calcauth}}\;\HSVar{j}\;\HV{h_1}\;\HSVar{ds1}\HSSym{\equiv} \HV{\hbox{\it \calcauth}}\;\HSVar{j}\;\HV{h_2}\;\HSVar{ds2}}
implies \ensuremath{\HV{h_1}\HSSym{\equiv} \HV{h_2}} and \ensuremath{\HSVar{ds1}\HSSym{\equiv} \HSVar{ds2}} (modulo hash collisions) can be used as the definition
for authenticators. We have done the proofs
in the remainder of this paper for the original definitions as well as the following, simpler,
\ensuremath{\HV{\hbox{\it \calcauth}}} function.

\begin{myhs}
\begin{hscode}\SaveRestoreHook
\column{B}{@{}>{\hspre}l<{\hspost}@{}}%
\column{3}{@{}>{\hspre}l<{\hspost}@{}}%
\column{E}{@{}>{\hspre}l<{\hspost}@{}}%
\>[B]{}\HV{\hbox{\it \calcauth}}\HSSym{::}\HSCon{Index}\HSSym{\to} \HSCon{Digest}\HSSym{\to} \HSSpecial{\HSSym{[\mskip1.5mu} }\HSCon{Digest}\HSSpecial{\HSSym{\mskip1.5mu]}}\HSSym{\to} \HSCon{Digest}{}\<[E]%
\\
\>[B]{}\HV{\hbox{\it \calcauth}}\;\HSVar{j}\;\HSVar{datumDig}\;\HSVar{lvlDigs}\HSSym{\mathrel{=}}{}\<[E]%
\\
\>[B]{}\hsindent{3}{}\<[3]%
\>[3]{}\HSVar{hash}\;\HSSpecial{(}\HSVar{concat}\;\HSSpecial{(}\HSVar{encode}\;\HSVar{j}\HSCon{\mathbin{:}}\HSVar{datumDig}\HSCon{\mathbin{:}}\HSVar{lvlDigs}\HSSpecial{)}\HSSpecial{)}{}\<[E]%
\ColumnHook
\end{hscode}\resethooks
\end{myhs}

\subsection{Advancement Proofs}

  As more data is appended to the log, a client
must be able to verify that the maintainer did not rewrite history, that is,
that the appended log entries are consistent with any previous log entries that the client has
already verified.  For this purpose, a maintainer constructs an advancement proof,
which can be checked by a client that already
knows and trusts the hash (authenticator) $h_i$ for the $i$th log entry
in order to verify that the hash $h_j$
for some $j > i$ is consistent with extending the log from position
$i$ to position $j$.

Depending on the use case, the
advancement proof could be sent in response to an explicit request, or the
sender could know what the recipient already knows and therefore what advancement proof
it requires.
Next, we examine how the hashing strategy described above enables
efficient construction and verification of these advancement proofs.

An advancement proof comprises a Merkle path~\cite{Merkle1988,Miraldo2018}
in the skiplog. For example, consider the skiplog in
\Cref{fig:skiplog-example}.  Suppose Alice knows and trusts a value (call it \ensuremath{\HV{a_{7}}}) for
the cumulative hash at position $7$, but has not yet verified that the log
advanced from index $7$ to $12$. To verify that the advancement
from 7 to 12 is consistent with the log she already trusts up to index 7,
Alice needs enough information in addition to the hashes she already trusts to compute a value for $h_{12}$.
\Cref{fig:skiplog-adv-proof} illustrates this information: circles
represent information Alice needs, squares represents information
she already has, and the bold links show the path from $12$ to $7$;
superfluous links have been erased for clarity.

  If Alice can obtain $h_4$, $h_6$, $h_{10}$, $h_{11}$ and the hashes
of the data in positions 12 and 8 ($d_{12}$ and $d_8$), then she can
compute a value for $h_{12}$ using \ensuremath{\HV{a_{7}}} (the value for index 7 that
she already trusts), along with the genesis hash $h_\star$. Note that
Alice's possession of a value for the hash at index 7 does
\emph{not} imply that she also knows $h_4$ or $h_6$: $a_{7}$ is
computed by hashing a value that depends on the data at index $7$ and
$h_6$, so Alice would have to invert the hash function to get $h_6$
from $h_7$.

\begin{myhs}
\begin{hscode}\SaveRestoreHook
\column{B}{@{}>{\hspre}l<{\hspost}@{}}%
\column{39}{@{}>{\hspre}l<{\hspost}@{}}%
\column{E}{@{}>{\hspre}l<{\hspost}@{}}%
\>[B]{}\HV{a_{12}}\HSSym{\mathrel{=}}\HV{\hbox{\it \calcauth}}\;\HSNumeral{12}\;\HV{d_{12}}\;\HSSpecial{\HSSym{[\mskip1.5mu} }\HV{h_{11}}\HSSpecial{,}\HV{h_{10}}\HSSpecial{,}{}\<[39]%
\>[39]{}\HV{\hbox{\it \calcauth}}\;\HSNumeral{8}\;\HV{d_{8}}\;\HSSpecial{\HSSym{[\mskip1.5mu} }\HV{a_{7}}\HSSpecial{,}\HV{h_{6}}\HSSpecial{,}\HV{h_{4}}\HSSpecial{,}\HV{h_\star}\HSSpecial{\HSSym{\mskip1.5mu]}}\HSSpecial{\HSSym{\mskip1.5mu]}}{}\<[E]%
\ColumnHook
\end{hscode}\resethooks
\end{myhs}

  In some contexts (such as in our motivating use case), Alice may
already know the expected value for $h_{12}$ (e.g., because consensus
has been reached on it).  In this case, she can compare her computed
value $a_{12}$ to $h_{12}$ and confirm that the advancement from index
7 to 12 is consistent with the log that led to $h_{12}$. We encode the
information Alice needs in the \ensuremath{\HT{\mathit{AdvProof}}} type.

\begin{myhs}
\begin{hscode}\SaveRestoreHook
\column{B}{@{}>{\hspre}l<{\hspost}@{}}%
\column{16}{@{}>{\hspre}c<{\hspost}@{}}%
\column{16E}{@{}l@{}}%
\column{19}{@{}>{\hspre}l<{\hspost}@{}}%
\column{E}{@{}>{\hspre}l<{\hspost}@{}}%
\>[B]{}\HSKeyword{data}\;\HT{\mathit{AdvProof}}{}\<[16]%
\>[16]{}\HSSym{\mathrel{=}}{}\<[16E]%
\>[19]{}\HSCon{Done}{}\<[E]%
\\
\>[16]{}\HSSym{\mid} {}\<[16E]%
\>[19]{}\HSCon{Hop}\;\HSCon{Index}\;\HSCon{Digest}\;\HSSpecial{\HSSym{[\mskip1.5mu} }\HSCon{Digest}\HSSpecial{\HSSym{\mskip1.5mu]}}\;\HSSpecial{\HSSym{[\mskip1.5mu} }\HSCon{Digest}\HSSpecial{\HSSym{\mskip1.5mu]}}\;\HT{\mathit{AdvProof}}{}\<[E]%
\ColumnHook
\end{hscode}\resethooks
\end{myhs}

  The structure of an advancement proof mimics a traversal through the
skip pointers in a skip list.  For example, an advancement proof from index 7 to index 12
sent to Alice can be represented by the following value \ensuremath{\HSVar{p}} of type
\ensuremath{\HT{\mathit{AdvProof}}}:

\begin{myhs}
\begin{hscode}\SaveRestoreHook
\column{B}{@{}>{\hspre}l<{\hspost}@{}}%
\column{E}{@{}>{\hspre}l<{\hspost}@{}}%
\>[B]{}\HSCon{Hop}\;\HSNumeral{12}\;\HV{d_{12}}\;\HSSpecial{\HSSym{[\mskip1.5mu} }\HV{h_{11}}\HSSpecial{,}\HV{h_{10}}\HSSpecial{\HSSym{\mskip1.5mu]}}\;\HSSpecial{\HSSym{[\mskip1.5mu} }\HSSpecial{\HSSym{\mskip1.5mu]}}\;\HSSpecial{(}\HSCon{Hop}\;\HSNumeral{8}\;\HV{d_{8}}\;\HSSpecial{\HSSym{[\mskip1.5mu} }\HSSpecial{\HSSym{\mskip1.5mu]}}\;\HSSpecial{\HSSym{[\mskip1.5mu} }\HV{h_{6}}\HSSpecial{,}\HV{h_{4}}\HSSpecial{,}\HV{h_\star}\HSSpecial{\HSSym{\mskip1.5mu]}}\;\HSCon{Done}\HSSpecial{)}{}\<[E]%
\ColumnHook
\end{hscode}\resethooks
\end{myhs}

Each \ensuremath{\HSCon{Hop}} carries all the information needed to build an
authenticator for the hop's source index using \ensuremath{\HV{\hbox{\it \calcauth}}}.  For example,
\ensuremath{\HSVar{p}}'s first hop (from index 12) contains the datum hash \ensuremath{\HV{d_{12}}} and
information sufficient to determine the authenticators for each
dependency of index 12, thus enabling the verifier to compute \ensuremath{\HV{a_{12}}}
using \ensuremath{\HV{\hbox{\it \calcauth}}}, as shown above.  The advancement proof contains two
lists of authenticators for each hop: one for dependencies of the
hop's source that are \emph{after} the hop's target index, and one for
those that are \emph{before} it.  In the example, the first list of the hop
from 12 contains authenticators for 11 and 10, which are after 8 (the
hop's target index), and there are no dependencies before 8 (note that
the authenticator for the genesis index \ensuremath{\HV{\star}} is not needed because
it is known by all in advance).

Finally the hop provides another advancement proof from the hop's
target to the final target of the advancement proof (7 in this case),
enabling recursive computation of the authenticator for the hop's
target (8 in the example).

This structure enables us to \emph{rebuild} an authenticator from an advancement
proof by recursively computing the authenticator for the hop's
target, and then using \ensuremath{\HV{\hbox{\it \calcauth}}} to combine this with the supplied authenticators
for the other dependencies.  When the recursion reaches the base case \ensuremath{\HSCon{Done}},
the authenticator value provided to the \ensuremath{\HSVar{rebuild}} function is used (\ensuremath{\HSSym{\plus} } is list concatenation).

\begin{myhs}
\begin{hscode}\SaveRestoreHook
\column{B}{@{}>{\hspre}l<{\hspost}@{}}%
\column{10}{@{}>{\hspre}l<{\hspost}@{}}%
\column{36}{@{}>{\hspre}l<{\hspost}@{}}%
\column{39}{@{}>{\hspre}l<{\hspost}@{}}%
\column{E}{@{}>{\hspre}l<{\hspost}@{}}%
\>[B]{}\HSVar{rebuild}{}\<[10]%
\>[10]{}\HSSym{::}\HT{\mathit{AdvProof}}\HSSym{\to} \HSCon{Digest}\HSSym{\to} \HSCon{Digest}{}\<[E]%
\\
\>[B]{}\HSVar{rebuild}\;{}\<[10]%
\>[10]{}\HSCon{Done}\;{}\<[36]%
\>[36]{}\HSVar{d}{}\<[39]%
\>[39]{}\HSSym{\mathrel{=}}\HSVar{d}{}\<[E]%
\\
\>[B]{}\HSVar{rebuild}\;{}\<[10]%
\>[10]{}\HSSpecial{(}\HSCon{Hop}\;\HSVar{j}\;\HSVar{datDig}\;\HSVar{af}\;\HSVar{bf}\;\HSVar{prf}\HSSpecial{)}\;{}\<[36]%
\>[36]{}\HSVar{d}{}\<[39]%
\>[39]{}\HSSym{\mathrel{=}}{}\<[E]%
\\
\>[10]{}\HV{\hbox{\it \calcauth}}\;\HSVar{j}\;\HSVar{datDig}\;\HSSpecial{(}\HSVar{af}\HSSym{\plus} \HSSpecial{\HSSym{[\mskip1.5mu} }\HSVar{rebuild}\;\HSVar{prf}\;\HSVar{d}\HSSpecial{\HSSym{\mskip1.5mu]}}\HSSym{\plus} \HSVar{bf}\HSSpecial{)}{}\<[E]%
\ColumnHook
\end{hscode}\resethooks
\end{myhs}

One can verify that \ensuremath{\HSVar{rebuild}\;\HSVar{p}\;\HV{a_{7}}} constructs the correct hash.  Note
that \ensuremath{\HSVar{rebuild}\;\HSCon{Done}\;\HV{a_{7}}} provides the hash Alice already trusts for
index 7.

Building an advancement proof is done by traversing the dependency graph of the log.
The smallest advancement proof is obtained by traversing
from position $j$ to $i < j$ via the
highest level hop $l$ such that $j - 2^{l-1} \geq i$;
this is facilitated by the \ensuremath{\HSVar{singleHopLevel}} function.

\begin{myhs}
\begin{hscode}\SaveRestoreHook
\column{B}{@{}>{\hspre}l<{\hspost}@{}}%
\column{3}{@{}>{\hspre}l<{\hspost}@{}}%
\column{17}{@{}>{\hspre}l<{\hspost}@{}}%
\column{E}{@{}>{\hspre}l<{\hspost}@{}}%
\>[B]{}\HSVar{singleHopLevel}{}\<[17]%
\>[17]{}\HSSym{::}\HSCon{Index}\HSSym{\to} \HSCon{Index}\HSSym{\to} \HSCon{Level}{}\<[E]%
\\
\>[B]{}\HSVar{singleHopLevel}\;{}\<[17]%
\>[17]{}\HSVar{from}\;\HSVar{to}\HSSym{\mathrel{=}}{}\<[E]%
\\
\>[B]{}\hsindent{3}{}\<[3]%
\>[3]{}\HSVar{min}\;\HSSpecial{(}\HSNumeral{1}\HSSym{+}\HSVar{floor}\;\HSSpecial{(}\HSVar{logBase}\;\HSNumeral{2}\;\HSSpecial{(}\HSVar{from}\HSSym{-}\HSVar{to}\HSSpecial{)}\HSSpecial{)}\HSSpecial{)}\;\HSSpecial{(}\HSVar{maxLvl}\;\HSVar{from}\HSSpecial{)}{}\<[E]%
\ColumnHook
\end{hscode}\resethooks
\end{myhs}

  The \ensuremath{\HV{\hbox{\it \mkadvType}}} function uses this to
traverse the dependency graph of a log
and construct an advancement from \ensuremath{\HSVar{j}} to \ensuremath{\HSVar{i}}:

\begin{myhs}
\begin{hscode}\SaveRestoreHook
\column{B}{@{}>{\hspre}l<{\hspost}@{}}%
\column{4}{@{}>{\hspre}l<{\hspost}@{}}%
\column{10}{@{}>{\hspre}l<{\hspost}@{}}%
\column{15}{@{}>{\hspre}l<{\hspost}@{}}%
\column{21}{@{}>{\hspre}l<{\hspost}@{}}%
\column{E}{@{}>{\hspre}l<{\hspost}@{}}%
\>[B]{}\HV{\hbox{\it \mkadvType}}\HSSym{::}\HSCon{Index}\HSSym{\to} \HSCon{Index}\HSSym{\to} \HSCon{Log}\;\HSVar{a}\HSSym{\to} \HT{\mathit{AdvProof}}{}\<[E]%
\\
\>[B]{}\HV{\hbox{\it \mkadvType}}\;\HSVar{i}\;\HSVar{j}\;\HSVar{log}\HSSym{\mathrel{=}}{}\<[E]%
\\
\>[B]{}\hsindent{4}{}\<[4]%
\>[4]{}\HSKeyword{if}\;\HSVar{i}\HSSym{\equiv} \HSVar{j}{}\<[E]%
\\
\>[B]{}\hsindent{4}{}\<[4]%
\>[4]{}\HSKeyword{then}\;\HSCon{Done}{}\<[E]%
\\
\>[B]{}\hsindent{4}{}\<[4]%
\>[4]{}\HSKeyword{else}\;{}\<[10]%
\>[10]{}\HSKeyword{let}\;{}\<[15]%
\>[15]{}\HT{\hbox{\it \authType}}\;\HSVar{dat}\;\HSVar{datDig}\HSSym{\mathrel{=}}\HSVar{lookup}\;\HSVar{j}\;\HSVar{log}{}\<[E]%
\\
\>[15]{}\HSVar{sh}{}\<[21]%
\>[21]{}\HSSym{\mathrel{=}}\HSSpecial{\HSSym{[\mskip1.5mu} }\HSVar{lookupHash}\;\HSVar{log}\;\HSSpecial{(}\HSVar{j}\HSSym{-}\HSNumeral{2}^{\HSVar{l}\HSSym{-}\HSNumeral{1}}\HSSpecial{)}\HSSym{\mid} \HSVar{l}\HSSym{\leftarrow} \HSSpecial{\HSSym{[\mskip1.5mu} }\HSNumeral{1}\HSSym{\mathinner{\ldotp\ldotp}}\HSVar{maxLvl}\;\HSVar{j}\HSSpecial{\HSSym{\mskip1.5mu]}}\HSSpecial{\HSSym{\mskip1.5mu]}}{}\<[E]%
\\
\>[15]{}\HSVar{hop}{}\<[21]%
\>[21]{}\HSSym{\mathrel{=}}\HSVar{singleHopLevel}\;\HSVar{j}\;\HSVar{i}{}\<[E]%
\\
\>[15]{}\HSVar{af}{}\<[21]%
\>[21]{}\HSSym{\mathrel{=}}\HSVar{take}\;\HSSpecial{(}\HSVar{hop}\HSSym{-}\HSNumeral{1}\HSSpecial{)}\;\HSVar{sh}{}\<[E]%
\\
\>[15]{}\HSVar{bf}{}\<[21]%
\>[21]{}\HSSym{\mathrel{=}}\HSVar{drop}\;\HSVar{hop}\;\HSVar{sh}{}\<[E]%
\\
\>[10]{}\HSKeyword{in}\;\HSCon{Hop}\;\HSVar{j}\;\HSVar{datDig}\;\HSVar{af}\;\HSVar{bf}\;\HSSpecial{(}\HV{\hbox{\it \mkadvType}}\;\HSVar{i}\;\HSSpecial{(}\HSVar{j}\HSSym{-}\HSNumeral{2}^{\HSVar{hop}\HSSym{-}\HSNumeral{1}}\HSSpecial{)}\;\HSVar{log}\HSSpecial{)}{}\<[E]%
\ColumnHook
\end{hscode}\resethooks
\end{myhs}

  Advancement proofs can also be used to construct membership proofs: a maintainer
can prove to a client that data \ensuremath{\HV{d_i}} is at position $i$, if the client
trusts the cumulative hash at position $j$, for $j > i$.  The
maintainer sends an advancement proof from $i$ to $j$, along with
a list of the authenticators of the dependencies of $i$; the
client can then compute and confirm the authenticator at $i$.

\begin{myhs}
\begin{hscode}\SaveRestoreHook
\column{B}{@{}>{\hspre}l<{\hspost}@{}}%
\column{3}{@{}>{\hspre}l<{\hspost}@{}}%
\column{13}{@{}>{\hspre}c<{\hspost}@{}}%
\column{13E}{@{}l@{}}%
\column{17}{@{}>{\hspre}l<{\hspost}@{}}%
\column{E}{@{}>{\hspre}l<{\hspost}@{}}%
\>[B]{}\HSKeyword{type}\;\HT{\mathit{MembershipProof}}\HSSym{\mathrel{=}}\HSSpecial{(}\HT{\mathit{AdvProof}}\HSSpecial{,}\HSSpecial{\HSSym{[\mskip1.5mu} }\HSCon{Digest}\HSSpecial{\HSSym{\mskip1.5mu]}}\HSSpecial{)}{}\<[E]%
\\[\blanklineskip]%
\>[B]{}\HSVar{isMemberAt}{}\<[13]%
\>[13]{}\HSSym{::}{}\<[13E]%
\>[17]{}\HSSpecial{(}\HSCon{Hashable}\;\HSVar{a}\HSSpecial{)}{}\<[E]%
\\
\>[13]{}\HSSym{\Rightarrow} {}\<[13E]%
\>[17]{}\HT{\mathit{MembershipProof}}\HSSym{\to} \HSCon{Index}\HSSym{\to} \HSVar{a}\HSSym{\to} \HSCon{Digest}{}\<[E]%
\\
\>[13]{}\HSSym{\to} {}\<[13E]%
\>[17]{}\HSCon{Bool}{}\<[E]%
\\
\>[B]{}\HSVar{isMemberAt}\;\HSSpecial{(}\HSVar{adv}\HSSpecial{,}\HSVar{ss}\HSSpecial{)}\;\HSVar{i}\;\HSVar{a}\;\HSVar{trustedRoot}\HSSym{\mathrel{=}}{}\<[E]%
\\
\>[B]{}\hsindent{3}{}\<[3]%
\>[3]{}\HSVar{rebuild}\;\HSVar{adv}\;\HSSpecial{(}\HV{\hbox{\it \calcauth}}\;\HSVar{i}\;\HSSpecial{(}\HSVar{hash}\;\HSVar{a}\HSSpecial{)}\;\HSVar{ss}\HSSpecial{)}\HSSym{\equiv} \HSVar{trustedRoot}{}\<[E]%
\ColumnHook
\end{hscode}\resethooks
\end{myhs}

  Constructing values of type \ensuremath{\HT{\mathit{MembershipProof}}} is trivial: We
construct an advancement proof and add the
necessary authenticators to it. Membership proofs can also
be used to establish relative order between elements.
Element $a$ is at an index smaller than element $b$ iff
there is a membership proof from the index of $b$ to the index of $a$.

\subsubsection{Well-Formed and Normalized Proofs}
\label{sec:wellformedness}

It is easy to construct values of type \ensuremath{\HT{\mathit{AdvProof}}} that
are not valid advancement proofs.  For example, here are two \emph{non-well-formed}
advancements from index 4 to index 12:

\begin{myhs}
\begin{hscode}\SaveRestoreHook
\column{B}{@{}>{\hspre}l<{\hspost}@{}}%
\column{E}{@{}>{\hspre}l<{\hspost}@{}}%
\>[B]{}\HV{w_1}\HSSym{\mathrel{=}}\HSCon{Hop}\;\HSNumeral{12}\;\HV{d_{12}}\;\HSSpecial{\HSSym{[\mskip1.5mu} }\HV{h_{11}}\HSSpecial{,}\HV{h_{10}}\HSSpecial{\HSSym{\mskip1.5mu]}}\;\HSSpecial{\HSSym{[\mskip1.5mu} }\HSSpecial{\HSSym{\mskip1.5mu]}}\;\HSSpecial{(}\HSCon{Hop}\;\HSNumeral{8}\;\HV{d_{8}}\;\HSSpecial{\HSSym{[\mskip1.5mu} }\HV{h_{7}}\HSSpecial{\HSSym{\mskip1.5mu]}}\;\HSSpecial{\HSSym{[\mskip1.5mu} }\HV{h_\star}\HSSpecial{\HSSym{\mskip1.5mu]}}\;\HSCon{Done}\HSSpecial{)}{}\<[E]%
\\[\blanklineskip]%
\>[B]{}\HV{w_2}\HSSym{\mathrel{=}}\HSCon{Hop}\;\HSNumeral{12}\;\HV{d_{12}}\;\HSSpecial{\HSSym{[\mskip1.5mu} }\HV{h_{11}}\HSSpecial{,}\HV{h_{10}}\HSSpecial{\HSSym{\mskip1.5mu]}}\;\HSSpecial{\HSSym{[\mskip1.5mu} }\HSSpecial{\HSSym{\mskip1.5mu]}}\;\HSSpecial{(}\HSCon{Hop}\;\HSNumeral{6}\;\HV{d_{6}}\;\HSSpecial{\HSSym{[\mskip1.5mu} }\HV{h_{5}}\HSSpecial{\HSSym{\mskip1.5mu]}}\;\HSSpecial{\HSSym{[\mskip1.5mu} }\HSSpecial{\HSSym{\mskip1.5mu]}}\;\HSCon{Done}\HSSpecial{)}{}\<[E]%
\ColumnHook
\end{hscode}\resethooks
\end{myhs}

  In \ensuremath{\HV{w_1}}, the \ensuremath{\HSCon{Hop}\;\HSNumeral{8}} step is missing an authenticator.  Recall that
the authenticator for position \ensuremath{\HSVar{i}} depends on \ensuremath{\HSVar{maxLvl}\;\HSVar{i}}
authenticators. We are supposed to compute one authenticator
recursively, which leaves the other \ensuremath{\HSVar{maxLvl}\;\HSVar{i}\HSSym{-}\HSNumeral{1}} of them to be
specified by the advancement proof.  Index 8 has level 4, so the total
number of authenticators in the \emph{before} and \emph{after} lists
should be exactly 3.  For \ensuremath{\HV{w_2}}, the first hop should be at level 2,
from index 12 to $12 - 2^2 \equiv 8$, but the proof declares that the
hop lands at index 6.

 Consequently, we must
identify what it means for a proof to be well formed---that
is, it has the correct number of authenticators and takes the correct
hop at every step.

\begin{myhs}
\begin{hscode}\SaveRestoreHook
\column{B}{@{}>{\hspre}l<{\hspost}@{}}%
\column{30}{@{}>{\hspre}l<{\hspost}@{}}%
\column{35}{@{}>{\hspre}c<{\hspost}@{}}%
\column{35E}{@{}l@{}}%
\column{39}{@{}>{\hspre}l<{\hspost}@{}}%
\column{E}{@{}>{\hspre}l<{\hspost}@{}}%
\>[B]{}\HSVar{wellformed}\HSSym{::}\HT{\mathit{AdvProof}}\HSSym{\to} \HSCon{Index}\HSSym{\to} \HSCon{Index}\HSSym{\to} \HSCon{Bool}{}\<[E]%
\\
\>[B]{}\HSVar{wellformed}\;\HSCon{Done}\;{}\<[30]%
\>[30]{}\HSVar{j}\;\HSVar{i}{}\<[35]%
\>[35]{}\HSSym{\mathrel{=}}{}\<[35E]%
\>[39]{}\HSSpecial{(}\HSVar{j}\HSSym{\equiv} \HSVar{i}\HSSpecial{)}{}\<[E]%
\\
\>[B]{}\HSVar{wellformed}\;\HSSpecial{(}\HSCon{Hop}\;\HSVar{k}\;\HSSym{\anonymous} \;\HSVar{af}\;\HSVar{bf}\;\HSVar{r}\HSSpecial{)}\;\HSVar{j}\;\HSVar{i}{}\<[35]%
\>[35]{}\HSSym{\mathrel{=}}{}\<[35E]%
\>[39]{}\HSSpecial{(}\HSVar{k}\HSSym{\equiv} \HSVar{j}\HSSpecial{)}{}\<[E]%
\\
\>[35]{}\HSSym{\mathrel{\wedge}}{}\<[35E]%
\>[39]{}\HSSpecial{(}\HSVar{length}\;\HSSpecial{(}\HSVar{af}\HSSym{\plus} \HSVar{bf}\HSSpecial{)}\HSSym{+}\HSNumeral{1}\HSSym{\equiv} \HSVar{maxLvl}\;\HSVar{j}\HSSpecial{)}{}\<[E]%
\\
\>[35]{}\HSSym{\mathrel{\wedge}}{}\<[35E]%
\>[39]{}\HSSpecial{(}\HSVar{wellformed}\;\HSVar{r}\;\HSSpecial{(}\HSVar{j}\HSSym{-}\HSNumeral{2}^{\HSVar{length}\;\HSVar{af}}\HSSpecial{)}\;\HSVar{i}\HSSpecial{)}{}\<[E]%
\ColumnHook
\end{hscode}\resethooks
\end{myhs}

  Well-formedness is a purely structural constraint
that is trivial to check in practice.  In our formal development, we
enforce the \ensuremath{\HSVar{wellformed}} invariant at the type level, which precludes
construction of non-well-formed proofs.

  Now, consider the following \ensuremath{\HT{\mathit{AdvProof}}} from 4 to 12:

\begin{myhs}
\begin{hscode}\SaveRestoreHook
\column{B}{@{}>{\hspre}l<{\hspost}@{}}%
\column{8}{@{}>{\hspre}l<{\hspost}@{}}%
\column{12}{@{}>{\hspre}l<{\hspost}@{}}%
\column{E}{@{}>{\hspre}l<{\hspost}@{}}%
\>[B]{}\HV{w_3}\HSSym{\mathrel{=}}{}\<[8]%
\>[8]{}\HSCon{Hop}\;\HSNumeral{12}\;\HV{d_{12}}\;\HSSpecial{\HSSym{[\mskip1.5mu} }\HV{h_{11}}\HSSpecial{\HSSym{\mskip1.5mu]}}\;\HSSpecial{\HSSym{[\mskip1.5mu} }\HV{h_{8}}\HSSpecial{\HSSym{\mskip1.5mu]}}\;\HSSpecial{(}\HSCon{Hop}\;\HSNumeral{10}\;\HV{d_{10}}\;\HSSpecial{\HSSym{[\mskip1.5mu} }\HV{h_{9}}\HSSpecial{\HSSym{\mskip1.5mu]}}\;\HSSpecial{\HSSym{[\mskip1.5mu} }\HSSpecial{\HSSym{\mskip1.5mu]}}{}\<[E]%
\\
\>[8]{}\hsindent{4}{}\<[12]%
\>[12]{}\HSSpecial{(}\HSCon{Hop}\;\HSNumeral{8}\;\HV{d_{8}}\;\HSSpecial{\HSSym{[\mskip1.5mu} }\HV{h_{7}}\HSSpecial{\HSSym{\mskip1.5mu]}}\;\HSSpecial{\HSSym{[\mskip1.5mu} }\HV{h_{4}}\HSSpecial{,}\HV{h_\star}\HSSpecial{\HSSym{\mskip1.5mu]}}\;\HSSpecial{(}\HSCon{Hop}\;\HSNumeral{6}\;\HV{d_{6}}\;\HSSpecial{\HSSym{[\mskip1.5mu} }\HV{h_{5}}\HSSpecial{\HSSym{\mskip1.5mu]}}\;\HSSpecial{\HSSym{[\mskip1.5mu} }\HSSpecial{\HSSym{\mskip1.5mu]}}\;\HSCon{Done}\HSSpecial{)}\HSSpecial{)}\HSSpecial{)}{}\<[E]%
\ColumnHook
\end{hscode}\resethooks
\end{myhs}

  Although \ensuremath{\HV{w_3}} is a well-formed proof, it takes more hops than necessary.
A \emph{normalized} proof takes the hop determined by \ensuremath{\HSVar{singleHopLevel}}
at each step, resulting in the shortest possible advancement proof, thus
reducing bandwidth and computation requirements for sending and verifying them.
Note that our results hold for advancement proofs formed by composition,
which are not always normalized.


\subsubsection{Size of Advancement Proofs}
\label{sec:advproofsize}

A normalized advancement proof
between indexes $i$ and $j > i$
has at most $2 \lceil \log_2 (1 + j - i) \rceil$ hops (we go up hop levels to get from $j$ to the
maximum-length hop that does not overshoot $i$, and each hop makes twice as much progress as the
previous; similarly, we then go down from that hop to reach $i$,
with each hop progressing at least half the way to $i$).
Each hop references at most $\lceil \log_2 j \rceil$ digests.
These observations yield
a bound of at most $2 \lceil \log_2 (1 + j - i) \rceil \lceil \log_2 j \rceil$ digests.

This bound is conservative: there may be fewer hops,
fewer digests per hop (and the maximum per hop is smaller for
smaller indexes), and each digest is included only once, even if it is
referenced by multiple hops in an advancement
proof (see the \ensuremath{\HSCon{View}} type in \Cref{sec:refinedblocks}).

Proving a tighter bound is left as future work.
However, using our Haskell specification
to examine all normalized advancement proofs between indexes less than $1,\!000$
yields several observations.
The advancement proof
from \ensuremath{\HSVar{i}\HSSym{\mathrel{=}}\HSNumeral{1}} to \ensuremath{\HSVar{j}\HSSym{\mathrel{=}}\HSNumeral{991}} is both the longest (visiting 17 indexes)
and the largest (85 digests).  About 40 digests are included on average,
which is about 1/4 of the number indicated by the conservative bound.
Avoiding duplicate digests (\Cref{sec:refinedblocks}) saves only about 3 digests on average,
but results in a worthwhile simplification to the algorithm and proofs.

\section{Verifying Key Properties Using Agda}
\label{sec:verifying}


  The original AAOSL work~\cite{Baker2003} presents manual proofs of
several properties, showing that they can be
violated only by an adversary that finds a collision for the
underlying cryptographic hash function, which is assumed to be
infeasible for a computationally bounded adversary.  We have
constructed an Agda proof of the main property specified by Maniatis and Baker,
called \emph{Evolutionary collision resistance of AAOSL membership
proofs} (Theorem 3).  Following the original naming, we call this property
\ensuremath{\HV{\textsc{evo-cr}}}.

  The essence of the \ensuremath{\HV{\textsc{evo-cr}}} property is that, if two clients have advanced their
logs to a point with the same digest, then
it is infeasible for a computationally bounded adversary to
convince them to accept different authenticators for the same earlier log
position, even if the clients advanced their logs via different
advancement proofs.  Furthermore, if the hop relation contains
a hop between every index (except the initial index) and its
predecessor---as with the hop relation defined
in~\Cref{sec:skiplog}---then there is a ``degenerate'' advancement proof
that visits every index between 0 and \ensuremath{\HSVar{j}}, for any \ensuremath{\HSVar{j}}. In this case,
\ensuremath{\HV{\textsc{evo-cr}}} implies that a computationally bounded adversary cannot
convince a client to accept a membership proof for index \ensuremath{\HSVar{i}} that is inconsistent
with a given log at index \ensuremath{\HSVar{i}} if the client has rebuilt the membership proof
to the same authenticator as that log at index $j \geq i$.

Our proof of \textsc{evo-cr} is based on several applications of a key
property that we prove first, called \ensuremath{\HV{\textsc{AgreeOnCommon}}} (and of a variant
of this property that we introduce later).  This property
states that, for any two advancement proofs into a new index $j$, if
rebuilding both proofs yields the same hash, then either these proofs
agree on the authenticator for every index that they both visit, or
there is a hash collision.

The overall proof is divided into two parts: an abstract model of a
\emph{class} of authenticated skip lists, and a concrete instantiation
thereof, which we prove meets the necessary requirements.  The
abstract model assumes a \ensuremath{\HSCon{DepRel}} (defined below).  This type implies
a dependency relation
satisfying several properties that are introduced below.  Our concrete
skiplog (described in \Cref{sec:skiplog}) is defined by instantiating
the abstract model with a \ensuremath{\HSCon{DepRel}} that implies the dependency relation
described in~\Cref{sec:skiplog}.  The same could be achieved for a wide variety of such
dependency relations.

  The \ensuremath{\HSCon{DepRel}} record defines the class of skiplogs for which we prove
\ensuremath{\HV{\textsc{AgreeOnCommon}}} and then \ensuremath{\HV{\textsc{evo-cr}}}.

\begin{myhs}
\begin{hscode}\SaveRestoreHook
\column{B}{@{}>{\hspre}l<{\hspost}@{}}%
\column{4}{@{}>{\hspre}l<{\hspost}@{}}%
\column{6}{@{}>{\hspre}l<{\hspost}@{}}%
\column{18}{@{}>{\hspre}c<{\hspost}@{}}%
\column{18E}{@{}l@{}}%
\column{19}{@{}>{\hspre}c<{\hspost}@{}}%
\column{19E}{@{}l@{}}%
\column{21}{@{}>{\hspre}l<{\hspost}@{}}%
\column{22}{@{}>{\hspre}l<{\hspost}@{}}%
\column{37}{@{}>{\hspre}l<{\hspost}@{}}%
\column{E}{@{}>{\hspre}l<{\hspost}@{}}%
\>[B]{}\HSKeyword{record}\;\HSCon{DepRel}\HSCon{\mathbin{:}}\HSCon{Set}\;\HSKeyword{where}{}\<[E]%
\\
\>[B]{}\hsindent{4}{}\<[4]%
\>[4]{}\HSKeyword{field}\;{}\<[E]%
\\
\>[4]{}\hsindent{2}{}\<[6]%
\>[6]{}\HSVar{maxlvl}{}\<[19]%
\>[19]{}\HSCon{\mathbin{:}}{}\<[19E]%
\>[22]{}\HT{\mathbb{N}}\HSSym{\to} \HT{\mathbb{N}}{}\<[E]%
\\
\>[4]{}\hsindent{2}{}\<[6]%
\>[6]{}\HV{\hbox{\it maxlvl\guydash{}z}}{}\<[19]%
\>[19]{}\HSCon{\mathbin{:}}{}\<[19E]%
\>[22]{}\HSVar{maxlvl}\;\HSNumeral{0}\HSSym{\equiv} \HSNumeral{0}{}\<[E]%
\\
\>[4]{}\hsindent{2}{}\<[6]%
\>[6]{}\HV{\hbox{\it maxlvl\guydash{}s}}{}\<[19]%
\>[19]{}\HSCon{\mathbin{:}}{}\<[19E]%
\>[22]{}\HSSpecial{(}\HSVar{m}\HSCon{\mathbin{:}}\HT{\mathbb{N}}\HSSpecial{)}\HSSym{\to} \HSNumeral{0}\HSSym{<}\HSVar{maxlvl}\;\HSSpecial{(}\HSVar{suc}\;\HSVar{m}\HSSpecial{)}{}\<[E]%
\\[\blanklineskip]%
\>[B]{}\hsindent{4}{}\<[4]%
\>[4]{}\HSCon{HopFrom}{}\<[18]%
\>[18]{}\HSCon{\mathbin{:}}{}\<[18E]%
\>[21]{}\HT{\mathbb{N}}\HSSym{\to} \HSCon{Set}{}\<[E]%
\\
\>[B]{}\hsindent{4}{}\<[4]%
\>[4]{}\HSCon{HopFrom}{}\<[18]%
\>[18]{}\HSSym{\mathrel{=}}{}\<[18E]%
\>[21]{}\HSCon{Fin}\HSSym{\mathbin{\circ}}\HSVar{maxlvl}{}\<[E]%
\\[\blanklineskip]%
\>[B]{}\hsindent{4}{}\<[4]%
\>[4]{}\HSKeyword{field}\;{}\<[E]%
\\
\>[4]{}\hsindent{2}{}\<[6]%
\>[6]{}\HV{\hbox{\it hop\guydash{}tgt}}{}\<[18]%
\>[18]{}\HSCon{\mathbin{:}}{}\<[18E]%
\>[21]{}\HSSpecial{\HSSym{\{\mskip1.5mu} }\HSVar{m}\HSCon{\mathbin{:}}\HT{\mathbb{N}}\HSSpecial{\HSSym{\mskip1.5mu\}}}\HSSym{\to} \HSCon{HopFrom}\;\HSVar{m}\HSSym{\to} \HT{\mathbb{N}}{}\<[E]%
\\[\blanklineskip]%
\>[4]{}\hsindent{2}{}\<[6]%
\>[6]{}\HV{\hbox{\it hop\guydash{}inj}}{}\<[18]%
\>[18]{}\HSCon{\mathbin{:}}{}\<[18E]%
\>[22]{}\HSSpecial{\HSSym{\{\mskip1.5mu} }\HSVar{m}\HSCon{\mathbin{:}}\HT{\mathbb{N}}\HSSpecial{\HSSym{\mskip1.5mu\}}}\;\HSSpecial{\HSSym{\{\mskip1.5mu} }\HV{h_1}\;\HV{h_2}\HSCon{\mathbin{:}}\HSCon{HopFrom}\;\HSVar{m}\HSSpecial{\HSSym{\mskip1.5mu\}}}{}\<[E]%
\\
\>[18]{}\HSSym{\to} {}\<[18E]%
\>[22]{}\HV{\hbox{\it hop\guydash{}tgt}}\;\HV{h_1}\HSSym{\equiv} \HV{\hbox{\it hop\guydash{}tgt}}\;\HV{h_2}\HSSym{\to} \HV{h_1}\HSSym{\equiv} \HV{h_2}{}\<[E]%
\\[\blanklineskip]%
\>[4]{}\hsindent{2}{}\<[6]%
\>[6]{}\HV{\hbox{\it hop\guydash{}\!<}}{}\<[18]%
\>[18]{}\HSCon{\mathbin{:}}{}\<[18E]%
\>[22]{}\HSSpecial{\HSSym{\{\mskip1.5mu} }\HSVar{m}\HSCon{\mathbin{:}}\HT{\mathbb{N}}\HSSpecial{\HSSym{\mskip1.5mu\}}}\;\HSSpecial{(}\HSVar{h}\HSCon{\mathbin{:}}\HSCon{HopFrom}\;\HSVar{m}\HSSpecial{)}\HSSym{\to} \HV{\hbox{\it hop\guydash{}tgt}}\;\HSVar{h}\HSSym{<}\HSVar{m}{}\<[E]%
\\[\blanklineskip]%
\>[4]{}\hsindent{2}{}\<[6]%
\>[6]{}\HV{\hbox{\it hop\guydash{}no\guydash{}cross}}{}\<[18]%
\>[18]{}\HSCon{\mathbin{:}}{}\<[18E]%
\>[22]{}\HSSpecial{\HSSym{\{\mskip1.5mu} }\HV{j_1}\;\HV{j_2}\HSCon{\mathbin{:}}\HT{\mathbb{N}}\HSSpecial{\HSSym{\mskip1.5mu\}}}\;{}\<[37]%
\>[37]{}\HSSpecial{\HSSym{\{\mskip1.5mu} }\HV{h_1}\HSCon{\mathbin{:}}\HSCon{HopFrom}\;\HV{j_1}\HSSpecial{\HSSym{\mskip1.5mu\}}}\;\HSSpecial{\HSSym{\{\mskip1.5mu} }\HV{h_2}\HSCon{\mathbin{:}}\HSCon{HopFrom}\;\HV{j_2}\HSSpecial{\HSSym{\mskip1.5mu\}}}{}\<[E]%
\\
\>[18]{}\HSSym{\to} {}\<[18E]%
\>[22]{}\HV{\hbox{\it hop\guydash{}tgt}}\;\HV{h_2}\HSSym{<}\HV{\hbox{\it hop\guydash{}tgt}}\;\HV{h_1}\HSSym{\to} \HV{\hbox{\it hop\guydash{}tgt}}\;\HV{h_1}\HSSym{<}\HV{j_2}{}\<[E]%
\\
\>[18]{}\HSSym{\to} {}\<[18E]%
\>[22]{}\HV{j_1}\HSSym{\leq} \HV{j_2}{}\<[E]%
\ColumnHook
\end{hscode}\resethooks
\end{myhs}

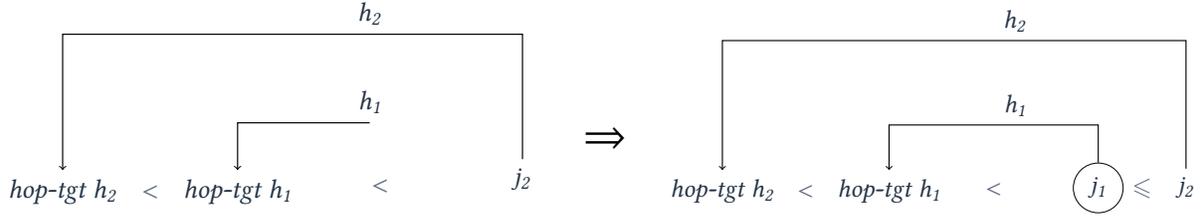
\begin{figure*}
\begin{center}
\resizebox{.9\textwidth}{!}{%
\begin{minipage}[t]{.40\textwidth}
\resizebox{\textwidth}{!}{%
\begin{tikzpicture}[ every node/.style={scale=1.6} ]
  \node                      (h1) {\ensuremath{\HV{h_1}}};
  \node [above = of h1]      (h2) {\ensuremath{\HV{h_2}}};
  \node [below left = of h1] (tgt1) {\ensuremath{\HV{\hbox{\it hop\guydash{}tgt}}\;\HV{h_1}}};
  \node [left = of tgt1]     (tgt2) {\ensuremath{\HV{\hbox{\it hop\guydash{}tgt}}\;\HV{h_2}}};
  \node [below right = of h1] (j1) {};
  \node [right = of j1]       (j2) {\ensuremath{\HV{j_2}}};

  \node (form) at ($ (tgt2)!0.5!(tgt1) $) {\ensuremath{\HSSym{<}}};
  \node (form2) at ($ (tgt1)!0.5!(j2)     $) {\ensuremath{\HSSym{<}}};

  \draw [line width=0.25mm, ->] (h1.south) -| (tgt1);
  \draw [line width=0.25mm, ->] (j2) |- (h2.south) -| (tgt2);
\end{tikzpicture}}
\end{minipage}%
\begin{minipage}[t]{.09\textwidth}
\begin{center}
\begin{tikzpicture}[ every node/.style={scale=1.6} ]
  \node (base)  at (0 , 0)   {}; 
  \node (base2) at (-.4 , 0) {}; 
  \node at (0,.8) {$\Rightarrow$};
\end{tikzpicture}
\end{center}
\end{minipage}%
\begin{minipage}[t]{.40\textwidth}
\resizebox{\textwidth}{!}{%
\begin{tikzpicture}[ every node/.style={scale=1.6} ]
  \node                      (h1) {\ensuremath{\HV{h_1}}};
  \node [above = of h1]      (h2) {\ensuremath{\HV{h_2}}};
  \node [below left = of h1] (tgt1) {\ensuremath{\HV{\hbox{\it hop\guydash{}tgt}}\;\HV{h_1}}};
  \node [left = of tgt1]     (tgt2) {\ensuremath{\HV{\hbox{\it hop\guydash{}tgt}}\;\HV{h_2}}};

  \node [draw,circle,below right = of h1] (j1) {\ensuremath{\HV{j_1}}};
  \node [right = of j1]       (j2) {\ensuremath{\HV{j_2}}};

  \node (form)  at ($ (tgt2)!0.5!(tgt1) $) {\ensuremath{\HSSym{<}}};
  \node (form2) at ($ (j1)!0.5!(j2)     $) {\ensuremath{\HSSym{\leq} }};
  \node (form3) at ($ (tgt1)!0.5!(j1)   $) {\ensuremath{\HSSym{<}}};

  \draw [line width=0.25mm, ->] (j1) |- (h1.south) -| (tgt1);
  \draw [line width=0.25mm, ->] (j2) |- (h2.south) -| (tgt2);
\end{tikzpicture}}
\end{minipage}}
\end{center}
\caption{Graphical representation of \ensuremath{\HV{\hbox{\it hop\guydash{}no\guydash{}cross}}}}
\label{fig:hopnocross}
\end{figure*}

The \ensuremath{\HSCon{DepRel}} definition requires a \ensuremath{\HSVar{maxlvl}} function, which
should return the number of hops from a given index, and a proof that
the level of 0 is 0, i.e., there are \emph{no} hops out of the initial
index.  For every index \ensuremath{\HSVar{i}} there are \ensuremath{\HSVar{maxlvl}\;\HSVar{i}} hops, encoded by the
\ensuremath{\HSCon{HopFrom}} datatype, and these hops have targets (\ensuremath{\HV{\hbox{\it hop\guydash{}tgt}}}). It is
important that \ensuremath{\HV{\hbox{\it hop\guydash{}tgt}}} is injective (\ensuremath{\HV{\hbox{\it hop\guydash{}inj}}}), but also that
taking a hop \emph{progresses}, that is, the target of a hop is always
smaller than the source (\ensuremath{\HV{\hbox{\it hop\guydash{}\!<}}}). It is also important that hops
never cross over each other, which is ensured by \ensuremath{\HV{\hbox{\it hop\guydash{}no\guydash{}cross}}}.  This
property is illustrated in \Cref{fig:hopnocross}; we examine it in
more detail when we describe how we ensure our concrete hop relation
respects it in \Cref{sec:concretemodel}.

  Constructing an inhabitant of \ensuremath{\HSCon{DepRel}} with \ensuremath{\HSVar{maxlvl}} as defined in \Cref{sec:skiplog}
and \ensuremath{\HV{\hbox{\it hop\guydash{}tgt}}\;\HSSpecial{\HSSym{\{\mskip1.5mu} }\HSVar{j}\HSSpecial{\HSSym{\mskip1.5mu\}}}\;\HSVar{h}} defined as $j - 2^h$ is mostly straightforward.  However, although one
can marvel at a geometric proof of the \ensuremath{\HV{\hbox{\it hop\guydash{}no\guydash{}cross}}} property for this hop relation on a napkin,
constructing a precise, machine checked proof of this property is a substantial challenge.
In the second part of our proof, presented in \Cref{sec:concretemodel}, we prove
that we can instantiate the abstract model with the hop relation used by our particular
implementation.

\subsection{Abstract Model}
\label{sec:refinedblocks}

  Our abstract model in Agda requires an arbitrary value of type
\ensuremath{\HSCon{DepRel}} as a module parameter, so the properties in this section
apply to any dependency relation implied by \ensuremath{\HSCon{DepRel}}'s requirements.  Next, we
introduce base notions necessary for constructing an AAOSL.


To avoid the inconsistency issues discussed in
\Cref{sec:wellformedness}, we represent an advancement proof using
a well-formed \emph{advancement path} that simply indicates which hops the
proof takes, along with the datum hashes of the source of each hop.

\begin{myhs}
\begin{hscode}\SaveRestoreHook
\column{B}{@{}>{\hspre}l<{\hspost}@{}}%
\column{3}{@{}>{\hspre}l<{\hspost}@{}}%
\column{9}{@{}>{\hspre}c<{\hspost}@{}}%
\column{9E}{@{}l@{}}%
\column{13}{@{}>{\hspre}l<{\hspost}@{}}%
\column{E}{@{}>{\hspre}l<{\hspost}@{}}%
\>[B]{}\HSKeyword{data}\;\HSCon{AdvPath}\HSCon{\mathbin{:}}\HT{\mathbb{N}}\HSSym{\to} \HT{\mathbb{N}}\HSSym{\to} \HSCon{Set}\;\HSKeyword{where}{}\<[E]%
\\
\>[B]{}\hsindent{3}{}\<[3]%
\>[3]{}\HSCon{Done}{}\<[9]%
\>[9]{}\HSCon{\mathbin{:}}{}\<[9E]%
\>[13]{}\HS{\forall}\;\HSSpecial{\HSSym{\{\mskip1.5mu} }\HSVar{i}\HSSpecial{\HSSym{\mskip1.5mu\}}}\HSSym{\to} \HSCon{AdvPath}\;\HSVar{i}\;\HSVar{i}{}\<[E]%
\\
\>[B]{}\hsindent{3}{}\<[3]%
\>[3]{}\HSCon{Hop}{}\<[9]%
\>[9]{}\HSCon{\mathbin{:}}{}\<[9E]%
\>[13]{}\HS{\forall}\;\HSSpecial{\HSSym{\{\mskip1.5mu} }\HSVar{j}\;\HSVar{i}\HSSpecial{\HSSym{\mskip1.5mu\}}}\HSSym{\to} \HSCon{Hash}\HSSym{\to} \HSSpecial{(}\HSVar{h}\HSCon{\mathbin{:}}\HSCon{HopFrom}\;\HSVar{j}\HSSpecial{)}{}\<[E]%
\\
\>[9]{}\HSSym{\to} {}\<[9E]%
\>[13]{}\HSCon{AdvPath}\;\HSSpecial{(}\HV{\hbox{\it hop\guydash{}tgt}}\;\HSVar{h}\HSSpecial{)}\;\HSVar{i}{}\<[E]%
\\
\>[9]{}\HSSym{\to} {}\<[9E]%
\>[13]{}\HSCon{AdvPath}\;\HSVar{j}\;\HSVar{i}{}\<[E]%
\ColumnHook
\end{hscode}\resethooks
\end{myhs}

A prover provides the authenticators associated with the indexes in an advancement
path in a separate \emph{view}, which we \emph{model} as a function from log indexes to hashes:

\begin{myhs}
\begin{hscode}\SaveRestoreHook
\column{B}{@{}>{\hspre}l<{\hspost}@{}}%
\column{E}{@{}>{\hspre}l<{\hspost}@{}}%
\>[B]{}\HSCon{View}\HSCon{\mathbin{:}}\HSCon{Set}{}\<[E]%
\\
\>[B]{}\HSCon{View}\HSSym{\mathrel{=}}\HT{\mathbb{N}}\HSSym{\to} \HSCon{Hash}{}\<[E]%
\ColumnHook
\end{hscode}\resethooks
\end{myhs}

This
way, a single authenticator is provided for each index, regardless of
how many times that index appears in the proof or is a dependency of
an index that appears in the proof. (A practical Haskell implementation
would represent a view as a partial map: \ensuremath{\HSCon{\HSCon{Data}.Map}\;\HSCon{Index}\;\HSCon{Hash}}; if the
map does not include an authenticator for an index required by the
advancement proof, then the advancement proof is rejected.)


  The \ensuremath{\HSVar{rebuild}} function in Agda operates over \ensuremath{\HSCon{View}}s instead
of receiving and returning a single hash. This is important because it
enables us to \emph{lookup} all rebuilt authenticators from a proof.
For example, take \ensuremath{\HSVar{a}\HSSym{\mathrel{=}}\HSCon{Hop}\;\HV{d_{8}}\;\HSNumeral{3}\;\HSSpecial{(}\HSCon{Hop}\;\HV{d_{4}}\;\HSNumeral{3}\;\HSCon{Done}\HSSpecial{)}}.  Calling
\ensuremath{\HSVar{rebuild}\;\HSVar{a}\;\HSVar{t}}, for some view \ensuremath{\HSVar{t}}, will return another view \ensuremath{\HSVar{t'}}, where
we can lookup the newly rebuilt value for 8 with \ensuremath{\HSVar{t'}\;\HSNumeral{8}}, and also
check the recursively rebuilt hash for 4, used in the computation of \ensuremath{\HSVar{t'}\;\HSNumeral{8}},
by calling \ensuremath{\HSVar{t'}\;\HSNumeral{4}}.

\begin{myhs}
\begin{hscode}\SaveRestoreHook
\column{B}{@{}>{\hspre}l<{\hspost}@{}}%
\column{20}{@{}>{\hspre}l<{\hspost}@{}}%
\column{30}{@{}>{\hspre}l<{\hspost}@{}}%
\column{E}{@{}>{\hspre}l<{\hspost}@{}}%
\>[B]{}\HSVar{rebuild}\HSCon{\mathbin{:}}\HS{\forall}\;\HSSpecial{\HSSym{\{\mskip1.5mu} }\HSVar{i}\;\HSVar{j}\HSSpecial{\HSSym{\mskip1.5mu\}}}\HSSym{\to} \HSCon{AdvPath}\;\HSVar{j}\;\HSVar{i}\HSSym{\to} \HSCon{View}\HSSym{\to} \HSCon{View}{}\<[E]%
\\
\>[B]{}\HSVar{rebuild}\;\HSCon{Done}\;{}\<[30]%
\>[30]{}\HSVar{view}\HSSym{\mathrel{=}}\HSVar{view}{}\<[E]%
\\
\>[B]{}\HSVar{rebuild}\;\HSSpecial{(}\HSCon{Hop}\;\HSSpecial{\HSSym{\{\mskip1.5mu} }\HSVar{j}\HSSym{\mathrel{=}}\HSVar{j}\HSSpecial{\HSSym{\mskip1.5mu\}}}\;\HSVar{x}\;\HSVar{h}\;\HSVar{a}\HSSpecial{)}\;{}\<[30]%
\>[30]{}\HSVar{view}\HSSym{\mathrel{=}}{}\<[E]%
\\
\>[B]{}\hsindent{20}{}\<[20]%
\>[20]{}\HSKeyword{let}\;\HSVar{view'}\HSSym{\mathrel{=}}\HSVar{rebuild}\;\HSVar{a}\;\HSVar{view}{}\<[E]%
\\
\>[B]{}\hsindent{20}{}\<[20]%
\>[20]{}\HSKeyword{in}\;\HSVar{insert}\;\HSVar{j}\;\HSSpecial{(}\HV{\hbox{\it \calcauth}}\;\HSVar{j}\;\HSVar{x}\;\HSVar{view'}\HSSpecial{)}\;\HSVar{view'}{}\<[E]%
\ColumnHook
\end{hscode}\resethooks
\end{myhs}

  Here, \ensuremath{\HSVar{insert}\;\HSVar{k}\;\HSVar{v}\;\HSVar{f}} inserts a new key-value pair into a map. Note how
we are computing the authenticator for \ensuremath{\HSVar{j}} using the view that results
from recursively rebuilding the proof.

  Next we look at encoding \emph{membership proofs}, which reuse the
constructions we have just seen.  Although similar, membership proofs
and advancement proofs work in different ``directions''. An
advancement proof is used to prove to someone who trusts the hash of
index $i$ that the skiplog can be extended to index $j > i$ in a way
that is consistent with the log up to index $i$.  Membership proofs,
on the other hand, prove to a verifier who trusts the hash of $j > i$,
that a given datum is at index $i$.

  As described above, when rebuilding an \emph{advancement proof} from
$i$ to $j$, a verifier already knows and trusts an authenticator for
index $i$.  In contrast, with \emph{membership proofs}, the prover
must provide sufficient additional information for the verifier to
compute the authenticator for index $i$.  The datum hash is provided
explicitly as the second component of the membership proof.  The third component ensures
that we do not construct a \ensuremath{\HSCon{MbrPath}} for index 0, which does not have
associated data. (Agda uses \ensuremath{\mathbin{\HT{\times}}} to express product types.)

\begin{myhs}
\begin{hscode}\SaveRestoreHook
\column{B}{@{}>{\hspre}l<{\hspost}@{}}%
\column{E}{@{}>{\hspre}l<{\hspost}@{}}%
\>[B]{}\HT{\mathit{MbrPath}}\HSCon{\mathbin{:}}\HT{\mathbb{N}}\HSSym{\to} \HT{\mathbb{N}}\HSSym{\to} \HSCon{Set}{}\<[E]%
\\
\>[B]{}\HT{\mathit{MbrPath}}\;\HSVar{j}\;\HSVar{i}\HSSym{\mathrel{=}}\HSCon{AdvPath}\;\HSVar{j}\;\HSVar{i}\;\mathbin{\HT{\times}}\;\HSCon{Hash}\;\mathbin{\HT{\times}}\;\HSVar{i}\HSSym{\not\equiv} \HSNumeral{0}{}\<[E]%
\\[\blanklineskip]%
\>[B]{}\HV{\hbox{\it datum\guydash{}hash}}\HSCon{\mathbin{:}}\HS{\forall}\;\HSSpecial{\HSSym{\{\mskip1.5mu} }\HSVar{j}\;\HSVar{i}\HSSpecial{\HSSym{\mskip1.5mu\}}}\HSSym{\to} \HT{\mathit{MbrPath}}\;\HSVar{j}\;\HSVar{i}\HSSym{\to} \HSCon{Hash}{}\<[E]%
\\
\>[B]{}\HV{\hbox{\it datum\guydash{}hash}}\;\HSSpecial{(}\HSSym{\anonymous} \HSSpecial{,}\HSVar{dat}\HSSpecial{,}\HSSym{\anonymous} \HSSpecial{)}\HSSym{\mathrel{=}}\HSVar{dat}{}\<[E]%
\\[\blanklineskip]%
\>[B]{}\HV{\hbox{\it mbr\guydash{}path}}\HSCon{\mathbin{:}}\HS{\forall}\;\HSSpecial{\HSSym{\{\mskip1.5mu} }\HSVar{j}\;\HSVar{i}\HSSpecial{\HSSym{\mskip1.5mu\}}}\HSSym{\to} \HT{\mathit{MbrPath}}\;\HSVar{j}\;\HSVar{i}\HSSym{\to} \HSCon{AdvPath}\;\HSVar{j}\;\HSVar{i}{}\<[E]%
\\
\>[B]{}\HV{\hbox{\it mbr\guydash{}path}}\;\HSSpecial{(}\HSVar{p}\HSSpecial{,}\HSSym{\anonymous} \HSSpecial{,}\HSSym{\anonymous} \HSSpecial{)}\HSSym{\mathrel{=}}\HSVar{p}{}\<[E]%
\ColumnHook
\end{hscode}\resethooks
\end{myhs}

In addition to a \ensuremath{\HT{\mathit{MbrPath}}}, the prover also provides a \ensuremath{\HSCon{View}} that
the verifier uses to rebuild the \ensuremath{\HT{\mathit{MbrPath}}}:

\begin{myhs}
\begin{hscode}\SaveRestoreHook
\column{B}{@{}>{\hspre}l<{\hspost}@{}}%
\column{22}{@{}>{\hspre}l<{\hspost}@{}}%
\column{E}{@{}>{\hspre}l<{\hspost}@{}}%
\>[B]{}\HSVar{insertAuth}\HSCon{\mathbin{:}}\HSCon{View}\HSSym{\to} \HT{\mathbb{N}}\HSSym{\to} \HSCon{Hash}\HSSym{\to} \HSCon{View}{}\<[E]%
\\
\>[B]{}\HSVar{insertAuth}\;\HSVar{t}\;\HSVar{i}\;\HSVar{d}\HSSym{\mathrel{=}}\HSVar{insert}\;\HSVar{i}\;\HSSpecial{(}\HV{\hbox{\it \calcauth}}\;\HSVar{i}\;\HSVar{d}\;\HSVar{t}\HSSpecial{)}\;\HSVar{t}{}\<[E]%
\\[\blanklineskip]%
\>[B]{}\HV{\mathit{rebuildMbr}}\HSCon{\mathbin{:}}\HS{\forall}\;\HSSpecial{\HSSym{\{\mskip1.5mu} }\HSVar{j}\;\HSVar{i}\HSSpecial{\HSSym{\mskip1.5mu\}}}\HSSym{\to} \HT{\mathit{MbrPath}}\;\HSVar{j}\;\HSVar{i}\HSSym{\to} \HSCon{View}\HSSym{\to} \HSCon{View}{}\<[E]%
\\
\>[B]{}\HV{\mathit{rebuildMbr}}\;\HSSpecial{\HSSym{\{\mskip1.5mu} }\HSVar{j}\HSSpecial{\HSSym{\mskip1.5mu\}}}\;\HSSpecial{\HSSym{\{\mskip1.5mu} }\HSVar{i}\HSSpecial{\HSSym{\mskip1.5mu\}}}\;\HSVar{mp}\;\HSVar{t}\HSSym{\mathrel{=}}{}\<[E]%
\\
\>[B]{}\hsindent{22}{}\<[22]%
\>[22]{}\HSVar{rebuild}\;\HSSpecial{(}\HV{\hbox{\it mbr\guydash{}path}}\;\HSVar{mp}\HSSpecial{)}\;\HSSpecial{(}\HSVar{insertAuth}\;\HSVar{t}\;\HSVar{i}\;\HSSpecial{(}\HV{\hbox{\it datum\guydash{}hash}}\;\HSVar{mp}\HSSpecial{)}\HSSpecial{)}{}\<[E]%
\ColumnHook
\end{hscode}\resethooks
\end{myhs}

Unlike with advancement proofs, the provided \ensuremath{\HSCon{View}} must also include
authenticators for each dependency of index $i$ that---together with the datum hash---enable the
verifier to compute the authenticator
for index $i$.

\paragraph{Reasoning about collision resistance}
\label{sec:collisionresist}

  The proofs presented in the next sections establish desired properties modulo hash
collisions.  Hence, it is important to model the
possibility that an adversary may find hash collisions.
One option for doing this in a constructive setting is to
carry around hash collisions evidenced in existentials~\cite{BrunTraytel2019}. We use
the \ensuremath{\HSCon{HashBroke}} datatype:

\begin{myhs}
\begin{hscode}\SaveRestoreHook
\column{B}{@{}>{\hspre}l<{\hspost}@{}}%
\column{E}{@{}>{\hspre}l<{\hspost}@{}}%
\>[B]{}\HSCon{HashBroke}\HSCon{\mathbin{:}}\HSCon{Set}{}\<[E]%
\\
\>[B]{}\HSCon{HashBroke}\HSSym{\mathrel{=}}\HT{\exists[}\HSVar{x}\HSSpecial{,}\HSVar{y}\HT{]}\;\HSSpecial{(}\HSVar{x}\HSSym{\not\equiv} \HSVar{y}\;\mathbin{\HT{\times}}\;\HSVar{hash}\;\HSVar{x}\HSSym{\equiv} \HSVar{hash}\;\HSVar{y}\HSSpecial{)}{}\<[E]%
\ColumnHook
\end{hscode}\resethooks
\end{myhs}

  This becomes necessary when reasoning about injectivity of authenticators,
which is central to our proofs. If two advancement proofs \ensuremath{\HV{a_{1}}\HSCon{\mathbin{:}}\HSCon{AdvPath}\;\HSVar{j}\;\HV{i_1}} and \ensuremath{\HV{a_{2}}\HSCon{\mathbin{:}}\HSCon{AdvPath}\;\HSVar{j}\;\HV{i_2}} rebuild using views \ensuremath{\HV{t_1}}
and \ensuremath{\HV{t_2}} respectively to the same authenticator at \ensuremath{\HSVar{j}} (i.e., \ensuremath{\HSVar{rebuild}\;\HV{a_{1}}\;\HV{t_1}\;\HSVar{j}\HSSym{\equiv} \HSVar{rebuild}\;\HV{a_{2}}\;\HV{t_2}\;\HSVar{j}}), then \emph{unless there is a hash collision}, \ensuremath{\HV{a_{1}}}
and \ensuremath{\HV{a_{2}}} provide the same datum hash for \ensuremath{\HSVar{j}} and both rebuilds use the
same authenticators for all dependencies of \ensuremath{\HSVar{j}}.

  This conclusion is reached via two injectivity lemmas applied
to the evidence that the rebuilt views agree at \ensuremath{\HSVar{j}}.  The first
establishes that (unless \ensuremath{\HSCon{HashBroke}}), the datum hashes passed to
\ensuremath{\HV{\hbox{\it \calcauth}}} to compute the authenticators for index \ensuremath{\HSVar{j}} for the
respective advancement proofs are equal:

\begin{myhs}
\begin{hscode}\SaveRestoreHook
\column{B}{@{}>{\hspre}l<{\hspost}@{}}%
\column{11}{@{}>{\hspre}l<{\hspost}@{}}%
\column{16}{@{}>{\hspre}c<{\hspost}@{}}%
\column{16E}{@{}l@{}}%
\column{19}{@{}>{\hspre}l<{\hspost}@{}}%
\column{E}{@{}>{\hspre}l<{\hspost}@{}}%
\>[B]{}\HV{\hbox{\it \calcauth\guydash{}inj\guydash{}1}}{}\<[16]%
\>[16]{}\HSCon{\mathbin{:}}{}\<[16E]%
\>[19]{}\HSSpecial{\HSSym{\{\mskip1.5mu} }\HSVar{j}\HSCon{\mathbin{:}}\HT{\mathbb{N}}\HSSpecial{\HSSym{\mskip1.5mu\}}}\;\HSSpecial{\HSSym{\{\mskip1.5mu} }\HV{h_1}\;\HV{h_2}\HSCon{\mathbin{:}}\HSCon{Hash}\HSSpecial{\HSSym{\mskip1.5mu\}}}\;\HSSpecial{\HSSym{\{\mskip1.5mu} }\HV{t_1}\;\HV{t_2}\HSCon{\mathbin{:}}\HSCon{View}\HSSpecial{\HSSym{\mskip1.5mu\}}}{}\<[E]%
\\
\>[B]{}\hsindent{11}{}\<[11]%
\>[11]{}\HSSym{\to} \HSVar{j}\HSSym{\not\equiv} \HSNumeral{0}\HSSym{\to} \HV{\hbox{\it \calcauth}}\;\HSVar{j}\;\HV{h_1}\;\HV{t_1}\HSSym{\equiv} \HV{\hbox{\it \calcauth}}\;\HSVar{j}\;\HV{h_2}\;\HV{t_2}{}\<[E]%
\\
\>[B]{}\hsindent{11}{}\<[11]%
\>[11]{}\HSSym{\to} \HSCon{Either}\;\HSCon{HashBroke}\;\HSSpecial{(}\HV{h_1}\HSSym{\equiv} \HV{h_2}\HSSpecial{)}{}\<[E]%
\ColumnHook
\end{hscode}\resethooks
\end{myhs}

 The \ensuremath{\HV{\hbox{\it \calcauth\guydash{}inj\guydash{}1}}} lemma is says that, if the \ensuremath{\HSVar{datumDig}} values
passed to two invocations of \ensuremath{\HV{\hbox{\it \calcauth}}} are different, but they return
the same hash, then we have a hash collision.  Its proof is somewhat
more involved, though, because \ensuremath{\HV{\hbox{\it \calcauth}}} uses concatenations and
encoding of natural numbers into bytestrings.  Therefore the proof
requires reasoning about injectivity of encoding and of concatenation
of fixed-size byte strings.  We make a simplifying assumption that
indexes are encoded into a fixed (but unspecified) number of bits.
This assumption is reasonable in practice and not difficult to relax,
but this would not add significant value to our proof.

Once we have \ensuremath{\HV{\hbox{\it \calcauth\guydash{}inj\guydash{}1}}} to determine that the datum hashes are equal (unless \ensuremath{\HSCon{HashBroke}}),
we define the other injectivity lemma to conclude that the lists of digests
provided to the \ensuremath{\HSVar{lvldigs}} arguments of the respective \ensuremath{\HV{\hbox{\it \calcauth}}} invocations are also equal:

\begin{myhs}
\begin{hscode}\SaveRestoreHook
\column{B}{@{}>{\hspre}l<{\hspost}@{}}%
\column{11}{@{}>{\hspre}l<{\hspost}@{}}%
\column{16}{@{}>{\hspre}l<{\hspost}@{}}%
\column{E}{@{}>{\hspre}l<{\hspost}@{}}%
\>[B]{}\HV{\hbox{\it \calcauth\guydash{}inj\guydash{}2}}{}\<[16]%
\>[16]{}\HSCon{\mathbin{:}}\HS{\forall}\;\HSSpecial{\HSSym{\{\mskip1.5mu} }\HSVar{j}\;\HSVar{h}\;\HV{t_1}\;\HV{t_2}\HSSpecial{\HSSym{\mskip1.5mu\}}}\HSSym{\to} \HV{\hbox{\it \calcauth}}\;\HSVar{j}\;\HSVar{h}\;\HV{t_1}\HSSym{\equiv} \HV{\hbox{\it \calcauth}}\;\HSVar{j}\;\HSVar{h}\;\HV{t_2}{}\<[E]%
\\
\>[B]{}\hsindent{11}{}\<[11]%
\>[11]{}\HSSym{\to} \HSCon{Either}\;\HSCon{HashBroke}\;\HSSpecial{(}\HSCon{Agree}\;\HV{t_1}\;\HV{t_2}\;\HSSpecial{(}\HSVar{depsof}\;\HSVar{j}\HSSpecial{)}\HSSpecial{)}{}\<[E]%
\ColumnHook
\end{hscode}\resethooks
\end{myhs}

  Where \ensuremath{\HSCon{Agree}\;\HV{t_1}\;\HV{t_2}\;\HSVar{xs}} states that \ensuremath{\HV{t_1}\;\HSVar{x}\HSSym{\equiv} \HV{t_2}\;\HSVar{x}} for every \ensuremath{\HSVar{x}\;\HS{\in}\;\HSVar{xs}} and
\ensuremath{\HSVar{depsof}} returns the dependencies of a given index \ensuremath{\HSVar{j}} by enumerating all hops from \ensuremath{\HSVar{j}}
and computing their target. In case \ensuremath{\HSVar{j}} is zero, it has no dependencies.

\begin{myhs}
\begin{hscode}\SaveRestoreHook
\column{B}{@{}>{\hspre}l<{\hspost}@{}}%
\column{17}{@{}>{\hspre}l<{\hspost}@{}}%
\column{E}{@{}>{\hspre}l<{\hspost}@{}}%
\>[B]{}\HSVar{depsof}\HSCon{\mathbin{:}}\HT{\mathbb{N}}\HSSym{\to} \HSCon{List}\;\HT{\mathbb{N}}{}\<[E]%
\\
\>[B]{}\HSVar{depsof}\;\HSNumeral{0}{}\<[17]%
\>[17]{}\HSSym{\mathrel{=}}\HSSpecial{\HSSym{[\mskip1.5mu} }\HSSpecial{\HSSym{\mskip1.5mu]}}{}\<[E]%
\\
\>[B]{}\HSVar{depsof}\;\HSSpecial{(}\HSVar{suc}\;\HSVar{i}\HSSpecial{)}{}\<[17]%
\>[17]{}\HSSym{\mathrel{=}}\HSVar{map}\;\HV{\hbox{\it hop\guydash{}tgt}}\;\HSSpecial{(}\HSVar{finsUpTo}\;\HSSpecial{(}\HSVar{lvlof}\;\HSSpecial{(}\HSVar{suc}\;\HSVar{i}\HSSpecial{)}\HSSpecial{)}\HSSpecial{)}{}\<[E]%
\ColumnHook
\end{hscode}\resethooks
\end{myhs}

   We have proven that both the original definition of \ensuremath{\HV{\hbox{\it \calcauth}}} used
by Maniatis and Baker~\cite{Baker2003} and the simpler variant
described in~\Cref{sec:skiplog} satisfy these two injectivity
lemmas. In fact, any definition of \ensuremath{\HV{\hbox{\it \calcauth}}} that satisfies \ensuremath{\HV{\hbox{\it \calcauth\guydash{}inj\guydash{}1}}}
and \ensuremath{\HV{\hbox{\it \calcauth\guydash{}inj\guydash{}2}}} could be used to construct an AAOSL that enjoys all the
properties we will explore next.

\subsection{Proving \ensuremath{\HV{\textsc{AgreeOnCommon}}}}

  The \ensuremath{\HV{\textsc{AgreeOnCommon}}} property states that, given two advancement
proofs into index \ensuremath{\HSVar{j}}, if both proofs rebuild to the same
hash, then either these proofs agree on the authenticators
of \emph{every} index that they visit in common,
or the adversary found a hash collision.  This
is stated in Agda as follows:

\begin{myhs}
\begin{hscode}\SaveRestoreHook
\column{B}{@{}>{\hspre}l<{\hspost}@{}}%
\column{3}{@{}>{\hspre}l<{\hspost}@{}}%
\column{13}{@{}>{\hspre}l<{\hspost}@{}}%
\column{14}{@{}>{\hspre}l<{\hspost}@{}}%
\column{E}{@{}>{\hspre}l<{\hspost}@{}}%
\>[B]{}\HV{\textsc{AgreeOnCommon}}{}\<[E]%
\\
\>[B]{}\hsindent{3}{}\<[3]%
\>[3]{}\HSCon{\mathbin{:}}\HS{\forall}\;{}\<[13]%
\>[13]{}\HSSpecial{\HSSym{\{\mskip1.5mu} }\HSVar{j}\;\HV{i_1}\;\HV{i_2}\HSSpecial{\HSSym{\mskip1.5mu\}}}\;\HSSpecial{\HSSym{\{\mskip1.5mu} }\HV{t_1}\;\HV{t_2}\HSCon{\mathbin{:}}\HSCon{View}\HSSpecial{\HSSym{\mskip1.5mu\}}}\;\HSSpecial{\HSSym{\{\mskip1.5mu} }\HV{a_1}\HSCon{\mathbin{:}}\HSCon{AdvPath}\;\HSVar{j}\;\HV{i_1}\HSSpecial{\HSSym{\mskip1.5mu\}}}\;\HSSpecial{\HSSym{\{\mskip1.5mu} }\HV{a_2}\HSCon{\mathbin{:}}\HSCon{AdvPath}\;\HSVar{j}\;\HV{i_2}\HSSpecial{\HSSym{\mskip1.5mu\}}}{}\<[E]%
\\
\>[B]{}\hsindent{3}{}\<[3]%
\>[3]{}\HSSym{\to} \HSVar{rebuild}\;\HV{a_1}\;\HV{t_1}\;\HSVar{j}\HSSym{\equiv} \HSVar{rebuild}\;\HV{a_2}\;\HV{t_2}\;\HSVar{j}{}\<[E]%
\\
\>[B]{}\hsindent{3}{}\<[3]%
\>[3]{}\HSSym{\to} \HSSpecial{\HSSym{\{\mskip1.5mu} }\HSVar{i}\HSCon{\mathbin{:}}\HT{\mathbb{N}}\HSSpecial{\HSSym{\mskip1.5mu\}}}\HSSym{\to} \HSVar{i}\;\HT{\epsilon_{\textsc{ap}}}\;\HV{a_1}\HSSym{\to} \HSVar{i}\;\HT{\epsilon_{\textsc{ap}}}\;\HV{a_2}{}\<[E]%
\\
\>[B]{}\hsindent{3}{}\<[3]%
\>[3]{}\HSSym{\to} \HSCon{Either}\;{}\<[14]%
\>[14]{}\HSCon{HashBroke}\;\HSSpecial{(}\HSVar{rebuild}\;\HV{a_1}\;\HV{t_1}\;\HSVar{i}\HSSym{\equiv} \HSVar{rebuild}\;\HV{a_2}\;\HV{t_2}\;\HSVar{i}\HSSpecial{)}{}\<[E]%
\ColumnHook
\end{hscode}\resethooks
\end{myhs}

Here, \ensuremath{\HT{\_}\HT{\epsilon_{\textsc{ap}}}\HT{\_}} encodes the relation of visited indexes of a given
advancement proof. A value of type \ensuremath{\HSVar{k}\;\HT{\epsilon_{\textsc{ap}}}\;\HSVar{a}} exists iff index \ensuremath{\HSVar{k}} is
visited by advancement proof \ensuremath{\HSVar{a}}:

\begin{myhs}
\begin{hscode}\SaveRestoreHook
\column{B}{@{}>{\hspre}l<{\hspost}@{}}%
\column{4}{@{}>{\hspre}l<{\hspost}@{}}%
\column{6}{@{}>{\hspre}l<{\hspost}@{}}%
\column{20}{@{}>{\hspre}l<{\hspost}@{}}%
\column{23}{@{}>{\hspre}l<{\hspost}@{}}%
\column{E}{@{}>{\hspre}l<{\hspost}@{}}%
\>[4]{}\HSKeyword{data}\;\HT{\_}\HT{\epsilon_{\textsc{ap}}}\HT{\_}\;\HSSpecial{(}\HSVar{k}\HSCon{\mathbin{:}}\HT{\mathbb{N}}\HSSpecial{)}\HSCon{\mathbin{:}}\HSSpecial{\HSSym{\{\mskip1.5mu} }\HSVar{j}\;\HSVar{i}\HSCon{\mathbin{:}}\HT{\mathbb{N}}\HSSpecial{\HSSym{\mskip1.5mu\}}}\HSSym{\to} \HSCon{AdvPath}\;\HSVar{j}\;\HSVar{i}\HSSym{\to} \HSCon{Set}\;\HSKeyword{where}{}\<[E]%
\\
\>[4]{}\hsindent{2}{}\<[6]%
\>[6]{}\HSVar{hereTgtDone}{}\<[20]%
\>[20]{}\HSCon{\mathbin{:}}\HSVar{k}\;\HT{\epsilon_{\textsc{ap}}}\;\HSSpecial{(}\HSCon{Done}\;\HSSpecial{\HSSym{\{\mskip1.5mu} }\HSVar{k}\HSSpecial{\HSSym{\mskip1.5mu\}}}\HSSpecial{)}{}\<[E]%
\\
\>[4]{}\hsindent{2}{}\<[6]%
\>[6]{}\HSVar{hereTgtHop}{}\<[20]%
\>[20]{}\HSCon{\mathbin{:}}{}\<[23]%
\>[23]{}\HSSpecial{\HSSym{\{\mskip1.5mu} }\HSVar{i}\HSCon{\mathbin{:}}\HT{\mathbb{N}}\HSSpecial{\HSSym{\mskip1.5mu\}}}\;\HSSpecial{\HSSym{\{\mskip1.5mu} }\HSVar{d}\HSCon{\mathbin{:}}\HSCon{Hash}\HSSpecial{\HSSym{\mskip1.5mu\}}}\;\HSSpecial{\HSSym{\{\mskip1.5mu} }\HSVar{h}\HSCon{\mathbin{:}}\HSCon{HopFrom}\;\HSVar{k}\HSSpecial{\HSSym{\mskip1.5mu\}}}{}\<[E]%
\\
\>[23]{}\HSSpecial{\HSSym{\{\mskip1.5mu} }\HSVar{a}\HSCon{\mathbin{:}}\HSCon{AdvPath}\;\HSSpecial{(}\HV{\hbox{\it hop\guydash{}tgt}}\;\HSVar{h}\HSSpecial{)}\;\HSVar{i}\HSSpecial{\HSSym{\mskip1.5mu\}}}{}\<[E]%
\\
\>[20]{}\HSSym{\to} \HSVar{k}\;\HT{\epsilon_{\textsc{ap}}}\;\HSSpecial{(}\HSCon{Hop}\;\HSVar{d}\;\HSVar{h}\;\HSVar{a}\HSSpecial{)}{}\<[E]%
\\[\blanklineskip]%
\>[4]{}\hsindent{2}{}\<[6]%
\>[6]{}\HSVar{step}{}\<[20]%
\>[20]{}\HSCon{\mathbin{:}}{}\<[23]%
\>[23]{}\HSSpecial{\HSSym{\{\mskip1.5mu} }\HSVar{i}\;\HSVar{j}\HSCon{\mathbin{:}}\HT{\mathbb{N}}\HSSpecial{\HSSym{\mskip1.5mu\}}}\;\HSSpecial{\HSSym{\{\mskip1.5mu} }\HSVar{d}\HSCon{\mathbin{:}}\HSCon{Hash}\HSSpecial{\HSSym{\mskip1.5mu\}}}\;\HSSpecial{\HSSym{\{\mskip1.5mu} }\HSVar{h}\HSCon{\mathbin{:}}\HSCon{HopFrom}\;\HSVar{j}\HSSpecial{\HSSym{\mskip1.5mu\}}}{}\<[E]%
\\
\>[23]{}\HSSpecial{\HSSym{\{\mskip1.5mu} }\HSVar{a}\HSCon{\mathbin{:}}\HSCon{AdvPath}\;\HSSpecial{(}\HV{\hbox{\it hop\guydash{}tgt}}\;\HSVar{h}\HSSpecial{)}\;\HSVar{i}\HSSpecial{\HSSym{\mskip1.5mu\}}}{}\<[E]%
\\
\>[20]{}\HSSym{\to} \HSVar{k}\HSSym{\not\equiv} \HSVar{j}\HSSym{\to} \HSVar{k}\;\HT{\epsilon_{\textsc{ap}}}\;\HSVar{a}\HSSym{\to} \HSVar{k}\;\HT{\epsilon_{\textsc{ap}}}\;\HSSpecial{(}\HSCon{Hop}\;\HSVar{d}\;\HSVar{h}\;\HSVar{a}\HSSpecial{)}{}\<[E]%
\ColumnHook
\end{hscode}\resethooks
\end{myhs}

  The proof of \ensuremath{\HV{\textsc{AgreeOnCommon}}} follows by induction on the
commonly visited indexes of the advancement proofs under scrutiny.
We present the proof in two parts. First, we explain how we encode this induction
principle in a datatype, \ensuremath{\HSCon{AOC}}, and why it provides
sufficient information to prove \ensuremath{\HV{\textsc{AgreeOnCommon}}}. Later we
prove that we can always construct a value of type \ensuremath{\HSCon{AOC}} for
the advancement proofs expected by \ensuremath{\HV{\textsc{AgreeOnCommon}}}.

\subsubsection*{Using the \ensuremath{\HSCon{AOC}} data type to prove \ensuremath{\HV{\textsc{AgreeOnCommon}}}}

   In this section we focus on how \ensuremath{\HSCon{AOC}} captures enough information
to conclude that two advancement proofs rebuild to the same hash at
\emph{every} index they both visit.  Intuitively, a value of type \ensuremath{\HSCon{AOC}\;\HV{t_1}\;\HV{t_2}\;\HV{a_{1}}\;\HV{a_{2}}} represents a list of all
indexes visited by both \ensuremath{\HV{a_{1}}} and \ensuremath{\HV{a_{2}}}, along with sufficient
evidence to prove that the
respective views obtained by rebuilding \ensuremath{\HV{a_{1}}} using \ensuremath{\HV{t_1}} and rebuilding
\ensuremath{\HV{a_{2}}} using \ensuremath{\HV{t_2}} agree at each of these indexes.
This can be seen in the type of \ensuremath{\HV{\hbox{\it aoc\guydash{}correct}}}, below.

\begin{myhs}
\begin{hscode}\SaveRestoreHook
\column{B}{@{}>{\hspre}l<{\hspost}@{}}%
\column{3}{@{}>{\hspre}l<{\hspost}@{}}%
\column{E}{@{}>{\hspre}l<{\hspost}@{}}%
\>[B]{}\HV{\hbox{\it aoc\guydash{}correct}}{}\<[E]%
\\
\>[B]{}\hsindent{3}{}\<[3]%
\>[3]{}\HSCon{\mathbin{:}}\HS{\forall}\;\HSSpecial{\HSSym{\{\mskip1.5mu} }\HSVar{j}\;\HV{i_1}\;\HV{i_2}\HSSpecial{\HSSym{\mskip1.5mu\}}}\;\HSSpecial{\HSSym{\{\mskip1.5mu} }\HV{a_{1}}\HSCon{\mathbin{:}}\HSCon{AdvPath}\;\HSVar{j}\;\HV{i_1}\HSSpecial{\HSSym{\mskip1.5mu\}}}\;\HSSpecial{\HSSym{\{\mskip1.5mu} }\HV{a_{2}}\HSCon{\mathbin{:}}\HSCon{AdvPath}\;\HSVar{j}\;\HV{i_2}\HSSpecial{\HSSym{\mskip1.5mu\}}}{}\<[E]%
\\
\>[B]{}\hsindent{3}{}\<[3]%
\>[3]{}\HSSym{\to} \HSSpecial{\HSSym{\{\mskip1.5mu} }\HV{t_1}\;\HV{t_2}\HSCon{\mathbin{:}}\HSCon{View}\HSSpecial{\HSSym{\mskip1.5mu\}}}{}\<[E]%
\\
\>[B]{}\hsindent{3}{}\<[3]%
\>[3]{}\HSSym{\to} \HSCon{AOC}\;\HV{t_1}\;\HV{t_2}\;\HV{a_{1}}\;\HV{a_{2}}{}\<[E]%
\\
\>[B]{}\hsindent{3}{}\<[3]%
\>[3]{}\HSSym{\to} \HSSpecial{\HSSym{\{\mskip1.5mu} }\HSVar{i}\HSCon{\mathbin{:}}\HT{\mathbb{N}}\HSSpecial{\HSSym{\mskip1.5mu\}}}\HSSym{\to} \HSVar{i}\;\HT{\epsilon_{\textsc{ap}}}\;\HV{a_{1}}\HSSym{\to} \HSVar{i}\;\HT{\epsilon_{\textsc{ap}}}\;\HV{a_{2}}{}\<[E]%
\\
\>[B]{}\hsindent{3}{}\<[3]%
\>[3]{}\HSSym{\to} \HSCon{Either}\;\HSCon{HashBroke}\;\HSSpecial{(}\HSVar{rebuild}\;\HV{a_{1}}\;\HV{t_1}\;\HSVar{i}\HSSym{\equiv} \HSVar{rebuild}\;\HV{a_{2}}\;\HV{t_2}\;\HSVar{i}\HSSpecial{)}{}\<[E]%
\ColumnHook
\end{hscode}\resethooks
\end{myhs}

  The proof of \ensuremath{\HV{\hbox{\it aoc\guydash{}correct}}} is a long but straightforward induction on
\ensuremath{\HSCon{AOC}} and \ensuremath{\HSVar{i}\;\HT{\epsilon_{\textsc{ap}}}\;\HV{a_{1}}} and \ensuremath{\HSVar{i}\;\HT{\epsilon_{\textsc{ap}}}\;\HV{a_{2}}}. The more intellectual step lies in
defining \ensuremath{\HSCon{AOC}} to have the right structure to make the rest of the proof
straightforward. Intuitively, \ensuremath{\HSCon{AOC}} is like a \ensuremath{\HSVar{zip}} over advancement proofs:
it puts in evidence the commonly visited indexes.
Its type signature reveals that it is a relation between \ensuremath{\HSCon{AdvPath}}s from a common index \ensuremath{\HSVar{j}}:

\begin{myhs}
\begin{hscode}\SaveRestoreHook
\column{B}{@{}>{\hspre}l<{\hspost}@{}}%
\column{5}{@{}>{\hspre}l<{\hspost}@{}}%
\column{E}{@{}>{\hspre}l<{\hspost}@{}}%
\>[B]{}\HSKeyword{data}\;\HSCon{AOC}\;\HSSpecial{(}\HV{t_1}\;\HV{t_2}\HSCon{\mathbin{:}}\HSCon{View}\HSSpecial{)}{}\<[E]%
\\
\>[B]{}\hsindent{5}{}\<[5]%
\>[5]{}\HSCon{\mathbin{:}}\HS{\forall}\;\HSSpecial{\HSSym{\{\mskip1.5mu} }\HV{i_1}\;\HV{i_2}\;\HSVar{j}\HSSpecial{\HSSym{\mskip1.5mu\}}}\HSSym{\to} \HSCon{AdvPath}\;\HSVar{j}\;\HV{i_1}\HSSym{\to} \HSCon{AdvPath}\;\HSVar{j}\;\HV{i_2}\HSSym{\to} \HSCon{Set}\;\HSKeyword{where}{}\<[E]%
\ColumnHook
\end{hscode}\resethooks
\end{myhs}

  Each of \ensuremath{\HSCon{AOC}}'s constructors represents one possible situation
with advancement proofs that share a common index.  There are three
base cases that handle pairs of advancement proofs \ensuremath{\HV{a_{1}}} and \ensuremath{\HV{a_{2}}} that
have one or two common indexes: (i) both proofs are \ensuremath{\HSCon{Done}}, (ii) \ensuremath{\HV{a_{1}}}
hops over \ensuremath{\HV{a_{2}}}, or (iii) \ensuremath{\HV{a_{2}}} hops over \ensuremath{\HV{a_{1}}}. These are captured by
the following \ensuremath{\HSCon{AOC}} constructors:

\begin{myhs}
\begin{hscode}\SaveRestoreHook
\column{B}{@{}>{\hspre}l<{\hspost}@{}}%
\column{3}{@{}>{\hspre}l<{\hspost}@{}}%
\column{11}{@{}>{\hspre}l<{\hspost}@{}}%
\column{14}{@{}>{\hspre}l<{\hspost}@{}}%
\column{E}{@{}>{\hspre}l<{\hspost}@{}}%
\>[3]{}\HT{PDoneDone}{}\<[14]%
\>[14]{}\HSCon{\mathbin{:}}\HSSpecial{\HSSym{\{\mskip1.5mu} }\HSVar{i}\HSCon{\mathbin{:}}\HT{\mathbb{N}}\HSSpecial{\HSSym{\mskip1.5mu\}}}\HSSym{\to} \HV{t_1}\;\HSVar{i}\HSSym{\equiv} \HV{t_2}\;\HSVar{i}{}\<[E]%
\\
\>[14]{}\HSSym{\to} \HSCon{AOC}\;\HV{t_1}\;\HV{t_2}\;\HSSpecial{\HSSym{\{\mskip1.5mu} }\HSVar{i}\HSSpecial{\HSSym{\mskip1.5mu\}}}\;\HSSpecial{\HSSym{\{\mskip1.5mu} }\HSVar{i}\HSSpecial{\HSSym{\mskip1.5mu\}}}\;\HSCon{Done}\;\HSCon{Done}{}\<[E]%
\\[\blanklineskip]%
\>[3]{}\HT{POverR}{}\<[14]%
\>[14]{}\HSCon{\mathbin{:}}\HS{\forall}\;\HSSpecial{\HSSym{\{\mskip1.5mu} }\HV{i_1}\;\HV{i_2}\;\HSVar{j}\HSSpecial{\HSSym{\mskip1.5mu\}}}\;\HSSpecial{\HSSym{\{\mskip1.5mu} }\HSVar{d}\HSCon{\mathbin{:}}\HSCon{Hash}\HSSpecial{\HSSym{\mskip1.5mu\}}}\;\HSSpecial{\HSSym{\{\mskip1.5mu} }\HSVar{h}\HSCon{\mathbin{:}}\HSCon{HopFrom}\;\HSVar{j}\HSSpecial{\HSSym{\mskip1.5mu\}}}{}\<[E]%
\\
\>[14]{}\HSSym{\to} \HSSpecial{(}\HV{a_{1}}\HSCon{\mathbin{:}}\HSCon{AdvPath}\;\HSSpecial{(}\HV{\hbox{\it hop\guydash{}tgt}}\;\HSVar{h}\HSSpecial{)}\;\HV{i_1}\HSSpecial{)}\;\HSSpecial{(}\HV{a_{2}}\HSCon{\mathbin{:}}\HSCon{AdvPath}\;\HSVar{j}\;\HV{i_2}\HSSpecial{)}{}\<[E]%
\\
\>[14]{}\HSSym{\to} \HV{\hbox{\it hop\guydash{}tgt}}\;\HSVar{h}\HSSym{\leq} \HV{i_2}{}\<[E]%
\\
\>[14]{}\HSSym{\to} \HSVar{rebuild}\;\HSSpecial{(}\HSCon{Hop}\;\HSVar{d}\;\HSVar{h}\;\HV{a_{1}}\HSSpecial{)}\;\HV{t_1}\;\HSVar{j}\HSSym{\equiv} \HSVar{rebuild}\;\HV{a_{2}}\;\HV{t_2}\;\HSVar{j}{}\<[E]%
\\
\>[14]{}\HSSym{\to} \HSCon{AOC}\;\HV{t_1}\;\HV{t_2}\;\HSSpecial{(}\HSCon{Hop}\;\HSVar{d}\;\HSVar{h}\;\HV{a_{1}}\HSSpecial{)}\;\HV{a_{2}}{}\<[E]%
\\[\blanklineskip]%
\>[3]{}\HT{POverL}{}\<[11]%
\>[11]{}\HSCon{\mathbin{:}}\HSComment{ -\! - symmetric}{}\<[E]%
\ColumnHook
\end{hscode}\resethooks
\end{myhs}

  The first constructor \ensuremath{\HT{PDoneDone}} represents pairs of advancement
proofs that trivially agree on their only common index because both
proofs are \ensuremath{\HSCon{Done}} and the two views agree at that index.

  In the second constructor, \ensuremath{\HT{POverR}}, the left
advancement path is not \ensuremath{\HSCon{Done}}: it takes a hop
\ensuremath{\HSCon{Hop}\;\HSVar{d}\;\HSVar{h}\;\HV{a_{1}}}. Moreover, \ensuremath{\HSVar{h}} hops \emph{over} \ensuremath{\HV{a_{2}}} (note that the constructor
requires evidence that
\ensuremath{\HV{\hbox{\it hop\guydash{}tgt}}\;\HSVar{h}\HSSym{\leq} \HV{i_2}}).  The diagram below illustrates this case.

\begin{center}
\resizebox{.35\textwidth}{!}{%
\begin{tikzpicture}[ every node/.style={scale=1.6} ]
  \node                      (a2) {\ensuremath{\HV{a_{2}}}};
  \node [below right = of a2] (j) {\ensuremath{\HSVar{j}}};
  \node [below left = of a2] (i2) {\ensuremath{\HV{i_2}}};
  \node [above left = of a2] (h)  {\ensuremath{\HSVar{h}}};
  \node [left = of i2]       (tgt){\ensuremath{\HV{\hbox{\it hop\guydash{}tgt}}\;\HSVar{h}}};
  \node [above left = of tgt] (a1) {\ensuremath{\HV{a_{1}}}};
  \node [below left = of a1] (i1)  {\ensuremath{\HV{i_1}}};

  \node (form) at ($ (i2)!0.3!(tgt) $) {\ensuremath{\HSSym{\leq} }};

  \draw [line width=0.25mm , ->] (j)   |- (a2.south) -| (i2);
  \draw [line width=0.25mm , ->] ($ (j.north) + (.1,0) $)
                             |- (h.south)
                             -| ($ (tgt.north) + (.1,0) $);
  \draw [line width=0.25mm , ->] (tgt) |- (a1.south) -| (i1);
\end{tikzpicture}}
\end{center}

   \ensuremath{\HT{POverR}} requires evidence that \ensuremath{\HSVar{rebuild}\;\HSSpecial{(}\HSCon{Hop}\;\HSVar{d}\;\HSVar{h}\;\HV{a_{1}}\HSSpecial{)}\;\HV{t_1}} agrees
with \ensuremath{\HSVar{rebuild}\;\HV{a_{2}}\;\HV{t_2}} on \ensuremath{\HSVar{j}}, which is sufficient to ensure that both
proofs rebuild to the same hash at \ensuremath{\HSVar{j}}.  It also requires evidence
that \ensuremath{\HV{\hbox{\it hop\guydash{}tgt}}\;\HSVar{h}\HSSym{\leq} \HV{i_2}}, which implies that the advancement proofs can
have at most one more commonly visited index besides \ensuremath{\HSVar{j}}, which is
\ensuremath{\HV{i_2}} when \ensuremath{\HV{\hbox{\it hop\guydash{}tgt}}\;\HSVar{h}\HSSym{\equiv} \HV{i_2}}.  In this case, because \ensuremath{\HV{i_2}} would then be a dependency of \ensuremath{\HSVar{j}}
(evidenced by the implicit hop \ensuremath{\HSVar{h}} provided to \ensuremath{\HT{POverR}}), applying
\ensuremath{\HV{\hbox{\it \calcauth\guydash{}inj\guydash{}1}}} and \ensuremath{\HV{\hbox{\it \calcauth\guydash{}inj\guydash{}2}}} to the hypothesis that \ensuremath{\HSCon{Hop}\;\HSVar{d}\;\HSVar{h}\;\HV{a_{1}}} and \ensuremath{\HV{a_{2}}} rebuild
the same authenticator at \ensuremath{\HSVar{j}}, yields \ensuremath{\HSCon{Agree}\;\HV{t_1}\;\HV{t_2}\;\HSSpecial{(}\HSVar{depsof}\;\HSVar{j}\HSSpecial{)}},
implying that the digests for \ensuremath{\HV{i_2}} in the two advancement proofs are
equal (again, unless \ensuremath{\HSCon{HashBroke}}).  This implies that the \ensuremath{\HSCon{View}}s
resulting from rebuilding the two advancement proofs used the same
authenticators for index \ensuremath{\HV{i_2}}; some simple lemmas ensure that
rebuilding hops to higher indexes do not modify the views at \ensuremath{\HV{i_2}}, so
the rebuilt views also agree at \ensuremath{\HV{i_2}}, as required.

  \ensuremath{\HT{POverL}} is symmetric to
\ensuremath{\HT{POverR}}, capturing the case in which the right
advancement proof hops over the entire left advancement proof.

  Next we examine the three inductive cases of \ensuremath{\HSCon{AOC}}.  The simplest is
\ensuremath{\HSCon{PCong}}, which is used when both proofs take the same hop.  \ensuremath{\HSCon{PCong}}
requires evidence \ensuremath{\HSCon{AOC}\;\HV{t_1}\;\HV{t_2}\;\HV{a_{1}}\;\HV{a_{2}}} that the rebuilt views agree at
all existing common indexes, as well as sufficient evidence to
conclude that \ensuremath{\HSVar{rebuild}\;\HSSpecial{(}\HSCon{Hop}\;\HSVar{d}\;\HSVar{h}\;\HV{a_{1}}\HSSpecial{)}\;\HV{t_1}\;\HSVar{j}} is equal to \ensuremath{\HSVar{rebuild}\;\HSSpecial{(}\HSCon{Hop}\;\HSVar{d}\;\HSVar{h}\;\HV{a_{2}}\HSSpecial{)}\;\HV{t_2}\;\HSVar{j}}.

\begin{myhs}
\begin{hscode}\SaveRestoreHook
\column{B}{@{}>{\hspre}l<{\hspost}@{}}%
\column{3}{@{}>{\hspre}l<{\hspost}@{}}%
\column{10}{@{}>{\hspre}l<{\hspost}@{}}%
\column{33}{@{}>{\hspre}l<{\hspost}@{}}%
\column{47}{@{}>{\hspre}c<{\hspost}@{}}%
\column{47E}{@{}l@{}}%
\column{E}{@{}>{\hspre}l<{\hspost}@{}}%
\>[3]{}\HSCon{PCong}{}\<[10]%
\>[10]{}\HSCon{\mathbin{:}}\HS{\forall}\;\HSSpecial{\HSSym{\{\mskip1.5mu} }\HV{i_1}\;\HV{i_2}\;\HSVar{j}\HSSpecial{\HSSym{\mskip1.5mu\}}}\;\HSSpecial{\HSSym{\{\mskip1.5mu} }\HSVar{d}\HSSpecial{\HSSym{\mskip1.5mu\}}}\;\HSSpecial{\HSSym{\{\mskip1.5mu} }\HSVar{h}\HSCon{\mathbin{:}}\HSCon{HopFrom}\;\HSVar{j}\HSSpecial{\HSSym{\mskip1.5mu\}}}{}\<[E]%
\\
\>[10]{}\HSSym{\to} \HSSpecial{(}\HV{a_{1}}\HSCon{\mathbin{:}}\HSCon{AdvPath}\;\HSSpecial{(}\HV{\hbox{\it hop\guydash{}tgt}}\;\HSVar{h}\HSSpecial{)}\;\HV{i_1}\HSSpecial{)}{}\<[E]%
\\
\>[10]{}\HSSym{\to} \HSSpecial{(}\HV{a_{2}}\HSCon{\mathbin{:}}\HSCon{AdvPath}\;\HSSpecial{(}\HV{\hbox{\it hop\guydash{}tgt}}\;\HSVar{h}\HSSpecial{)}\;\HV{i_2}\HSSpecial{)}{}\<[E]%
\\
\>[10]{}\HSSym{\to} \HSCon{Agree}\;\HSSpecial{(}\HSVar{rebuild}\;\HV{a_{1}}\;\HV{t_1}\HSSpecial{)}\;\HSSpecial{(}\HSVar{rebuild}\;\HV{a_{2}}\;\HV{t_2}\HSSpecial{)}\;\HSSpecial{(}\HSVar{depsof}\;\HSVar{j}\HSSpecial{)}{}\<[E]%
\\
\>[10]{}\HSSym{\to} \HSCon{AOC}\;\HV{t_1}\;\HV{t_2}\;{}\<[33]%
\>[33]{}\HV{a_{1}}\;{}\<[47]%
\>[47]{}\HV{a_{2}}{}\<[47E]%
\\
\>[10]{}\HSSym{\to} \HSCon{AOC}\;\HV{t_1}\;\HV{t_2}\;\HSSpecial{(}\HSCon{Hop}\;\HSVar{d}\;\HSVar{h}\;{}\<[33]%
\>[33]{}\HV{a_{1}}\HSSpecial{)}\;\HSSpecial{(}\HSCon{Hop}\;\HSVar{d}\;\HSVar{h}\;{}\<[47]%
\>[47]{}\HV{a_{2}}\HSSpecial{)}{}\<[47E]%
\ColumnHook
\end{hscode}\resethooks
\end{myhs}


  Finally, the case in which \ensuremath{\HV{a_2}} is extended to index \ensuremath{\HSVar{j}} by a hop
\ensuremath{\HV{h_2}} and \ensuremath{\HV{a_1}} is extended by a \emph{different} path to index \ensuremath{\HSVar{j}}
is handled by the \ensuremath{\HSCon{PMeetR}} constructor; \ensuremath{\HSCon{PMeetL}} being its symmetric counterpart.
This situation is captured in the diagram below.

\begin{center}
\resizebox{.35\textwidth}{!}{%
\begin{tikzpicture}[ every node/.style={scale=1.6} ]
  \node (j) {\ensuremath{\HSVar{j}}};
  \node [left = of j] (tgt1) {\ensuremath{\HV{\hbox{\it hop\guydash{}tgt}}\;\HV{h_1}}};
  \node [left = of tgt1] (tgt2) {\ensuremath{\HV{\hbox{\it hop\guydash{}tgt}}\;\HV{h_2}}};
  \node (form) at ($ (tgt1)!0.5!(tgt2) $) {\ensuremath{\HSSym{<}}};
  \node (h1) at ($ (tgt1)!0.5!(j) + (0,1.6) $) {\ensuremath{\HV{h_1}}};
  \node (h2) at ($ (tgt2)!0.5!(j) + (0,3) $) {\ensuremath{\HV{h_2}}};
  \node (e) at ($(form) + (0,1.5) $) {\ensuremath{\HSVar{e}}};

  \draw [line width=0.25mm , ->] ($ (j.north) - (.1,0) $) |- (h1.south) -| ($ (tgt1.north) + (.05,0) $);
  \draw [line width=0.25mm , ->] (j) |- (h2.south) -| (tgt2);
  \draw [line width=0.25mm , ->] ($ (tgt1.north) - (.1,0) $) |- (e.south) -| ($ (tgt2.north) + (0.15,0) $);

  \node (a1) at ($ (tgt2) + (-1.6,1.2) $) {\ensuremath{\HV{a_{1}}}};
  \node (a2) at ($ (a1) + (0, 1.4) $) {\ensuremath{\HV{a_{2}}}};

  \draw [line width=0.25mm , ->] ($ (tgt2.north) - (.3,0) $) |- ($ (a1.south) - (1,0) $);
  \draw [line width=0.25mm , ->] ($ (tgt2.north) - (.2,0) $) |- ($ (a2.south) - (1,0) $);
\end{tikzpicture}}
\end{center}

The idea here is that we can look at the advancement proof that takes the smaller hop
as a composition of two advancement proofs making evident the
next common index.

\begin{myhs}
\begin{hscode}\SaveRestoreHook
\column{B}{@{}>{\hspre}l<{\hspost}@{}}%
\column{3}{@{}>{\hspre}l<{\hspost}@{}}%
\column{11}{@{}>{\hspre}l<{\hspost}@{}}%
\column{18}{@{}>{\hspre}l<{\hspost}@{}}%
\column{25}{@{}>{\hspre}l<{\hspost}@{}}%
\column{36}{@{}>{\hspre}l<{\hspost}@{}}%
\column{E}{@{}>{\hspre}l<{\hspost}@{}}%
\>[3]{}\HSCon{PMeetR}{}\<[11]%
\>[11]{}\HSCon{\mathbin{:}}\HS{\forall}\;\HSSpecial{\HSSym{\{\mskip1.5mu} }\HSVar{j}\;\HV{i_1}\;\HV{i_2}\;\HSVar{d}\HSSpecial{\HSSym{\mskip1.5mu\}}}\;\HSSpecial{\HSSym{\{\mskip1.5mu} }\HV{h_1}\;\HV{h_2}\HSCon{\mathbin{:}}\HSCon{HopFrom}\;\HSVar{j}\HSSpecial{\HSSym{\mskip1.5mu\}}}{}\<[E]%
\\
\>[11]{}\HSSym{\to} \HSSpecial{(}\HSVar{e}{}\<[18]%
\>[18]{}\HSCon{\mathbin{:}}\HSCon{AdvPath}\;\HSSpecial{(}\HV{\hbox{\it hop\guydash{}tgt}}\;\HV{h_1}\HSSpecial{)}\;\HSSpecial{(}\HV{\hbox{\it hop\guydash{}tgt}}\;\HV{h_2}\HSSpecial{)}\HSSpecial{)}{}\<[E]%
\\
\>[11]{}\HSSym{\to} \HSSpecial{(}\HV{a_{1}}\HSCon{\mathbin{:}}\HSCon{AdvPath}\;\HSSpecial{(}\HV{\hbox{\it hop\guydash{}tgt}}\;\HV{h_2}\HSSpecial{)}\;\HV{i_1}\HSSpecial{)}{}\<[E]%
\\
\>[11]{}\HSSym{\to} \HSSpecial{(}\HV{a_{2}}\HSCon{\mathbin{:}}\HSCon{AdvPath}\;\HSSpecial{(}\HV{\hbox{\it hop\guydash{}tgt}}\;\HV{h_2}\HSSpecial{)}\;\HV{i_2}\HSSpecial{)}{}\<[E]%
\\
\>[11]{}\HSSym{\to} \HV{\hbox{\it hop\guydash{}tgt}}\;\HV{h_2}\HSSym{<}\HV{\hbox{\it hop\guydash{}tgt}}\;\HV{h_1}{}\<[E]%
\\
\>[11]{}\HSSym{\to} \HSCon{Agree}\;\HSSpecial{(}\HSVar{rebuild}\;\HSSpecial{(}\HSVar{e}\;\mathbin{\HV{\oplus}}\;\HV{a_{1}}\HSSpecial{)}\;\HV{t_1}\HSSpecial{)}\;\HSSpecial{(}\HSVar{rebuild}\;\HV{a_{2}}\;\HV{t_2}\HSSpecial{)}\;\HSSpecial{(}\HSVar{depsof}\;\HSVar{j}\HSSpecial{)}{}\<[E]%
\\
\>[11]{}\HSSym{\to} \HSCon{AOC}\;\HV{t_1}\;\HV{t_2}\;\HV{a_{1}}\;\HV{a_{2}}{}\<[E]%
\\
\>[11]{}\HSSym{\to} \HSCon{AOC}\;\HV{t_1}\;\HV{t_2}\;{}\<[25]%
\>[25]{}\HSSpecial{(}\HSCon{Hop}\;\HSVar{d}\;\HV{h_1}\;{}\<[36]%
\>[36]{}\HSSpecial{(}\HSVar{e}\;\mathbin{\HV{\oplus}}\;\HV{a_{1}}\HSSpecial{)}\HSSpecial{)}\;\HSSpecial{(}\HSCon{Hop}\;\HSVar{d}\;\HV{h_2}\;\HV{a_{2}}\HSSpecial{)}{}\<[E]%
\\[\blanklineskip]%
\>[3]{}\HSCon{PMeetL}{}\<[11]%
\>[11]{}\HSCon{\mathbin{:}}\HSComment{ -\! - symmetric}{}\<[E]%
\ColumnHook
\end{hscode}\resethooks
\end{myhs}

  Here, \ensuremath{\HV{\_\!\oplus\!\_}} composes advancement proofs.

\begin{myhs}
\begin{hscode}\SaveRestoreHook
\column{B}{@{}>{\hspre}l<{\hspost}@{}}%
\column{15}{@{}>{\hspre}l<{\hspost}@{}}%
\column{E}{@{}>{\hspre}l<{\hspost}@{}}%
\>[B]{}\HV{\_\!\oplus\!\_}{}\<[15]%
\>[15]{}\HSCon{\mathbin{:}}\HS{\forall}\;\HSSpecial{\HSSym{\{\mskip1.5mu} }\HSVar{i}\;\HSVar{j}\;\HSVar{k}\HSSpecial{\HSSym{\mskip1.5mu\}}}\HSSym{\to} \HSCon{AdvPath}\;\HSVar{j}\;\HSVar{k}\HSSym{\to} \HSCon{AdvPath}\;\HSVar{k}\;\HSVar{i}\HSSym{\to} \HSCon{AdvPath}\;\HSVar{j}\;\HSVar{i}{}\<[E]%
\ColumnHook
\end{hscode}\resethooks
\end{myhs}

  \ensuremath{\HSCon{PMeetR}} and \ensuremath{\HSCon{PMeetL}} are similar to \ensuremath{\HSCon{PCong}} in that they all require an \ensuremath{\HSCon{AOC}\;\HV{t_1}\;\HV{t_2}\;\HV{a_{1}}\;\HV{a_{2}}}, which provides
evidence that the smaller advancement proofs agree at each of their
common indexes.  They also all require sufficient evidence to prove
that the authenticator computed for the new common index \ensuremath{\HSVar{j}} is the
same following either of the new advancement proofs to index \ensuremath{\HSVar{j}}.


  One important observation is that rebuilding an advancement proof never
interferes with previous indexes. That is, rebuilding a path \ensuremath{\HSVar{e}\HSCon{\mathbin{:}}\HSCon{AdvPath}\;\HV{j_1}\;\HV{j_2}} on a view
that was obtained from rebuilding \ensuremath{\HSVar{a}\HSCon{\mathbin{:}}\HSCon{AdvPath}\;\HV{j_2}\;\HSVar{i}} on a view \ensuremath{\HSVar{t}} will not change any
authenticator already computed for any \ensuremath{\HSVar{k}\HSSym{\leq} \HSVar{i}}. This is witnessed by the \ensuremath{\HV{\hbox{\it rebuild\guydash{}}\oplus}} property below,
which guarantees that the inductive hypothesis provided by \ensuremath{\HSCon{AOC}\;\HV{t_1}\;\HV{t_2}\;\HV{a_{1}}\;\HV{a_{2}}} in \ensuremath{\HSCon{PMeetR}}
is not falsified by rebuilding \ensuremath{\HSSpecial{(}\HSVar{e}\;\mathbin{\HV{\oplus}}\;\HV{a_{1}}\HSSpecial{)}} on top of \ensuremath{\HV{t_1}}.

\begin{myhs}
\begin{hscode}\SaveRestoreHook
\column{B}{@{}>{\hspre}l<{\hspost}@{}}%
\column{15}{@{}>{\hspre}l<{\hspost}@{}}%
\column{25}{@{}>{\hspre}l<{\hspost}@{}}%
\column{E}{@{}>{\hspre}l<{\hspost}@{}}%
\>[B]{}\HV{\hbox{\it rebuild\guydash{}}\oplus}{}\<[15]%
\>[15]{}\HSCon{\mathbin{:}}\HS{\forall}\;{}\<[25]%
\>[25]{}\HSSpecial{\HSSym{\{\mskip1.5mu} }\HV{j_1}\;\HV{j_2}\;\HSVar{i}\;\HSVar{k}\;\HSVar{t}\HSSpecial{\HSSym{\mskip1.5mu\}}}\;\HSSpecial{(}\HSVar{e}\HSCon{\mathbin{:}}\HSCon{AdvPath}\;\HV{j_1}\;\HV{j_2}\HSSpecial{)}\;\HSSpecial{(}\HV{a_1}\HSCon{\mathbin{:}}\HSCon{AdvPath}\;\HV{j_2}\;\HSVar{i}\HSSpecial{)}{}\<[E]%
\\
\>[15]{}\HSSym{\to} \HSVar{k}\;\HT{\epsilon_{\textsc{ap}}}\;\HV{a_1}{}\<[E]%
\\
\>[15]{}\HSSym{\to} \HSVar{rebuild}\;\HSSpecial{(}\HSVar{e}\;\mathbin{\HV{\oplus}}\;\HV{a_1}\HSSpecial{)}\;\HSVar{t}\;\HSVar{k}\HSSym{\equiv} \HSVar{rebuild}\;\HV{a_1}\;\HSVar{t}\;\HSVar{k}{}\<[E]%
\ColumnHook
\end{hscode}\resethooks
\end{myhs}

This property is proved by a simple induction using the fact that
computing authenticators for indexes higher than \ensuremath{\HV{j_2}} does not modify
the provided view at \ensuremath{\HV{j_2}} or any lower index.   Because \ensuremath{\HSVar{k}\;\HT{\epsilon_{\textsc{ap}}}\;\HV{a_1}} and
\ensuremath{\HV{a_1}} is an \ensuremath{\HSCon{AdvPath}\;\HV{j_2}\;\HSVar{i}}, \ensuremath{\HSVar{k}\HSSym{\leq} \HV{j_2}}; thus the two rebuilds agree at \ensuremath{\HSVar{k}}.

Using \ensuremath{\HV{\hbox{\it rebuild\guydash{}}\oplus}}, we prove \ensuremath{\HV{\hbox{\textsc{AgreeOnCommon}\guydash{}$\in$}}}, a variant of \ensuremath{\HV{\textsc{AgreeOnCommon}}}
that allows \ensuremath{\HV{a_{2}}} to be an \ensuremath{\HSCon{AdvPath}\;\HV{j_2}\;\HV{i_2}} for any index \ensuremath{\HV{j_2}} that is visited by \ensuremath{\HV{a_{1}}}
(i.e., \ensuremath{\HV{j_2}\;\HT{\epsilon_{\textsc{ap}}}\;\HV{a_{1}}}), not just the source index of \ensuremath{\HV{a_{1}}} as required by \ensuremath{\HV{\textsc{AgreeOnCommon}}}.
\ensuremath{\HV{\hbox{\textsc{AgreeOnCommon}\guydash{}$\in$}}} requires that the two advancement proofs rebuild to the same hash at
their common index \ensuremath{\HV{j_2}}.

\subsubsection*{Constructing an \ensuremath{\HSCon{AOC}} inhabitant}

  To conclude the proof of \ensuremath{\HV{\textsc{AgreeOnCommon}}}, we must
construct an \ensuremath{\HSCon{AOC}\;\HV{t_1}\;\HV{t_2}\;\HV{a_{1}}\;\HV{a_{2}}} given \ensuremath{\HSVar{rebuild}\;\HV{a_{1}}\;\HV{t_1}\;\HSVar{j}\HSSym{\equiv} \HSVar{rebuild}\;\HV{a_{2}}\;\HV{t_2}\;\HSVar{j}}, where \ensuremath{\HV{a_{1}}} and \ensuremath{\HV{a_{2}}} are advancement proofs to index \ensuremath{\HSVar{j}}. This is
accomplished by the \ensuremath{\HV{\textsc{aoc}}} lemma.

\begin{myhs}
\begin{hscode}\SaveRestoreHook
\column{B}{@{}>{\hspre}l<{\hspost}@{}}%
\column{6}{@{}>{\hspre}l<{\hspost}@{}}%
\column{E}{@{}>{\hspre}l<{\hspost}@{}}%
\>[B]{}\HV{\textsc{aoc}}{}\<[6]%
\>[6]{}\HSCon{\mathbin{:}}\HS{\forall}\;\HSSpecial{\HSSym{\{\mskip1.5mu} }\HV{i_1}\;\HV{i_2}\;\HSVar{j}\HSSpecial{\HSSym{\mskip1.5mu\}}}\;\HSSpecial{(}\HV{t_1}\;\HV{t_2}\HSCon{\mathbin{:}}\HSCon{View}\HSSpecial{)}\;\HSSpecial{(}\HV{a_{1}}\HSCon{\mathbin{:}}\HSCon{AdvPath}\;\HSVar{j}\;\HV{i_1}\HSSpecial{)}\;\HSSpecial{(}\HV{a_{2}}\HSCon{\mathbin{:}}\HSCon{AdvPath}\;\HSVar{j}\;\HV{i_2}\HSSpecial{)}{}\<[E]%
\\
\>[6]{}\HSSym{\to} \HV{i_1}\HSSym{\leq} \HV{i_2}\HSComment{ -\! - w.l.o.g}{}\<[E]%
\\
\>[6]{}\HSSym{\to} \HSVar{rebuild}\;\HV{a_{1}}\;\HV{t_1}\;\HSVar{j}\HSSym{\equiv} \HSVar{rebuild}\;\HV{a_{2}}\;\HV{t_2}\;\HSVar{j}{}\<[E]%
\\
\>[6]{}\HSSym{\to} \HSCon{Either}\;\HSCon{HashBroke}\;\HSSpecial{(}\HSCon{AOC}\;\HV{t_1}\;\HV{t_2}\;\HV{a_{1}}\;\HV{a_{2}}\HSSpecial{)}{}\<[E]%
\ColumnHook
\end{hscode}\resethooks
\end{myhs}

First,
we pattern match on \ensuremath{\HV{a_{1}}} and \ensuremath{\HV{a_{2}}} and compare the \ensuremath{\HV{\hbox{\it hop\guydash{}tgt}}} (if any)
of the hops taken. This enables us to understand which constructor
of \ensuremath{\HSCon{AOC}} must be used.
In most cases we can easily extract the evidence needed
for the relevant \ensuremath{\HSCon{AOC}} constructor via case analysis on the structure
of the two proofs and the injectivity lemmas discussed above,
applied to the hypothesis that \ensuremath{\HSVar{rebuild}\;\HV{a_{1}}\;\HV{t_1}\;\HSVar{j}\HSSym{\equiv} \HSVar{rebuild}\;\HV{a_{2}}\;\HV{t_2}\;\HSVar{j}}.

However, we face an additional challenge when the two advancement proofs take different hops, but
neither hop passes the other's entire proof.
These cases require the \ensuremath{\HSCon{PMeetR}} or \ensuremath{\HSCon{PMeetL}} constructor,
and we must \emph{split} the proof that took the shorter hop into two
composable pieces in order to obtain an
advancement proof between the targets of the hops taken by the
respective advancement proofs.  This is needed for the \ensuremath{\HSVar{e}} argument to the constructor.
Because of the \ensuremath{\HV{\hbox{\it hop\guydash{}no\guydash{}cross}}} requirement on \ensuremath{\HSCon{DepRel}} we can always perform
this \ensuremath{\HSVar{split}} as evidenced by the lemma below. No hop can cross from an index strictly between
\ensuremath{\HV{\hbox{\it hop\guydash{}tgt}}\;\HV{h_2}} and \ensuremath{\HV{\hbox{\it hop\guydash{}tgt}}\;\HV{h_1}} to an index lower than or equal to \ensuremath{\HV{\hbox{\it hop\guydash{}tgt}}\;\HV{h_1}} without
passing through \ensuremath{\HV{\hbox{\it hop\guydash{}tgt}}\;\HV{h_1}} first.

\begin{myhs}
\begin{hscode}\SaveRestoreHook
\column{B}{@{}>{\hspre}l<{\hspost}@{}}%
\column{15}{@{}>{\hspre}l<{\hspost}@{}}%
\column{26}{@{}>{\hspre}l<{\hspost}@{}}%
\column{E}{@{}>{\hspre}l<{\hspost}@{}}%
\>[B]{}\HV{\hbox{\it split\guydash{}}\oplus}{}\<[15]%
\>[15]{}\HSCon{\mathbin{:}}\HS{\forall}\;\HSSpecial{\HSSym{\{\mskip1.5mu} }\HV{j_1}\;\HV{j_2}\;\HSVar{i}\HSSpecial{\HSSym{\mskip1.5mu\}}}\;\HSSpecial{\HSSym{\{\mskip1.5mu} }\HV{h_1}\HSCon{\mathbin{:}}\HSCon{HopFrom}\;\HV{j_1}\HSSpecial{\HSSym{\mskip1.5mu\}}}\;\HSSpecial{\HSSym{\{\mskip1.5mu} }\HV{h_2}\HSCon{\mathbin{:}}\HSCon{HopFrom}\;\HV{j_2}\HSSpecial{\HSSym{\mskip1.5mu\}}}{}\<[E]%
\\
\>[15]{}\HSSym{\to} \HV{j_2}\HSSym{\leq} \HV{j_1}\HSSym{\to} \HV{\hbox{\it hop\guydash{}tgt}}\;\HV{h_1}\HSSym{<}\HV{\hbox{\it hop\guydash{}tgt}}\;\HV{h_2}\HSSym{\to} \HSVar{i}\HSSym{\leq} \HV{\hbox{\it hop\guydash{}tgt}}\;\HV{h_1}{}\<[E]%
\\
\>[15]{}\HSSym{\to} \HSSpecial{(}\HSVar{a}\HSCon{\mathbin{:}}\HSCon{AdvPath}\;\HSSpecial{(}\HV{\hbox{\it hop\guydash{}tgt}}\;\HV{h_2}\HSSpecial{)}\;\HSVar{i}\HSSpecial{)}{}\<[E]%
\\
\>[15]{}\HSSym{\to} \HT{\exists[}{}\<[26]%
\>[26]{}\HSVar{x}\HSCon{\mathbin{:}}\HSCon{AdvPath}\;\HSSpecial{(}\HV{\hbox{\it hop\guydash{}tgt}}\;\HV{h_2}\HSSpecial{)}\;\HSSpecial{(}\HV{\hbox{\it hop\guydash{}tgt}}\;\HV{h_1}\HSSpecial{)}\HSSpecial{,}\HSVar{y}\HT{]}\;{}\<[E]%
\\
\>[26]{}\HSSpecial{(}\HSVar{a}\HSSym{\equiv} \HSVar{x}\;\mathbin{\HV{\oplus}}\;\HSVar{y}\HSSpecial{)}{}\<[E]%
\ColumnHook
\end{hscode}\resethooks
\end{myhs}

\subsection{Proving \textsc{evo-cr}}

  With \ensuremath{\HV{\textsc{AgreeOnCommon}}} out of the way, we move to the key correctness property \textsc{evocr}~\cite{Baker2003},
which states that
a computationally bounded adversary cannot produce two advancement proofs \ensuremath{\HV{a_{1}}}
and \ensuremath{\HV{a_{2}}} that rebuild to the same authenticator
at some index $j$, and also provide two membership proofs from
\ensuremath{\HV{s_1}} to \ensuremath{\HSVar{tgt}} and \ensuremath{\HV{s_2}} to \ensuremath{\HSVar{tgt}} that prove a conflicting claim at
\ensuremath{\HSVar{tgt}}, given that \ensuremath{\HV{a_{1}}} visits \ensuremath{\HV{s_1}} and \ensuremath{\HV{a_{2}}} visits \ensuremath{\HV{s_2}}.

\begin{figure*}
\begin{center}
\begin{tikzpicture}[
  stepref/.style = {scale=0.4, circle, draw, fill=white, inner sep=.5mm}
]
  \node (j) {\ensuremath{\HSVar{j}}};
  \node [left = 2cm of j] (hs2) {};
  \node [left = 3cm of hs2] (hs1) {};
  \node [left = 3cm of hs1] (tgt) {\ensuremath{\HSVar{tgt}}};

  \node [below = of hs2] (s2) {\ensuremath{\HV{s_2}}};
  \node [above = of hs1] (s1) {\ensuremath{\HV{s_1}}};
  \node (m) at ($ (hs1)!0.5!(hs2) + (0, 2cm) $) {\ensuremath{\HV{\mathcal{M}}}};

  \node (i1) at ($ (tgt|-s1) - (2cm, 0) $) {\ensuremath{\HV{i_1}}};
  \node (i2) at ($ (tgt|-s2) - (1cm, 0) $) {\ensuremath{\HV{i_2}}};

  \node (hr) at ($ (tgt|-s1)!0.5!(s1) $) {};
  \node (r) at (hr|-m) {\ensuremath{\HV{\mathcal{R}}}};

  \draw[very thick, ->] (i1) -- node[near start, above] {\ensuremath{\HV{a_{12}}}} (s1);
  \draw[very thick, ->] (s1) to[out=0, in=130] node[near end, above] {\ensuremath{\HV{a_{11}}}} (j);
  \draw[very thick, gray, ->] (i2) -- node[midway, below] {\ensuremath{\HV{a_{22}}}} (s2);
  \draw[very thick, gray, ->] (s2) to[out=0, in=-120] node[midway, below] {\ensuremath{\HV{a_{21}}}} (j);

  \draw[->] ($ (tgt) + (3mm,  1.3mm) $) -| node[pos=0.6, above right] {\ensuremath{\HV{m_{1}}}} (s1);
  \draw[gray, ->] ($ (tgt) + (3mm, -1.3mm) $) -| node[pos=0.5, above] {\ensuremath{\HV{m_{2}}}} (s2);

  \draw[densely dotted] (r.south) -- ($ (tgt-|r) - (0, 1.3mm) $);
  \draw[densely dotted] (m.south) -- (s2-|m);

  \node[stepref] at (s1-|m) {1};
  \node[stepref] at (s2-|m) {1};
  \node[stepref] at ($ (tgt-|m) - (0, 1.3mm) $) {4};

  \node[stepref] at (s1-|r) {5};
  \node[stepref] at ($ (tgt-|r) - (0, 1.3mm) $) {5};
  \node[stepref] at ($ (tgt-|r) + (0, 1.3mm) $) {6};

\end{tikzpicture}
\end{center}
\caption{Graphical representation \ensuremath{\HV{\textsc{evo-cr}}} property~\cite{Baker2003}. Thick lines represent
the advancement paths, thin lines represent the membership proofs, dotted lines represent
points that must be visited by all proofs they intersect and the numbered dots reference the proof
step that justifies the intersection.}
\label{fig:evocr}
\end{figure*}
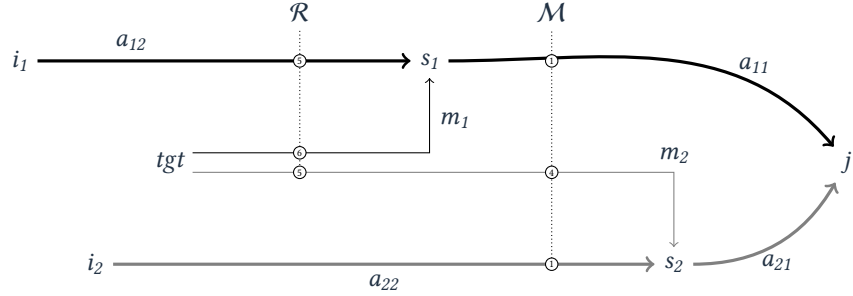

To enable a rigorous proof of this property, we require a precise statement of it,
which is provided by the type of \ensuremath{\HV{\textsc{evo-cr}}}
and illustrated in \Cref{fig:evocr}.

\begin{myhs}
\begin{hscode}\SaveRestoreHook
\column{B}{@{}>{\hspre}l<{\hspost}@{}}%
\column{8}{@{}>{\hspre}l<{\hspost}@{}}%
\column{E}{@{}>{\hspre}l<{\hspost}@{}}%
\>[B]{}\HV{\textsc{evo-cr}}{}\<[8]%
\>[8]{}\HSCon{\mathbin{:}}\HS{\forall}\;\HSSpecial{\HSSym{\{\mskip1.5mu} }\HSVar{j}\;\HV{i_1}\;\HV{i_2}\HSSpecial{\HSSym{\mskip1.5mu\}}}\;\HSSpecial{\HSSym{\{\mskip1.5mu} }\HV{t_1}\;\HV{t_2}\HSCon{\mathbin{:}}\HSCon{View}\HSSpecial{\HSSym{\mskip1.5mu\}}}{}\<[E]%
\\
\>[8]{}\HSSym{\to} \HSSpecial{(}\HV{a_{1}}\HSCon{\mathbin{:}}\HSCon{AdvPath}\;\HSVar{j}\;\HV{i_1}\HSSpecial{)}\;\HSSpecial{(}\HV{a_{2}}\HSCon{\mathbin{:}}\HSCon{AdvPath}\;\HSVar{j}\;\HV{i_2}\HSSpecial{)}{}\<[E]%
\\
\>[8]{}\HSSym{\to} \HSVar{rebuild}\;\HV{a_{1}}\;\HV{t_1}\;\HSVar{j}\HSSym{\equiv} \HSVar{rebuild}\;\HV{a_{2}}\;\HV{t_2}\;\HSVar{j}{}\<[E]%
\\
\>[8]{}\HSSym{\to} \HS{\forall}\;\HSSpecial{\HSSym{\{\mskip1.5mu} }\HV{s_1}\;\HV{s_2}\;\HSVar{tgt}\HSSpecial{\HSSym{\mskip1.5mu\}}}\;\HSSpecial{\HSSym{\{\mskip1.5mu} }\HV{u_1}\;\HV{u_2}\HSCon{\mathbin{:}}\HSCon{View}\HSSpecial{\HSSym{\mskip1.5mu\}}}{}\<[E]%
\\
\>[8]{}\HSSym{\to} \HSSpecial{(}\HV{m_{1}}\HSCon{\mathbin{:}}\HT{\mathit{MbrPath}}\;\HV{s_1}\;\HSVar{tgt}\HSSpecial{)}\;\HSSpecial{(}\HV{m_{2}}\HSCon{\mathbin{:}}\HT{\mathit{MbrPath}}\;\HV{s_2}\;\HSVar{tgt}\HSSpecial{)}{}\<[E]%
\\
\>[8]{}\HSSym{\to} \HV{s_1}\;\HT{\epsilon_{\textsc{ap}}}\;\HV{a_{1}}\HSSym{\to} \HV{s_2}\;\HT{\epsilon_{\textsc{ap}}}\;\HV{a_{2}}{}\<[E]%
\\
\>[8]{}\HSSym{\to} \HV{s_1}\HSSym{\leq} \HV{s_2}\HSSym{\to} \HV{i_1}\HSSym{\leq} \HSVar{tgt}\HSSym{\to} \HV{i_2}\HSSym{\leq} \HSVar{tgt}{}\<[E]%
\\
\>[8]{}\HSSym{\to} \HV{\mathit{rebuildMbr}}\;\HV{m_{1}}\;\HV{u_1}\;\HV{s_1}\HSSym{\equiv} \HSVar{rebuild}\;\HV{a_{1}}\;\HV{t_1}\;\HV{s_1}{}\<[E]%
\\
\>[8]{}\HSSym{\to} \HV{\mathit{rebuildMbr}}\;\HV{m_{2}}\;\HV{u_2}\;\HV{s_2}\HSSym{\equiv} \HSVar{rebuild}\;\HV{a_{2}}\;\HV{t_2}\;\HV{s_2}{}\<[E]%
\\
\>[8]{}\HSSym{\to} \HSCon{Either}\;\HSCon{HashBroke}\;\HSSpecial{(}\HV{\hbox{\it datum\guydash{}hash}}\;\HV{m_{1}}\HSSym{\equiv} \HV{\hbox{\it datum\guydash{}hash}}\;\HV{m_{2}}\HSSpecial{)}{}\<[E]%
\ColumnHook
\end{hscode}\resethooks
\end{myhs}

When translating the informal statement of this property~\cite{Baker2003} to a precise one,
we originally~\cite{Miraldo2021} required \ensuremath{\HSVar{tgt}} to be visited by both \ensuremath{\HV{a_{1}}} and \ensuremath{\HV{a_{2}}}.
This was a mistake, because the purpose of a membership proof is
to enable proving that a log entry that has not necessarily been verified
previously is consistent with an advancement
proof that has been verified: if \ensuremath{\HSVar{tgt}} were required to be in the advancement proof, then it
would already have been verified.  Therefore, in this extended version, we present the
proof of the stronger \ensuremath{\HV{\textsc{evo-cr}}} property stated above.


The bulk of the proof of \ensuremath{\HV{\textsc{evo-cr}}} comprises proving that the authenticators computed for
index \ensuremath{\HSVar{tgt}} when rebuilding \ensuremath{\HV{m_{1}}} and \ensuremath{\HV{m_{2}}} are equal, enabling us to invoke \ensuremath{\HV{\hbox{\it \calcauth\guydash{}inj\guydash{}1}}}
to establish that (unless there is a hash collision), the two membership proofs \ensuremath{\HV{m_{1}}} and \ensuremath{\HV{m_{2}}}
claim the same datum at index \ensuremath{\HSVar{tgt}}.  To achieve this, we must work back from the assumptions
that \ensuremath{\HV{m_{1}}} (respectively \ensuremath{\HV{m_{2}}}) rebuilds to an index \ensuremath{\HV{s_1}} (respectively \ensuremath{\HV{s_2}}) visited by \ensuremath{\HV{a_{1}}} (respectively \ensuremath{\HV{a_{2}}}),
consistent at these indexes with rebuilding \ensuremath{\HV{a_{1}}} and \ensuremath{\HV{a_{2}}}, and that \ensuremath{\HV{a_{1}}} and \ensuremath{\HV{a_{2}}} build to the
same hash at their common index \ensuremath{\HSVar{j}}.

The first step is to split \ensuremath{\HV{a_{1}}} into \ensuremath{\HV{a_{11}}\;\mathbin{\HV{\oplus}}\;\HV{a_{12}}}, and similarly for \ensuremath{\HV{a_{2}}},
as depicted in \Cref{fig:evocr}.  Next, following the original manual proof~\cite{Baker2003}, we
identify two special indexes, each of which is common to three advancement proofs.

The first special index, which we call \ensuremath{\HV{\mathcal{M}}} (consistent with index \ensuremath{\HSVar{m}} in
\cite[Fig. 4]{Baker2003}), is required to be common to
\ensuremath{\HV{a_{22}}}, \ensuremath{\HV{m_{2}}} and \ensuremath{\HV{a_{11}}}.  If, \ensuremath{\HV{s_1}\HSSym{\mathrel{=}}\HV{s_2}}, then that index serves the purpose,
because every \ensuremath{\HSCon{AdvPath}} contains its starting and ending indexes.  Otherwise,
\ensuremath{\HV{s_1}\HSSym{<}\HV{s_2}}.  In that case, \ensuremath{\HV{\mathcal{M}}} is chosen to be the \emph{last} index in \ensuremath{\HV{a_{11}}} \emph{before} \ensuremath{\HV{s_2}}
(note that it exists because \ensuremath{\HV{s_1}\HSSym{<}\HV{s_2}}, so \ensuremath{\HV{s_1}} could be \ensuremath{\HV{\mathcal{M}}} if there is no index in \ensuremath{\HV{a_{11}}}
between \ensuremath{\HV{s_1}} and \ensuremath{\HV{s_2}}).  The function \ensuremath{\HV{\hbox{\it{last\guydash{}bef}}}} determines \ensuremath{\HV{\mathcal{M}}}:

\begin{myhs}
\begin{hscode}\SaveRestoreHook
\column{B}{@{}>{\hspre}l<{\hspost}@{}}%
\column{10}{@{}>{\hspre}c<{\hspost}@{}}%
\column{10E}{@{}l@{}}%
\column{13}{@{}>{\hspre}l<{\hspost}@{}}%
\column{E}{@{}>{\hspre}l<{\hspost}@{}}%
\>[B]{}\HV{\hbox{\it{last\guydash{}bef}}}{}\<[10]%
\>[10]{}\HSCon{\mathbin{:}}{}\<[10E]%
\>[13]{}\HS{\forall}\;\HSSpecial{\HSSym{\{\mskip1.5mu} }\HSVar{j}\;\HSVar{i}\;\HSVar{k}\HSSpecial{\HSSym{\mskip1.5mu\}}}\;\HSSpecial{(}\HSVar{a}\HSCon{\mathbin{:}}\HSCon{AdvPath}\;\HSVar{j}\;\HSVar{i}\HSSpecial{)}\;\HSSpecial{(}\HSVar{ik}\HSCon{\mathbin{:}}\HSVar{i}\HSSym{<}\HSVar{k}\HSSpecial{)}\;\HSSpecial{(}\HSVar{kj}\HSCon{\mathbin{:}}\HSVar{k}\HSSym{\leq} \HSVar{j}\HSSpecial{)}\HSSym{\to} \HT{\mathbb{N}}{}\<[E]%
\ColumnHook
\end{hscode}\resethooks
\end{myhs}

We define \ensuremath{\HV{\mathcal{M}}} as \ensuremath{\HV{\hbox{\it{last\guydash{}bef}}}\;\HV{a_{11}}\;\HSVar{prf}}, where \ensuremath{\HSVar{prf}} is evidence that \ensuremath{\HV{s_1}\HSSym{<}\HV{s_2}}.
A correctness lemma establishes that \ensuremath{\HV{\mathcal{M}}\;\HT{\epsilon_{\textsc{ap}}}\;\HV{a_{11}}}:

\begin{myhs}
\begin{hscode}\SaveRestoreHook
\column{B}{@{}>{\hspre}l<{\hspost}@{}}%
\column{17}{@{}>{\hspre}l<{\hspost}@{}}%
\column{20}{@{}>{\hspre}l<{\hspost}@{}}%
\column{E}{@{}>{\hspre}l<{\hspost}@{}}%
\>[B]{}\HV{\hbox{\it{last\guydash{}bef\guydash{}correct}}}{}\<[17]%
\>[17]{}\HSCon{\mathbin{:}}{}\<[20]%
\>[20]{}\HS{\forall}\;\HSSpecial{\HSSym{\{\mskip1.5mu} }\HSVar{j}\;\HSVar{i}\;\HSVar{k}\HSSpecial{\HSSym{\mskip1.5mu\}}}\;\HSSpecial{(}\HSVar{a}\HSCon{\mathbin{:}}\HSCon{AdvPath}\;\HSVar{j}\;\HSVar{i}\HSSpecial{)}\;\HSSpecial{(}\HSVar{ik}\HSCon{\mathbin{:}}\HSVar{i}\HSSym{<}\HSVar{k}\HSSpecial{)}\;\HSSpecial{(}\HSVar{kj}\HSCon{\mathbin{:}}\HSVar{k}\HSSym{\leq} \HSVar{j}\HSSpecial{)}{}\<[E]%
\\
\>[17]{}\HSSym{\to} \HV{\hbox{\it{last\guydash{}bef}}}\;\HSVar{a}\;\HSVar{ik}\;\HSVar{kj}\;\HT{\epsilon_{\textsc{ap}}}\;\HSVar{a}{}\<[E]%
\ColumnHook
\end{hscode}\resethooks
\end{myhs}

We also prove that \ensuremath{\HV{\mathcal{M}}\;\HT{\epsilon_{\textsc{ap}}}\;\HV{a_{22}}} and \ensuremath{\HV{\mathcal{M}}\;\HT{\epsilon_{\textsc{ap}}}\;\HV{\hbox{\it mbr\guydash{}path}}\;\HV{m_{2}}}, via two applications
of the following lemma, which is based on Lemma 5 of the original manual proof~\cite{Baker2003}, hence the name:

\begin{myhs}
\begin{hscode}\SaveRestoreHook
\column{B}{@{}>{\hspre}l<{\hspost}@{}}%
\column{9}{@{}>{\hspre}c<{\hspost}@{}}%
\column{9E}{@{}l@{}}%
\column{13}{@{}>{\hspre}l<{\hspost}@{}}%
\column{E}{@{}>{\hspre}l<{\hspost}@{}}%
\>[B]{}\HSVar{lemma5}{}\<[9]%
\>[9]{}\HSCon{\mathbin{:}}{}\<[9E]%
\>[13]{}\HS{\forall}\;\HSSpecial{\HSSym{\{\mskip1.5mu} }\HSVar{j}\;\HSVar{i}\;\HSVar{k}\HSSpecial{\HSSym{\mskip1.5mu\}}}\;\HSSpecial{(}\HSVar{a}\HSCon{\mathbin{:}}\HSCon{AdvPath}\;\HSVar{j}\;\HSVar{i}\HSSpecial{)}\;\HSSpecial{(}\HSVar{ik}\HSCon{\mathbin{:}}\HSVar{i}\HSSym{<}\HSVar{k}\HSSpecial{)}\;\HSSpecial{(}\HSVar{kj}\HSCon{\mathbin{:}}\HSVar{k}\HSSym{\leq} \HSVar{j}\HSSpecial{)}{}\<[E]%
\\
\>[9]{}\HSSym{\to} {}\<[9E]%
\>[13]{}\HS{\forall}\;\HSSpecial{\HSSym{\{\mskip1.5mu} }\HV{i_0}\HSSpecial{\HSSym{\mskip1.5mu\}}}\;\HSSpecial{(}\HSVar{b}\HSCon{\mathbin{:}}\HSCon{AdvPath}\;\HSVar{k}\;\HV{i_0}\HSSpecial{)}\HSSym{\to} \HV{i_0}\HSSym{\leq} \HSVar{i}{}\<[E]%
\\
\>[9]{}\HSSym{\to} {}\<[9E]%
\>[13]{}\HV{\hbox{\it{last\guydash{}bef}}}\;\HSVar{a}\;\HSVar{ik}\;\HSVar{kj}\;\HT{\epsilon_{\textsc{ap}}}\;\HSVar{b}{}\<[E]%
\ColumnHook
\end{hscode}\resethooks
\end{myhs}

Lemma 5 ensures that, given an advancement proof \ensuremath{\HSVar{a}} from \ensuremath{\HSVar{j}} to \ensuremath{\HSVar{i}} and another
advancement proof \ensuremath{\HSVar{b}} from \ensuremath{\HSVar{k}} to \ensuremath{\HV{i_0}}, where \ensuremath{\HSVar{i}\HSSym{<}\HSVar{k}\HSSym{\leq} \HSVar{j}} and \ensuremath{\HV{i_0}\HSSym{\leq} \HSVar{i}}, the last index
in \ensuremath{\HSVar{a}} before index \ensuremath{\HSVar{k}} is also in \ensuremath{\HSVar{b}}.  The key to proving
this lemma is the \ensuremath{\HV{\hbox{\it hop\guydash{}no\guydash{}cross}}} property: because \ensuremath{\HSVar{b}} is an \ensuremath{\HSCon{AdvPath}\;\HSVar{k}\;\HV{i_0}}, \ensuremath{\HV{\hbox{\it{last\guydash{}bef}}}\;\HSVar{a}\;\HSVar{ik}\;\HSVar{kj}\HSSym{<}\HSVar{k}},
and \ensuremath{\HV{i_0}\HSSym{\leq} \HSVar{i}\HSSym{\leq} \HV{\hbox{\it{last\guydash{}bef}}}\;\HSVar{a}\;\HSVar{ik}\;\HSVar{kj}}, \ensuremath{\HSVar{b}} must pass index \ensuremath{\HV{\hbox{\it{last\guydash{}bef}}}\;\HSVar{a}\;\HSVar{ik}\;\HSVar{kj}}; due to
\ensuremath{\HV{\hbox{\it hop\guydash{}no\guydash{}cross}}}, it cannot do so without visiting index \ensuremath{\HV{\hbox{\it{last\guydash{}bef}}}\;\HSVar{a}\;\HSVar{ik}\;\HSVar{kj}}.

Similarly, we identify the second special index \ensuremath{\HV{\mathcal{R}}} (corresponding to \ensuremath{\HSVar{r}} in \cite[Fig. 4]{Baker2003}) and prove that \ensuremath{\HV{\mathcal{R}}\;\HT{\epsilon_{\textsc{ap}}}\;\HV{\hbox{\it mbr\guydash{}path}}\;\HV{m_{1}}}, \ensuremath{\HV{\mathcal{R}}\;\HT{\epsilon_{\textsc{ap}}}\;\HV{\hbox{\it mbr\guydash{}path}}\;\HV{m_{2}}} and
\ensuremath{\HV{\mathcal{R}}\;\HT{\epsilon_{\textsc{ap}}}\;\HV{a_{12}}}.  If \ensuremath{\HSVar{tgt}\HSSym{\mathrel{=}}\HV{s_1}}, then that index serves the purpose of \ensuremath{\HV{\mathcal{R}}}, again because every
\ensuremath{\HSCon{AdvPath}} includes its endpoints.  Otherwise, \ensuremath{\HSVar{tgt}\HSSym{<}\HV{s_1}}, and we show that the \emph{first}
index in \ensuremath{\HV{a_{12}}} \emph{after} \ensuremath{\HSVar{tgt}} can serve as \ensuremath{\HV{\mathcal{R}}}, using similar machinery as above
(\ensuremath{\HV{\hbox{\it{first\guydash{}aft}}}}, \ensuremath{\HV{\hbox{\it{first\guydash{}aft\guydash{}correct}}}} and \ensuremath{\HV{lemma5^{\prime}}}, not shown).

Having established the two special indexes \ensuremath{\HV{\mathcal{M}}} and \ensuremath{\HV{\mathcal{R}}}, and showing that each of them is common to
three advancement proofs, we then proceed to prove that
the authenticators computed for index \ensuremath{\HSVar{tgt}} when rebuilding \ensuremath{\HV{m_{1}}} and \ensuremath{\HV{m_{2}}} are equal.  Our proof
is based on several applications of \ensuremath{\HV{\textsc{AgreeOnCommon}}}
and its variant \ensuremath{\HV{\hbox{\textsc{AgreeOnCommon}\guydash{}$\in$}}}, as well as some smaller lemmas including \ensuremath{\HV{\hbox{\it rebuild\guydash{}}\oplus}}.
Next we look at the individual proof steps together with an illustration on how they
are related, in \Cref{fig:evocr}.

\newcommand{\optionalLabel}[1]{}

\begin{enumerate}
\itemsep1.5em
 \item First, we notice that \ensuremath{\HV{a_{1}}\HSSym{\mathrel{=}}\HV{a_{11}}\;\mathbin{\HV{\oplus}}\;\HV{a_{12}}} and \ensuremath{\HV{a_{2}}\HSSym{\mathrel{=}}\HV{a_{21}}\;\mathbin{\HV{\oplus}}\;\HV{a_{22}}} arrive and
        agree at \ensuremath{\HSVar{j}}: \ensuremath{\HSVar{rebuild}\;\HV{a_{1}}\;\HV{t_2}\;\HSVar{j}\HSSym{\equiv} \HSVar{rebuild}\;\HV{a_{2}}\;\HV{t_2}\;\HSVar{j}}.
        Now define \ensuremath{\HV{\mathcal{M}}} as \ensuremath{\HV{\hbox{\it{last\guydash{}bef}}}\;\HV{a_{11}}\;\HSVar{hyp}}, where \ensuremath{\HSVar{hyp}} is the hypothesis that \ensuremath{\HV{s_1}\HSSym{<}\HV{s_2}},
        we have that \ensuremath{\HV{\mathcal{M}}\;\HT{\epsilon_{\textsc{ap}}}\;\HV{a_{22}}} and \ensuremath{\HV{\mathcal{M}}\;\HT{\epsilon_{\textsc{ap}}}\;\HV{a_{11}}}. A first application of \ensuremath{\HV{\textsc{AgreeOnCommon}}} yields:
        \[ \ensuremath{\HSVar{rebuild}\;\HV{a_{1}}\;\HV{t_1}\;\HV{\mathcal{M}}\HSSym{\equiv} \HSVar{rebuild}\;\HV{a_{2}}\;\HV{t_2}\;\HV{\mathcal{M}}} \]                                        \optionalLabel{(1) M-a1a2}

 \item  Another application of \ensuremath{\HV{\textsc{AgreeOnCommon}}} with the hypothesis that \ensuremath{\HV{\mathit{rebuildMbr}}\;\HV{m_{2}}\;\HV{u_2}\;\HV{s_2}\HSSym{\equiv} \HSVar{rebuild}\;\HV{a_{2}}\;\HV{t_2}\;\HV{s_2}} and with \ensuremath{\HV{\hbox{\it rebuild\guydash{}}\oplus}\;\HV{a_{21}}\;\HV{a_{22}}}, yields:
        \[ \ensuremath{\HSVar{rebuild}\;\HV{a_{22}}\;\HV{t_2}\;\HV{\mathcal{M}}\HSSym{\equiv} \HV{\mathit{rebuildMbr}}\;\HV{m_{2}}\;\HV{u_2}\;\HV{\mathcal{M}}} \]                                    \optionalLabel{(2) M-a22m2}

 \item Combining (1) and (2) above with another application of \ensuremath{\HV{\hbox{\it rebuild\guydash{}}\oplus}} yields
        \[ \ensuremath{\HSVar{rebuild}\;\HSSpecial{(}\HV{a_{11}}\;\mathbin{\HV{\oplus}}\;\HV{a_{12}}\HSSpecial{)}\;\HV{t_1}\;\HV{\mathcal{M}}\HSSym{\equiv} \HV{\mathit{rebuildMbr}}\;\HV{m_{2}}\;\HV{u_2}\;\HV{\mathcal{M}}} \]                         \optionalLabel{(3) M-a1m2}

 \item Now we split \ensuremath{\HV{\hbox{\it mbr\guydash{}path}}\;\HV{m_{2}}} at index \ensuremath{\HV{\mathcal{M}}}, yielding \ensuremath{\HV{m_{21}}} and \ensuremath{\HV{m_{22}}} (not labelled in \Cref{fig:evocr}) such that
        \ensuremath{\HV{\hbox{\it mbr\guydash{}path}}\;\HV{m_{2}}\HSSym{\equiv} \HV{m_{21}}\;\mathbin{\HV{\oplus}}\;\HV{m_{22}}}.  We then prove via (3) and another application of \ensuremath{\HV{\hbox{\it rebuild\guydash{}}\oplus}} that
       \[ \ensuremath{\HSVar{rebuild}\;\HSSpecial{(}\HV{a_{11}}\;\mathbin{\HV{\oplus}}\;\HV{a_{12}}\HSSpecial{)}\;\HV{t_1}\;\HV{\mathcal{M}}\HSSym{\equiv} \HSVar{rebuild}\;\HV{m_{22}}\;\HV{{u_2}^{\prime}}\;\HV{\mathcal{M}}} \]                       \optionalLabel{(4) M-a1m22}
        Here, \ensuremath{\HV{{u_2}^{\prime}}\HSSym{\mathrel{=}}\HSVar{insertAuth}\;\HV{u_2}\;\HSVar{tgt}\;\HSSpecial{(}\HV{\hbox{\it datum\guydash{}hash}}\;\HV{m_{2}}\HSSpecial{)}}.
        Thus, we have established that rebuilding \ensuremath{\HV{a_{1}}} yields the same hash at \ensuremath{\HV{\mathcal{M}}} as rebuilding \ensuremath{\HV{m_{2}}}, but rebuilding only the \ensuremath{\HV{m_{22}}} component.

 \item Having established in (3) that \ensuremath{\HV{a_{1}}\HSSym{\mathrel{=}}\HV{a_{11}}\;\mathbin{\HV{\oplus}}\;\HV{a_{12}}} rebuilds to the same hash as \ensuremath{\HV{m_{2}}} at index \ensuremath{\HV{\mathcal{M}}},
        we can now show that they also agree at index \ensuremath{\HV{\mathcal{R}}} (recall that \ensuremath{\HV{\mathcal{R}}\;\HT{\epsilon_{\textsc{ap}}}\;\HV{a_{12}}} and \ensuremath{\HV{\mathcal{R}}\;\HT{\epsilon_{\textsc{ap}}}\;\HV{m_{2}}}).
        To do so,
        we apply \ensuremath{\HV{\hbox{\textsc{AgreeOnCommon}\guydash{}$\in$}}} to \ensuremath{\HV{a_{1}}} and \ensuremath{\HV{m_{22}}} using (4), yielding:
        \[ \ensuremath{\HSVar{rebuild}\;\HSSpecial{(}\HV{a_{11}}\;\mathbin{\HV{\oplus}}\;\HV{a_{12}}\HSSpecial{)}\;\HV{t_1}\;\HV{\mathcal{R}}\HSSym{\equiv} \HSVar{rebuild}\;\HV{m_{22}}\;\HV{{u_2}^{\prime}}\;\HV{\mathcal{R}}} \]                      \optionalLabel{(5) R-a1m22}
        This invocation of \ensuremath{\HV{\hbox{\textsc{AgreeOnCommon}\guydash{}$\in$}}} requires
        evidence that \ensuremath{\HV{\mathcal{R}}\;\HT{\epsilon_{\textsc{ap}}}\;\HV{m_{22}}}, which is proved using
        some simple lemmas (not shown) that reason about whether an index
        in an \ensuremath{\HSCon{AdvPath}} that is
        formed by composing two \ensuremath{\HSCon{AdvPath}}s using \ensuremath{\mathbin{\HV{\oplus}}} is in the first or second \ensuremath{\HSCon{AdvPath}},
        depending on its index.

 \item Analogous to (2), we again invoke \ensuremath{\HV{\textsc{AgreeOnCommon}}} using the hypothesis that
       \ensuremath{\HV{\mathit{rebuildMbr}}\;\HV{m_{1}}\;\HV{u_1}\;\HV{s_1}\HSSym{\equiv} \HSVar{rebuild}\;\HV{a_{1}}\;\HV{t_1}\;\HV{s_1}} in combination with \ensuremath{\HV{\hbox{\it rebuild\guydash{}}\oplus}} to show
        that rebuilding \ensuremath{\HV{a_{12}}} and \ensuremath{\HV{m_{1}}} yield the same hash at index \ensuremath{\HV{\mathcal{R}}}.
        \[ \ensuremath{\HSVar{rebuild}\;\HV{a_{12}}\;\HV{t_1}\;\HV{\mathcal{R}}\HSSym{\equiv} \HV{\mathit{rebuildMbr}}\;\HV{m_{1}}\;\HV{u_1}\;\HV{\mathcal{R}}} \]                                    \optionalLabel{(6) R-a12m1}

 \item Next, we split \ensuremath{\HV{m_{1}}} at index \ensuremath{\HV{\mathcal{R}}}, resulting in \ensuremath{\HV{\hbox{\it mbr\guydash{}path}}\;\HV{m_{1}}\HSSym{\equiv} \HV{m_{11}}\;\mathbin{\HV{\oplus}}\;\HV{m_{12}}} (not labelled in~\Cref{fig:evocr}).
       Then, combining (5) and (6) and using \ensuremath{\HV{\hbox{\it rebuild\guydash{}}\oplus}} twice, we establish that
       rebuilding \ensuremath{\HV{m_{22}}} and \ensuremath{\HV{m_{12}}} both result in the same hash at index \ensuremath{\HV{\mathcal{R}}}:
        \[ \ensuremath{\HSVar{rebuild}\;\HV{m_{22}}\;\HV{{u_2}^{\prime}}\;\HV{\mathcal{R}}\HSSym{\equiv} \HSVar{rebuild}\;\HV{m_{12}}\;\HV{{u_1}^{\prime}}\;\HV{\mathcal{R}}} \]                            \optionalLabel{(7) R-m22m12}
       (Here, \ensuremath{\HV{{u_1}^{\prime}}} is defined analogously to \ensuremath{\HV{{u_2}^{\prime}}}.)
       At last we know that the rebuilds of \ensuremath{\HV{m_{12}}} and \ensuremath{\HV{m_{22}}} agree at an index (\ensuremath{\HV{\mathcal{R}}}) that they share; next
       we work back from there to show that they also agree at \ensuremath{\HSVar{tgt}}.

  \item By applying \ensuremath{\HV{\hbox{\textsc{AgreeOnCommon}\guydash{}$\in$}}} using (7), we can conclude that the views achieved by rebuilding
        \ensuremath{\HV{m_{22}}} and \ensuremath{\HV{m_{12}}} agree at index \ensuremath{\HSVar{tgt}}.
        \[ \ensuremath{\HSVar{rebuild}\;\HV{m_{22}}\;\HV{{u_2}^{\prime}}\;\HSVar{tgt}\HSSym{\equiv} \HSVar{rebuild}\;\HV{m_{12}}\;\HV{{u_1}^{\prime}}\;\HSVar{tgt}} \]                        \optionalLabel{(8) tgt-m22m12}

  \item Applying \ensuremath{\HV{\hbox{\it rebuild\guydash{}}\oplus}} twice to (8) yields
        \ensuremath{\HSVar{rebuild}\;\HSSpecial{(}\HV{m_{21}}\;\mathbin{\HV{\oplus}}\;\HV{m_{22}}\HSSpecial{)}\;\HV{{u_2}^{\prime}}\;\HSVar{tgt}\HSSym{\equiv} \HSVar{rebuild}\;\HSSpecial{(}\HV{m_{11}}\;\mathbin{\HV{\oplus}}\;\HV{m_{12}}\HSSpecial{)}\;\HV{{u_1}^{\prime}}\;\HSVar{tgt}}.       \optionalLabel{(9) tgt-m1m2}

  \item Finally, the observation that rebuilding an
        \ensuremath{\HSCon{AdvPath}\;\HSVar{j}\;\HSVar{i}} using view \ensuremath{\HSVar{t}} does not modify \ensuremath{\HSVar{t}} at \ensuremath{\HSVar{i}} establishes that the
        authenticators computed for index \ensuremath{\HSVar{tgt}} when rebuilding \ensuremath{\HV{m_{1}}} and \ensuremath{\HV{m_{2}}} are equal.
\end{enumerate}

  This concludes the abstract properties. We have shown how any value
of \ensuremath{\HSCon{DepRel}} will yield a skiplog that enjoys \ensuremath{\HV{\textsc{AgreeOnCommon}}} and \ensuremath{\HV{\textsc{evo-cr}}}.
The next step is proving that we can inhabit the \ensuremath{\HSCon{DepRel}} datatype
with the skiplog described in \Cref{sec:skiplog}.

\subsection{Instantiating the Abstract Model}
\label{sec:concretemodel}

Having proved that any instantiation of \ensuremath{\HSCon{DepRel}}
enjoys the \ensuremath{\HV{\textsc{AgreeOnCommon}}} and \ensuremath{\HV{\textsc{evo-cr}}} properties, we now instantiate \ensuremath{\HSCon{DepRel}}
with the hop relation defined in \Cref{sec:skiplog}: we
define \ensuremath{\HSVar{maxlvl}} to be equivalent to the
definition presented there (see below) and define hop targets as follows.

\begin{myhs}
\begin{hscode}\SaveRestoreHook
\column{B}{@{}>{\hspre}l<{\hspost}@{}}%
\column{E}{@{}>{\hspre}l<{\hspost}@{}}%
\>[B]{}\HV{\hbox{\it hop\guydash{}tgt}}\HSCon{\mathbin{:}}\HSSpecial{\HSSym{\{\mskip1.5mu} }\HSVar{j}\HSCon{\mathbin{:}}\HT{\mathbb{N}}\HSSpecial{\HSSym{\mskip1.5mu\}}}\HSSym{\to} \HSCon{HopFrom}\;\HSVar{j}\HSSym{\to} \HT{\mathbb{N}}{}\<[E]%
\\
\>[B]{}\HV{\hbox{\it hop\guydash{}tgt}}\;\HSSpecial{\HSSym{\{\mskip1.5mu} }\HSVar{j}\HSSpecial{\HSSym{\mskip1.5mu\}}}\;\HSVar{h}\HSSym{\mathrel{=}}\HSVar{j}\;\HS{\dotminus}\;\HSSpecial{(}\HSNumeral{2}^{\HV{\mathit{to}\mathbb{N}}\;\HSVar{h}}\HSSpecial{)}{}\<[E]%
\ColumnHook
\end{hscode}\resethooks
\end{myhs}

  We use Agda's \ensuremath{\HS{\dotminus}} operator, which truncates negative subtraction results to
zero.  Reasoning is simplified by a lemma proving that hop targets never ``overshoot'' zero.

In this section, we explain how we provide the remaining properties
required to prove that the hop relation arising from the definitions of
\ensuremath{\HSVar{maxlvl}} and \ensuremath{\HV{\hbox{\it hop\guydash{}tgt}}}
satisfies the requirements of \ensuremath{\HSCon{DepRel}}.
Most of them are straightforward.
However, proving that this relation satisfies \ensuremath{\HV{\hbox{\it hop\guydash{}no\guydash{}cross}}} is quite challenging.
The first step towards a manageable proof is
to define suitable induction principles over the hops we seek to analyze.

In our case, we want to prove a number of lemmas over the structure
that arises from establishing $j - 2^l$ as a dependency of $j$,
for $l$ less than the largest power of two that divides $j$.
Because $j - 2^{l+1} = (j - 2^l) - 2^l$, we can form hops at level
\ensuremath{\HSVar{l}\HSSym{+}\HSNumeral{1}} by composing adjacent hops at level \ensuremath{\HSVar{l}}.

This structure is encoded in a simple Agda datatype:

\begin{myhs}
\begin{hscode}\SaveRestoreHook
\column{B}{@{}>{\hspre}l<{\hspost}@{}}%
\column{4}{@{}>{\hspre}l<{\hspost}@{}}%
\column{8}{@{}>{\hspre}l<{\hspost}@{}}%
\column{E}{@{}>{\hspre}l<{\hspost}@{}}%
\>[B]{}\HSKeyword{data}\;\HSCon{H}\HSCon{\mathbin{:}}\HT{\mathbb{N}}\HSSym{\to} \HT{\mathbb{N}}\HSSym{\to} \HT{\mathbb{N}}\HSSym{\to} \HSCon{Set}\;\HSKeyword{where}{}\<[E]%
\\
\>[B]{}\hsindent{4}{}\<[4]%
\>[4]{}\HSVar{hz}{}\<[8]%
\>[8]{}\HSCon{\mathbin{:}}\HS{\forall}\;\HSVar{x}\HSSym{\to} \HSCon{H}\;\HSNumeral{0}\;\HSSpecial{(}\HSVar{suc}\;\HSVar{x}\HSSpecial{)}\;\HSVar{x}{}\<[E]%
\\
\>[B]{}\hsindent{4}{}\<[4]%
\>[4]{}\HSVar{hs}{}\<[8]%
\>[8]{}\HSCon{\mathbin{:}}\HS{\forall}\;\HSSpecial{\HSSym{\{\mskip1.5mu} }\HSVar{l}\;\HSVar{x}\;\HSVar{y}\;\HSVar{z}\HSSpecial{\HSSym{\mskip1.5mu\}}}{}\<[E]%
\\
\>[8]{}\HSSym{\to} \HSCon{H}\;\HSVar{l}\;\HSVar{x}\;\HSVar{y}\HSSym{\to} \HSCon{H}\;\HSVar{l}\;\HSVar{y}\;\HSVar{z}{}\<[E]%
\\
\>[8]{}\HSSym{\to} \HSVar{suc}\;\HSVar{l}\HSSym{<}\HSVar{maxLvl}\;\HSVar{x}{}\<[E]%
\\
\>[8]{}\HSSym{\to} \HSCon{H}\;\HSSpecial{(}\HSVar{suc}\;\HSVar{l}\HSSpecial{)}\;\HSVar{x}\;\HSVar{z}{}\<[E]%
\ColumnHook
\end{hscode}\resethooks
\end{myhs}

  A value \ensuremath{\HSVar{h}} of type \ensuremath{\HSCon{H}\;\HSVar{l}\;\HSVar{j}\;\HSVar{i}} proves the existence of a hop at level
\ensuremath{\HSVar{l}} connecting \ensuremath{\HSVar{j}} and \ensuremath{\HSVar{i}}.  The constraint that \ensuremath{\HSVar{suc}\;\HSVar{l}\HSSym{<}\HSVar{maxLvl}\;\HSVar{x}}
ensures that hops are created from index \ensuremath{\HSVar{x}} only for levels up to \ensuremath{\HSVar{maxLvl}\;\HSVar{x}}.
Without this constraint, for example,
a hop from index 6 to index 2 could be constructed, which
would cross the hop from 0 to 4.  As neither of these hops is nested
within the other, this would violate the \ensuremath{\HV{\hbox{\it hop\guydash{}no\guydash{}cross}}} property, thus
preventing a proof that our hop relation inhabits \ensuremath{\HSCon{DepRel}}.

Another aspect that complicated our earlier direct proof
efforts was the difficulty of dealing inductively with the original
recursive definition of the \ensuremath{\HSVar{maxLvl}} function, defined in Agda as:

\begin{myhs}
\begin{hscode}\SaveRestoreHook
\column{B}{@{}>{\hspre}l<{\hspost}@{}}%
\column{10}{@{}>{\hspre}l<{\hspost}@{}}%
\column{13}{@{}>{\hspre}l<{\hspost}@{}}%
\column{E}{@{}>{\hspre}l<{\hspost}@{}}%
\>[B]{}\HSVar{maxLvl}\HSCon{\mathbin{:}}\HT{\mathbb{N}}\HSSym{\to} \HT{\mathbb{N}}{}\<[E]%
\\
\>[B]{}\HSVar{maxLvl}\;\HSNumeral{0}\HSSym{\mathrel{=}}\HSNumeral{0}{}\<[E]%
\\
\>[B]{}\HSVar{maxLvl}\;\HSSpecial{(}\HSVar{suc}\;\HSVar{n}\HSSpecial{)}\;\HSKeyword{with}\;\HSNumeral{2}\HSSym{|?}\HSSpecial{(}\HSVar{suc}\;\HSVar{n}\HSSpecial{)}{}\<[E]%
\\
\>[B]{}\HSSym{...|}\HSVar{no}\;{}\<[10]%
\>[10]{}\HSSym{\anonymous} {}\<[13]%
\>[13]{}\HSSym{\mathrel{=}}\HSNumeral{1}{}\<[E]%
\\
\>[B]{}\HSSym{...|}\HSVar{yes}\;\HSVar{e}{}\<[13]%
\>[13]{}\HSSym{\mathrel{=}}\HSVar{suc}\;\HSSpecial{(}\HSVar{maxLvl}\;\HSSpecial{(}\HSVar{quotient}\;\HSVar{e}\HSSpecial{)}\HSSpecial{)}{}\<[E]%
\ColumnHook
\end{hscode}\resethooks
\end{myhs}

We tamed this complexity by observing that every non-zero natural number can be uniquely represented as
the product of a power of two and an odd number, and then using a non-inductive
definition of \ensuremath{\HSVar{maxLvl}} that is isomorphic to our original definition.
This is known as a \emph{view} type in the literature~\cite{McBride2004}.

\begin{myhs}
\begin{hscode}\SaveRestoreHook
\column{B}{@{}>{\hspre}l<{\hspost}@{}}%
\column{3}{@{}>{\hspre}l<{\hspost}@{}}%
\column{9}{@{}>{\hspre}l<{\hspost}@{}}%
\column{E}{@{}>{\hspre}l<{\hspost}@{}}%
\>[B]{}\HSKeyword{data}\;\HSCon{EvenOdd}\HSCon{\mathbin{:}}\HT{\mathbb{N}}\HSSym{\to} \HSCon{Set}\;\HSKeyword{where}{}\<[E]%
\\
\>[B]{}\hsindent{3}{}\<[3]%
\>[3]{}\HSVar{zero}{}\<[9]%
\>[9]{}\HSCon{\mathbin{:}}\HSCon{EvenOdd}\;\HSVar{zero}{}\<[E]%
\\
\>[B]{}\hsindent{3}{}\<[3]%
\>[3]{}\HSVar{nz}{}\<[9]%
\>[9]{}\HSCon{\mathbin{:}}\HS{\forall}\;\HSSpecial{\HSSym{\{\mskip1.5mu} }\HSVar{n}\HSSpecial{\HSSym{\mskip1.5mu\}}}\;\HSVar{l}\;\HSVar{d}\HSSym{\to} \HSCon{Odd}\;\HSVar{d}\HSSym{\to} \HSVar{n}\HSSym{\equiv} 2^{\HSVar{l}}\HSSym{*}\HSVar{d}\HSSym{\to} \HSCon{EvenOdd}\;\HSVar{n}{}\<[E]%
\\[\blanklineskip]%
\>[B]{}\HSVar{to}\HSCon{\mathbin{:}}\HSSpecial{(}\HSVar{n}\HSCon{\mathbin{:}}\HT{\mathbb{N}}\HSSpecial{)}\HSSym{\to} \HSCon{EvenOdd}\;\HSVar{n}{}\<[E]%
\ColumnHook
\end{hscode}\resethooks
\end{myhs}

  This enables us to write a non-inductive version of \ensuremath{\HSVar{maxLvl}} that simply
extracts this largest power of two.

\begin{myhs}
\begin{hscode}\SaveRestoreHook
\column{B}{@{}>{\hspre}l<{\hspost}@{}}%
\column{24}{@{}>{\hspre}l<{\hspost}@{}}%
\column{E}{@{}>{\hspre}l<{\hspost}@{}}%
\>[B]{}\HSVar{maxLvl'}\HSCon{\mathbin{:}}\HS{\forall}\;\HSSpecial{\HSSym{\{\mskip1.5mu} }\HSVar{n}\HSSpecial{\HSSym{\mskip1.5mu\}}}\HSSym{\to} \HSCon{EvenOdd}\;\HSVar{n}\HSSym{\to} \HT{\mathbb{N}}{}\<[E]%
\\
\>[B]{}\HSVar{maxLvl'}\;\HSVar{zero}{}\<[24]%
\>[24]{}\HSSym{\mathrel{=}}\HSVar{zero}{}\<[E]%
\\
\>[B]{}\HSVar{maxLvl'}\;\HSSpecial{(}\HSVar{nz}\;\HSVar{l}\;\HSSym{\anonymous} \;\HSSym{\anonymous} \;\HSSym{\anonymous} \HSSpecial{)}{}\<[24]%
\>[24]{}\HSSym{\mathrel{=}}\HSVar{suc}\;\HSVar{l}{}\<[E]%
\ColumnHook
\end{hscode}\resethooks
\end{myhs}

  The definition of \ensuremath{\HSVar{maxLvl'}} is provably equivalent to the recursive version \ensuremath{\HSVar{maxLvl}}
used in the specification. The proof is an easy induction on \ensuremath{\HSVar{n}}, and
the lemma is used frequently throughout the proofs in our model and
has been invaluable in helping us complete the proof.

\begin{myhs}
\begin{hscode}\SaveRestoreHook
\column{B}{@{}>{\hspre}l<{\hspost}@{}}%
\column{E}{@{}>{\hspre}l<{\hspost}@{}}%
\>[B]{}\HV{maxLvl\!\equiv\!maxLvl}\HSCon{\mathbin{:}}\HS{\forall}\;\HSVar{n}\HSSym{\to} \HSVar{maxLvl}\;\HSVar{n}\HSSym{\equiv} \HSVar{maxLvl'}\;\HSSpecial{(}\HSVar{to}\;\HSVar{n}\HSSpecial{)}{}\<[E]%
\ColumnHook
\end{hscode}\resethooks
\end{myhs}

   Next, we must be able to inhabit the type \ensuremath{\HSCon{H}}, witnessing the existence of
hops from a given index $j$ into $j - 2^l$, as long as $l$ is less than \ensuremath{\HSVar{maxLvl}\;\HSVar{j}}.
This proves that the datatype is
inhabited and meets the criteria we expect: i.e., it connects $j$ and its
dependencies one power of two away.

\begin{myhs}
\begin{hscode}\SaveRestoreHook
\column{B}{@{}>{\hspre}l<{\hspost}@{}}%
\column{E}{@{}>{\hspre}l<{\hspost}@{}}%
\>[B]{}\HV{\hbox{\it h\guydash{}correct}}\HSCon{\mathbin{:}}\HS{\forall}\;\HSVar{j}\;\HSVar{l}\HSSym{\to} \HSVar{l}\HSSym{<}\HSVar{maxLvl}\;\HSVar{j}\HSSym{\to} \HSCon{H}\;\HSVar{l}\;\HSVar{j}\;\HSSpecial{(}\HSVar{j}\HSSym{-}2^{\HSVar{l}}\HSSpecial{)}{}\<[E]%
\ColumnHook
\end{hscode}\resethooks
\end{myhs}

  Conversely, given \ensuremath{\HSVar{h}\HSCon{\mathbin{:}}\HSCon{H}\;\HSVar{l}\;\HSVar{j}\;\HSVar{i}}, we can prove that $i$ is
a $2^l$ away from $j$, which shows that the \ensuremath{\HSCon{H}} datatype
encodes exactly the required structure.

\begin{myhs}
\begin{hscode}\SaveRestoreHook
\column{B}{@{}>{\hspre}l<{\hspost}@{}}%
\column{E}{@{}>{\hspre}l<{\hspost}@{}}%
\>[B]{}\HV{\hbox{\it h\guydash{}univ}_i}\HSCon{\mathbin{:}}\HS{\forall}\;\HSSpecial{\HSSym{\{\mskip1.5mu} }\HSVar{l}\;\HSVar{i}\;\HSVar{j}\HSSpecial{\HSSym{\mskip1.5mu\}}}\HSSym{\to} \HSCon{H}\;\HSVar{l}\;\HSVar{j}\;\HSVar{i}\HSSym{\to} \HSVar{i}\HSSym{\equiv} \HSVar{j}\HSSym{-}2^{\HSVar{l}}{}\<[E]%
\ColumnHook
\end{hscode}\resethooks
\end{myhs}

A central lemma used in the proof that our hop relation
satisfies the \ensuremath{\HV{\hbox{\it hop\guydash{}no\guydash{}cross}}} property
is that all the indexes that
a hop skips over have a level lower than the level of the hop.
In other words, if a hop from $j$ to $i$ at level $l$
hops over index $k$, then \ensuremath{\HSVar{maxLvl}\;\HSVar{k}} is at most $l$.

\begin{myhs}
\begin{hscode}\SaveRestoreHook
\column{B}{@{}>{\hspre}l<{\hspost}@{}}%
\column{E}{@{}>{\hspre}l<{\hspost}@{}}%
\>[B]{}\HV{\hbox{\it h\guydash{}lvl\guydash{}mid}}\HSCon{\mathbin{:}}\HS{\forall}\;\HSSpecial{\HSSym{\{\mskip1.5mu} }\HSVar{l}\;\HSVar{j}\;\HSVar{i}\HSSpecial{\HSSym{\mskip1.5mu\}}}\;\HSVar{k}\HSSym{\to} \HSCon{H}\;\HSVar{l}\;\HSVar{j}\;\HSVar{i}\HSSym{\to} \HSVar{i}\HSSym{<}\HSVar{k}\HSSym{\to} \HSVar{k}\HSSym{<}\HSVar{j}\HSSym{\to} \HSVar{maxLvl}\;\HSVar{k}\HSSym{\leq} \HSVar{l}{}\<[E]%
\ColumnHook
\end{hscode}\resethooks
\end{myhs}

Next, we discuss how we prove
that our hop relation satisfies the \ensuremath{\HV{\hbox{\it hop\guydash{}no\guydash{}cross}}} property required by \ensuremath{\HSCon{DepRel}}.
Recalling the definition of this property, as illustrated in~\Cref{fig:hopnocross},
it would suffice to prove the following property:
\begin{myhs}
\begin{hscode}\SaveRestoreHook
\column{B}{@{}>{\hspre}l<{\hspost}@{}}%
\column{11}{@{}>{\hspre}l<{\hspost}@{}}%
\column{E}{@{}>{\hspre}l<{\hspost}@{}}%
\>[B]{}\HSVar{nocross'}{}\<[11]%
\>[11]{}\HSCon{\mathbin{:}}\HS{\forall}\;\HSSpecial{\HSSym{\{\mskip1.5mu} }\HV{l_1}\;\HV{i_1}\;\HV{j_1}\;\HV{l_2}\;\HV{i_2}\;\HV{j_2}\HSSpecial{\HSSym{\mskip1.5mu\}}}\;\HSSpecial{(}\HV{h_1}\HSCon{\mathbin{:}}\HSCon{H}\;\HV{l_1}\;\HV{j_1}\;\HV{i_1}\HSSpecial{)}\;\HSSpecial{(}\HV{h_2}\HSCon{\mathbin{:}}\HSCon{H}\;\HV{l_2}\;\HV{j_2}\;\HV{i_2}\HSSpecial{)}{}\<[E]%
\\
\>[11]{}\HSSym{\to} \HV{i_2}\HSSym{<}\HV{i_1}\HSSym{\to} \HSCon{Either}\;\HSSpecial{(}\HV{j_2}\HSSym{\leq} \HV{i_1}\HSSpecial{)}\;\HSSpecial{(}\HV{j_1}\HSSym{\leq} \HV{j_2}\HSSpecial{)}{}\<[E]%
\ColumnHook
\end{hscode}\resethooks
\end{myhs}

  This property captures the intuition of hops \emph{not crossing}:
given any two hops that do not share an index (w.l.o.g. \ensuremath{\HV{i_2}\HSSym{<}\HV{i_1}}),
either they do not overlap or one is contained within the other.
(Note that two hops that share an index do not cross.)
  Our earlier attempts to
prove \ensuremath{\HSVar{nocross'}} directly became unmanageable due to the combined
effects of dealing with arithmetic nuances,
seemingly minor details such as subtraction in Agda truncating to zero
to ensure the result is always a natural, and the fact that
recursive
calls do not directly inform us about the relationship between \ensuremath{\HV{l_1}} and \ensuremath{\HV{l_2}}.

Therefore, we instead prove \ensuremath{\HSVar{hopcross}} via a stronger property and an auxiliary lemma:

\begin{myhs}
\begin{hscode}\SaveRestoreHook
\column{B}{@{}>{\hspre}l<{\hspost}@{}}%
\column{13}{@{}>{\hspre}l<{\hspost}@{}}%
\column{E}{@{}>{\hspre}l<{\hspost}@{}}%
\>[B]{}\HSVar{nocross}{}\<[13]%
\>[13]{}\HSCon{\mathbin{:}}\HS{\forall}\;\HSSpecial{\HSSym{\{\mskip1.5mu} }\HV{l_1}\;\HV{i_1}\;\HV{j_1}\;\HV{l_2}\;\HV{i_2}\;\HV{j_2}\HSSpecial{\HSSym{\mskip1.5mu\}}}\;\HSSpecial{(}\HV{h_1}\HSCon{\mathbin{:}}\HSCon{H}\;\HV{l_1}\;\HV{j_1}\;\HV{i_1}\HSSpecial{)}\;\HSSpecial{(}\HV{h_2}\HSCon{\mathbin{:}}\HSCon{H}\;\HV{l_2}\;\HV{j_2}\;\HV{i_2}\HSSpecial{)}{}\<[E]%
\\
\>[13]{}\HSSym{\to} \HV{i_2}\HSSym{<}\HV{i_1}\HSSym{\to} \HSCon{Either}\;\HSSpecial{(}\HV{j_2}\HSSym{\leq} \HV{i_1}\HSSpecial{)}\;\HSSpecial{(}\HV{h_1}\;\HT{\subseteq}\;\HV{h_2}\HSSpecial{)}{}\<[E]%
\\[\blanklineskip]%
\>[B]{}\HV{\subseteq\!\!-\!src\!-\!\!\leq}\HSCon{\mathbin{:}}\HS{\forall}\;\HSSpecial{\HSSym{\{\mskip1.5mu} }\HV{l_1}\;\HV{i_1}\;\HV{j_1}\;\HV{l_2}\;\HV{i_2}\;\HV{j_2}\HSSpecial{\HSSym{\mskip1.5mu\}}}\;\HSSpecial{(}\HV{h_1}\HSCon{\mathbin{:}}\HSCon{H}\;\HV{l_1}\;\HV{j_1}\;\HV{i_1}\HSSpecial{)}\;\HSSpecial{(}\HV{h_2}\HSCon{\mathbin{:}}\HSCon{H}\;\HV{l_2}\;\HV{j_2}\;\HV{i_2}\HSSpecial{)}{}\<[E]%
\\
\>[B]{}\hsindent{13}{}\<[13]%
\>[13]{}\HSSym{\to} \HV{h_1}\;\HT{\subseteq}\;\HV{h_2}\HSSym{\to} \HV{j_1}\HSSym{\leq} \HV{j_2}{}\<[E]%
\ColumnHook
\end{hscode}\resethooks
\end{myhs}

  The auxiliary lemma is proved via a straightforward induction on hops, using the following definition of \ensuremath{\HT{\subseteq}}.



\begin{myhs}
\begin{hscode}\SaveRestoreHook
\column{B}{@{}>{\hspre}l<{\hspost}@{}}%
\column{3}{@{}>{\hspre}l<{\hspost}@{}}%
\column{4}{@{}>{\hspre}l<{\hspost}@{}}%
\column{11}{@{}>{\hspre}l<{\hspost}@{}}%
\column{18}{@{}>{\hspre}l<{\hspost}@{}}%
\column{21}{@{}>{\hspre}l<{\hspost}@{}}%
\column{47}{@{}>{\hspre}l<{\hspost}@{}}%
\column{E}{@{}>{\hspre}l<{\hspost}@{}}%
\>[B]{}\HSKeyword{data}\;\HT{\_\!\subseteq\!\_}\HSCon{\mathbin{:}}\HS{\forall}\;\HSSpecial{\HSSym{\{\mskip1.5mu} }\HV{l_1}\;\HV{i_1}\;\HV{j_1}\;\HV{l_2}\;\HV{i_2}\;\HV{j_2}\HSSpecial{\HSSym{\mskip1.5mu\}}}\HSSym{\to} \HSCon{H}\;\HV{l_1}\;\HV{j_1}\;\HV{i_1}\HSSym{\to} \HSCon{H}\;\HV{l_2}\;\HV{j_2}\;\HV{i_2}\HSSym{\to} \HSCon{Set}{}\<[E]%
\\
\>[B]{}\hsindent{3}{}\<[3]%
\>[3]{}\HSKeyword{where}{}\<[E]%
\\
\>[3]{}\hsindent{1}{}\<[4]%
\>[4]{}\HSVar{here}{}\<[11]%
\>[11]{}\HSCon{\mathbin{:}}\HS{\forall}\;\HSSpecial{\HSSym{\{\mskip1.5mu} }\HSVar{l}\;\HSVar{i}\;\HSVar{j}\HSSpecial{\HSSym{\mskip1.5mu\}}}\;\HSSpecial{(}\HSVar{h}\HSCon{\mathbin{:}}\HSCon{H}\;\HSVar{l}\;\HSVar{j}\;\HSVar{i}\HSSpecial{)}\HSSym{\to} \HSVar{h}\;\HT{\subseteq}\;\HSVar{h}{}\<[E]%
\\[\blanklineskip]%
\>[3]{}\hsindent{1}{}\<[4]%
\>[4]{}\HSVar{left}{}\<[11]%
\>[11]{}\HSCon{\mathbin{:}}\HS{\forall}\;{}\<[21]%
\>[21]{}\HSSpecial{\HSSym{\{\mskip1.5mu} }\HV{l_1}\;\HV{i_1}\;\HV{j_1}\;\HV{l_2}\;\HV{i_2}\;\HSVar{w}\;\HV{j_2}\HSSpecial{\HSSym{\mskip1.5mu\}}}\;\HSSpecial{(}\HSVar{h}{}\<[47]%
\>[47]{}\HSCon{\mathbin{:}}\HSCon{H}\;\HV{l_1}\;\HV{j_1}\;\HV{i_1}\HSSpecial{)}\;{}\<[E]%
\\
\>[21]{}\HSSpecial{(}\HV{w_0}\HSCon{\mathbin{:}}\HSCon{H}\;\HV{l_2}\;\HV{j_2}\;\HSVar{w}\HSSpecial{)}\;\HSSpecial{(}\HV{w_1}\HSCon{\mathbin{:}}\HSCon{H}\;\HV{l_2}\;\HSVar{w}\;\HV{i_2}\HSSpecial{)}{}\<[E]%
\\
\>[11]{}\HSSym{\to} \HSSpecial{(}\HSVar{p}{}\<[18]%
\>[18]{}\HSCon{\mathbin{:}}\HSVar{suc}\;\HV{l_2}\HSSym{<}\HSVar{maxLvl}\;\HV{j_2}\HSSpecial{)}{}\<[E]%
\\
\>[11]{}\HSSym{\to} \HSVar{h}\;\HT{\subseteq}\;\HV{w_0}\HSSym{\to} \HSVar{h}\;\HT{\subseteq}\;\HSSpecial{(}\HSVar{hs}\;\HV{w_0}\;\HV{w_1}\;\HSVar{p}\HSSpecial{)}{}\<[E]%
\\[\blanklineskip]%
\>[3]{}\hsindent{1}{}\<[4]%
\>[4]{}\HSVar{right}{}\<[11]%
\>[11]{}\HSCon{\mathbin{:}}\HSSym{...}\HSComment{ -\! - analogous to left}{}\<[E]%
\ColumnHook
\end{hscode}\resethooks
\end{myhs}

This definition establishes that every hop contains itself, and then recursively defines a hop \ensuremath{\HSVar{h}}
as being contained by another if the latter is the composition of two adjacent hops, one of which
contains \ensuremath{\HSVar{h}}.  However, this simplified description is too inclusive.  Observe that the \ensuremath{\HSVar{left}} and \ensuremath{\HSVar{right}} constructors
require constraints to enable the construction of legitimate hops using the \ensuremath{\HSVar{hs}} constructor of \ensuremath{\HSCon{H}}.

With these definitions established, we can describe the proof
that our hop relation satisfies the \ensuremath{\HSVar{nocross}} property, which
proceeds by case analysis on hops.  To facilitate this case analysis,
we define an auxiliary relation that categorizes
the relationship between two
hops \ensuremath{\HV{h_1}} and \ensuremath{\HV{h_2}} into one of five possible situations, depending
          on the relative indexes of the hops' sources and targets:

\begin{enumerate}
\item The hops are equal, i.e.:
      \begin{center}
      \resizebox{.23\textwidth}{!}{%
      \begin{tikzpicture}[ every node/.style={scale=1.6} ]
        \node                      (h2) {\ensuremath{\HV{h_1}\HSSym{\equiv} \HV{h_2}}};
        \node [below left  = of h2] (tgt2) {\ensuremath{\HV{\hbox{\it hop\guydash{}tgt}}\;\HV{h_2}}};
        \node [below right = of h2] (j2) {\ensuremath{\HV{j_2}}};
        \draw [line width=0.25mm, ->] (j2) |- (h2.south) -| (tgt2);
      \end{tikzpicture}}
    \end{center}
\item The hops are disjoint, with \ensuremath{\HV{\hbox{\it hop\guydash{}src}}\;\HV{h_1}\HSSym{\leq} \HV{\hbox{\it hop\guydash{}tgt}}\;\HV{h_2}}, i.e.:
      \begin{center}
      \resizebox{.4\textwidth}{!}{%
      \begin{tikzpicture}[ every node/.style={scale=1.6} ]
        \node                       (h1) {\ensuremath{\HV{h_1}}};
        \node [below left  = of h1] (tgt1) {\ensuremath{\HV{\hbox{\it hop\guydash{}tgt}}\;\HV{h_1}}};
        \node [below right = of h1] (j1) {\ensuremath{\HV{j_1}}};

        \node [right = of j1] (tgt2) {\ensuremath{\HV{\hbox{\it hop\guydash{}tgt}}\;\HV{h_2}}};
        \node [above right = of tgt2] (h2) {\ensuremath{\HV{h_2}}};
        \node [below right = of h2] (j2) {\ensuremath{\HV{j_2}}};

        \node (form) at ($ (j1)!0.3!(tgt2) $) {\ensuremath{\HSSym{\leq} }};

        \draw [line width=0.25mm, ->] (j2) |- (h2.south) -| (tgt2);
        \draw [line width=0.25mm, ->] (j1) |- (h1.south) -| (tgt1);
      \end{tikzpicture}}
      \end{center}
\item The hops are disjoint, with \ensuremath{\HV{\hbox{\it hop\guydash{}src}}\;\HV{h_2}\HSSym{\leq} \HV{\hbox{\it hop\guydash{}tgt}}\;\HV{h_1}}, i.e.:
      \begin{center}
      \resizebox{.4\textwidth}{!}{%
      \begin{tikzpicture}[ every node/.style={scale=1.6} ]
        \node                       (h2) {\ensuremath{\HV{h_2}}};
        \node [below left  = of h2] (tgt2) {\ensuremath{\HV{\hbox{\it hop\guydash{}tgt}}\;\HV{h_2}}};
        \node [below right = of h2] (j2) {\ensuremath{\HV{j_2}}};

        \node [right = of j2] (tgt1) {\ensuremath{\HV{\hbox{\it hop\guydash{}tgt}}\;\HV{h_1}}};
        \node [above right = of tgt1] (h1) {\ensuremath{\HV{h_1}}};
        \node [below right = of h1] (j1) {\ensuremath{\HV{j_1}}};

        \node (form) at ($ (j2)!0.3!(tgt1) $) {\ensuremath{\HSSym{\leq} }};

        \draw [line width=0.15mm, ->] (j1) |- (h1.south) -| (tgt1);
        \draw [line width=0.15mm, ->] (j2) |- (h2.south) -| (tgt2);
      \end{tikzpicture}}
      \end{center}

\item The hops are nested but not equal, with \ensuremath{\HV{h_2}} being the shorter hop, i.e.:
\begin{center}
\resizebox{.35\textwidth}{!}{%
\begin{tikzpicture}[ every node/.style={scale=1.6} ]
  \node                      (h2) {\ensuremath{\HV{h_2}}};
  \node [above = of h2]      (h1) {\ensuremath{\HV{h_1}}};
  \node [below left = of h2] (tgt2) {\ensuremath{\HV{\hbox{\it hop\guydash{}tgt}}\;\HV{h_2}}};
  \node [left = of tgt2]     (tgt1) {\ensuremath{\HV{\hbox{\it hop\guydash{}tgt}}\;\HV{h_1}}};
  \node [below right = of h2] (j2) {\ensuremath{\HV{j_2}}};
  \node [right = of j2]       (j1) {\ensuremath{\HV{j_1}}};

  \node (form)  at ($ (tgt1)!0.5!(tgt2) $) {\ensuremath{\HSSym{\leq} }};
  \node (form2) at ($ (j2)!0.5!(j1) $) {\ensuremath{\HSSym{<}}};

  \draw [line width=0.25mm, ->] (j2) |- (h2.south) -| (tgt2);
  \draw [line width=0.25mm, ->] (j1) |- (h1.south) -| (tgt1);
\end{tikzpicture}}
\end{center}
\item The hops are nested but not equal, with \ensuremath{\HV{h_1}} being the shorter hop, i.e.:
\begin{center}
\resizebox{.35\textwidth}{!}{%
\begin{tikzpicture}[ every node/.style={scale=1.6} ]
  \node                      (h1) {\ensuremath{\HV{h_1}}};
  \node [above = of h1]      (h2) {\ensuremath{\HV{h_2}}};
  \node [below left = of h1] (tgt1) {\ensuremath{\HV{\hbox{\it hop\guydash{}tgt}}\;\HV{h_1}}};
  \node [left = of tgt1]     (tgt2) {\ensuremath{\HV{\hbox{\it hop\guydash{}tgt}}\;\HV{h_2}}};
  \node [below right = of h1] (j1) {\ensuremath{\HV{j_1}}};
  \node [right = of j1]       (j2) {\ensuremath{\HV{j_2}}};

  \node (form)  at ($ (tgt2)!0.5!(tgt1) $) {\ensuremath{\HSSym{\leq} }};
  \node (form1) at ($ (j1)!0.5!(j2) $) {\ensuremath{\HSSym{<}}};

  \draw [line width=0.25mm, ->] (j1) |- (h1.south) -| (tgt1);
  \draw [line width=0.25mm, ->] (j2) |- (h2.south) -| (tgt2);
\end{tikzpicture}}
\end{center}
\end{enumerate}

          It is interesting to note that the definition of the categorization mechanism (not shown) is mutually recursive with the definition
of \ensuremath{\HSVar{nocross}}: it uses \ensuremath{\HSVar{nocross}} to distinguish some cases.  Agda's termination checker confirms there is no circular reasoning here
because it can prove that such recursive uses of \ensuremath{\HSVar{nocross}} are on lower level hops, implying that the recursion must terminate.

The proof of \ensuremath{\HSVar{nocross}\;\HSVar{h}\;\HSVar{v}} proceeds by induction on \ensuremath{\HSVar{v}}.
If \ensuremath{\HV{\hbox{\it hop\guydash{}tgt}}\;\HSVar{h}\HSSym{\leq} \HV{\hbox{\it hop\guydash{}tgt}}\;\HSVar{v}}, then \ensuremath{\HSVar{nocross}} holds vacuously.
Otherwise, \ensuremath{\HV{\hbox{\it hop\guydash{}tgt}}\;\HSVar{v}\HSSym{<}\HV{\hbox{\it hop\guydash{}tgt}}\;\HSVar{h}}, so if one hop is nested within
the other, then it is \ensuremath{\HSVar{h}} that is contained within \ensuremath{\HSVar{v}}, the longer hop.

  If \ensuremath{\HSVar{v}} is a hop at level zero, we are done because any hop that
 hops over \ensuremath{\HSVar{v}} must contain it.  In case
\ensuremath{\HSVar{v}} is a composite hop \ensuremath{\HSVar{hs}\;\HV{v_0}\;\HV{v_1}\;\HSVar{p}}, we use the categorization mechanism to
determine the relationship between \ensuremath{\HSVar{h}} and \ensuremath{\HV{v_0}}; in some cases we then use
the categorization mechanism again to determine the relationship between
\ensuremath{\HSVar{h}} and \ensuremath{\HV{v_1}}.
It is straightforward in all cases except one
to use information from the categorization mechanism to prove that:
\begin{itemize}
\item the case cannot arise; or
\item \ensuremath{\HV{\hbox{\it hop\guydash{}src}}\;\HSVar{v}\HSSym{\leq} \HV{\hbox{\it hop\guydash{}tgt}}\;\HSVar{h}}, implying that \ensuremath{\HSVar{nocross}} holds via the ``left'' alternative; or
\item \ensuremath{\HSVar{h}\;\HT{\subseteq}\;\HV{v_0}} or \ensuremath{\HSVar{h}\;\HT{\subseteq}\;\HV{v_1}}, implying that \ensuremath{\HSVar{h}\;\HT{\subseteq}\;\HSVar{v}}, so \ensuremath{\HSVar{nocross}} holds via the ``right'' alternative.
\end{itemize}

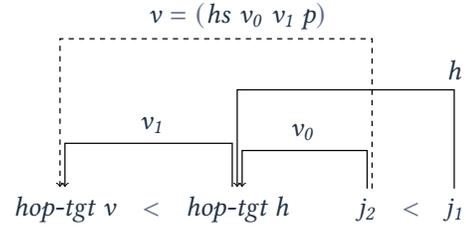
\begin{figure}
\begin{center}
\resizebox{.35\textwidth}{!}{%
\begin{tikzpicture}[ every node/.style={scale=1.6} ]
  \node (j1) {\ensuremath{\HV{j_1}}};
  \node [left = of j1] (j2) {\ensuremath{\HV{j_2}}};
  \node [left = of j2] (tgt1) {\ensuremath{\HV{\hbox{\it hop\guydash{}tgt}}\;\HSVar{h}}};
  \node [left = of tgt1] (tgt2) {\ensuremath{\HV{\hbox{\it hop\guydash{}tgt}}\;\HSVar{v}}};

  \node (form)  at ($ (j2)!0.5!(j1) $) {\ensuremath{\HSSym{<}}};
  \node (form2) at ($ (tgt2)!0.5!(tgt1) $) {\ensuremath{\HSSym{<}}};

  \node [above = 2cm of j1] (h1) {\ensuremath{\HSVar{h}}};
  \draw [line width=0.25mm, ->] (j1) |- (h1.south) -| (tgt1);

  \node (aux) at ($ (j2)!0.5!(tgt1) $) {};
  \node [above = of aux] (v0) {\ensuremath{\HV{v_0}}};
  \node [above = of form2] (v1) {\ensuremath{\HV{v_1}}};

  \draw [line width=0.25mm, ->] (j2) |- (v0.south)
        -| ($ (tgt1.north) + (.1,0) $);
  \draw [line width=0.25mm, ->] ($ (tgt1.north) - (.1,0) $)
        |- (v1.south)
        -| (tgt2);

  \node [above = 3cm of tgt1] (v) {\ensuremath{\HSVar{v}\HSSym{\mathrel{=}}\HSSpecial{(}\HSVar{hs}\;\HV{v_0}\;\HV{v_1}\;\HSVar{p}\HSSpecial{)}}};

  \draw [dashed, line width=0.25mm, ->]
    ($ (j2.north) + (.1,0) $) |- (v.south) -| ($ (tgt2.north) - (.1,0) $);
\end{tikzpicture}}
\end{center}
\caption{There exists no \ensuremath{\HSVar{p}} with type \ensuremath{\HSVar{suc}\;\HSVar{l}\HSSym{<}\HSVar{maxLvl}\;\HV{j_2}} when in
this situation. Hence, the composite hop \ensuremath{\HSVar{hs}\;\HV{v_0}\;\HV{v_1}\;\HSVar{p}} cannot be built.}
\label{fig:impossiblehop}
\end{figure}

The difficult case is
illustrated in \Cref{fig:impossiblehop}.  In this case, \ensuremath{\HV{\hbox{\it hop\guydash{}tgt}}\;\HSVar{h}} coincides
with the point at which \ensuremath{\HV{v_0}} and \ensuremath{\HV{v_1}} meet, so there is no opportunity
for the categorization mechanism to eliminate the case using a recursive instance of \ensuremath{\HSVar{nocross}} against \ensuremath{\HV{v_0}} or \ensuremath{\HV{v_1}}.
It simply tells us that \ensuremath{\HV{v_1}} is disjoint from \ensuremath{\HSVar{h}} and that \ensuremath{\HSVar{h}} contains \ensuremath{\HV{v_0}}.

The \ensuremath{\HV{\hbox{\it h\guydash{}lvl\guydash{}mid}}} lemma is invaluable here: it proves that this case is impossible.
Suppose \ensuremath{\HV{v_0}} and \ensuremath{\HV{v_1}} are hops at level \ensuremath{\HSVar{l}}, so \ensuremath{\HSVar{v}} is at level \ensuremath{\HSVar{suc}\;\HSVar{l}}.
The contradiction arises from \ensuremath{\HSVar{p}}, which is (supposedly)
a proof that \ensuremath{\HSVar{suc}\;\HSVar{l}\HSSym{<}\HSVar{maxLvl}\;\HV{j_2}}.  Because \ensuremath{\HSVar{h}} hops \emph{over} \ensuremath{\HV{j_2}},
\ensuremath{\HV{\hbox{\it h\guydash{}lvl\guydash{}mid}}} implies that \ensuremath{\HSVar{maxLvl}\;\HV{j_2}} is at most \ensuremath{\HSVar{suc}\;\HSVar{l}}.

Having proved these properties about our hop relation,
we are ready to use it to instantiate the abstract model with
our particular definition of \ensuremath{\HSCon{DepRel}}.  The properties
required by \ensuremath{\HSCon{DepRel}} other than \ensuremath{\HV{\hbox{\it hop\guydash{}no\guydash{}cross}}} are straightforward
to prove; some of them are shown below.

For \ensuremath{\HV{\hbox{\it hop\guydash{}no\guydash{}cross}}}, first we define a simple convenience
function to witness the existence of hops:

\begin{myhs}
\begin{hscode}\SaveRestoreHook
\column{B}{@{}>{\hspre}l<{\hspost}@{}}%
\column{E}{@{}>{\hspre}l<{\hspost}@{}}%
\>[B]{}\HSVar{getHop}\HSCon{\mathbin{:}}\HS{\forall}\;\HSSpecial{\HSSym{\{\mskip1.5mu} }\HSVar{j}\HSSpecial{\HSSym{\mskip1.5mu\}}}\;\HSSpecial{(}\HSVar{h}\HSCon{\mathbin{:}}\HSCon{Fin}\;\HSSpecial{(}\HSVar{maxLvl}\;\HSVar{j}\HSSpecial{)}\HSSpecial{)}\HSSym{\to} \HSCon{H}\;\HSSpecial{(}\HV{\mathit{to}\mathbb{N}}\;\HSVar{h}\HSSpecial{)}\;\HSVar{j}\;\HSSpecial{(}\HSVar{j}\HSSym{-}\HSNumeral{2}^{\HV{\mathit{to}\mathbb{N}}\;\HSVar{h}}\HSSpecial{)}{}\<[E]%
\\
\>[B]{}\HSVar{getHop}\;\HSSpecial{\HSSym{\{\mskip1.5mu} }\HSVar{j}\HSSpecial{\HSSym{\mskip1.5mu\}}}\;\HSVar{h}\HSSym{\mathrel{=}}\HV{\hbox{\it h\guydash{}correct}}\;\HSVar{j}\;\HSSpecial{(}\HV{\mathit{to}\mathbb{N}}\;\HSVar{h}\HSSpecial{)}\;\HSSpecial{(}\HV{\mathit{to}\mathbb{N}\!\!<\!\!n}\;\HSVar{h}\HSSpecial{)}{}\<[E]%
\ColumnHook
\end{hscode}\resethooks
\end{myhs}

  Then we prove the \ensuremath{\HV{\hbox{\it hop\guydash{}no\guydash{}cross}}} property
using \ensuremath{\HSVar{getHop}} and \ensuremath{\HSVar{nocross}}.  We pattern match on the result of
\ensuremath{\HSVar{nocross}} and discharge the \ensuremath{\HT{\mathit{Left}}} branch as impossible
with simple arithmetic (not shown).
The other branch carries a proof that one hop is entirely contained
within the other, from which we extract a value of the desired type.

\begin{myhs}
\begin{hscode}\SaveRestoreHook
\column{B}{@{}>{\hspre}l<{\hspost}@{}}%
\column{3}{@{}>{\hspre}l<{\hspost}@{}}%
\column{9}{@{}>{\hspre}l<{\hspost}@{}}%
\column{11}{@{}>{\hspre}l<{\hspost}@{}}%
\column{17}{@{}>{\hspre}l<{\hspost}@{}}%
\column{18}{@{}>{\hspre}l<{\hspost}@{}}%
\column{19}{@{}>{\hspre}c<{\hspost}@{}}%
\column{19E}{@{}l@{}}%
\column{22}{@{}>{\hspre}l<{\hspost}@{}}%
\column{32}{@{}>{\hspre}l<{\hspost}@{}}%
\column{E}{@{}>{\hspre}l<{\hspost}@{}}%
\>[B]{}\HSVar{skiplog}\HSCon{\mathbin{:}}\HSCon{DepRel}{}\<[E]%
\\
\>[B]{}\HSVar{skiplog}\HSSym{\mathrel{=}}\HSKeyword{record}{}\<[E]%
\\
\>[B]{}\hsindent{3}{}\<[3]%
\>[3]{}\HSSpecial{\HSSym{\{\mskip1.5mu} }\HSVar{maxlvl}{}\<[18]%
\>[18]{}\HSSym{\mathrel{=}}\HSVar{maxLvl}{}\<[E]%
\\
\>[B]{}\hsindent{3}{}\<[3]%
\>[3]{}\HSSpecial{;}\HV{\hbox{\it maxlvl\guydash{}z}}{}\<[18]%
\>[18]{}\HSSym{\mathrel{=}}\HSVar{refl}{}\<[E]%
\\
\>[B]{}\hsindent{3}{}\<[3]%
\>[3]{}\HSSpecial{;}\HV{\hbox{\it hop\guydash{}tgt}}{}\<[17]%
\>[17]{}\HSSym{\mathrel{=}}\HS{\lambda}\;\HSSpecial{\HSSym{\{\mskip1.5mu} }\HSVar{m}\HSSpecial{\HSSym{\mskip1.5mu\}}}\;\HSVar{h}\HSSym{\to} \HSVar{m}\;\HS{\dotminus}\;\HSSpecial{(}\HSNumeral{2}^{\HV{\mathit{to}\mathbb{N}}\;\HSVar{h}}\HSSpecial{)}{}\<[E]%
\\
\>[B]{}\hsindent{3}{}\<[3]%
\>[3]{}\HSSpecial{;}\HV{\hbox{\it hop\guydash{}inj}}{}\<[17]%
\>[17]{}\HSSym{\mathrel{=}}\HS{\dots}\HSComment{ -\! - simple Agda exercises}{}\<[E]%
\\
\>[B]{}\hsindent{3}{}\<[3]%
\>[3]{}\HSSpecial{;}\HV{\hbox{\it hop\guydash{}\!<}}{}\<[17]%
\>[17]{}\HSSym{\mathrel{=}}\HS{\dots}{}\<[E]%
\\
\>[B]{}\hsindent{3}{}\<[3]%
\>[3]{}\HSSpecial{;}\HV{\hbox{\it hop\guydash{}no\guydash{}cross}}{}\<[17]%
\>[17]{}\HSSym{\mathrel{=}}\HS{\lambda}\;\HSSpecial{\HSSym{\{\mskip1.5mu} }\HV{h_1}\HSSym{\mathrel{=}}\HV{h_1}\HSSpecial{\HSSym{\mskip1.5mu\}}}\;\HSSpecial{\HSSym{\{\mskip1.5mu} }\HV{h_2}\HSSpecial{\HSSym{\mskip1.5mu\}}}\;\HSVar{p1}\;\HSVar{p2}\HSSym{\to} {}\<[E]%
\\
\>[3]{}\hsindent{6}{}\<[9]%
\>[9]{}\HSKeyword{case}\;\HSVar{nocross}\;\HSSpecial{(}\HSVar{getHop}\;\HV{h_1}\HSSpecial{)}\;\HSSpecial{(}\HSVar{getHop}\;\HV{h_2}\HSSpecial{)}\;\HSKeyword{of}{}\<[E]%
\\
\>[9]{}\hsindent{2}{}\<[11]%
\>[11]{}\HS{\lambda}\;{}\<[19]%
\>[19]{}\HSSpecial{\HSSym{\{\mskip1.5mu} }{}\<[19E]%
\>[22]{}\HT{\mathit{Left}}\;\HSVar{imp}{}\<[32]%
\>[32]{}\HSSym{\to} \HS{\dots}{}\<[E]%
\\
\>[19]{}\HSSpecial{;}{}\<[19E]%
\>[22]{}\HT{\mathit{Right}}\;\HSVar{r}{}\<[32]%
\>[32]{}\HSSym{\to} \HV{\subseteq\!\!-\!src\!-\!\!\leq}\;\HSSpecial{(}\HSVar{getHop}\;\HV{h_1}\HSSpecial{)}\;\HSSpecial{(}\HSVar{getHop}\;\HV{h_2}\HSSpecial{)}\;\HSVar{r}\HSSpecial{\HSSym{\mskip1.5mu\}}}\HSSpecial{\HSSym{\mskip1.5mu\}}}{}\<[E]%
\ColumnHook
\end{hscode}\resethooks
\end{myhs}


\section{Related and Future Work}
\label{sec:relwork}


Pugh's original skip lists~\cite{Pugh1990} supported insertion,
necessitating the use of probabilistic ``heights'' (analogous to our
deterministic \ensuremath{\HSVar{maxLvl}} function) for elements added to the list.
Authenticated skip lists proposed by Goodrich and
Tamassia~\cite{Goodrich2000} similarly required probabilistic heights,
and depended on the use of a commutative hash function.  Maniatis and
Baker~\cite{Baker2003} proposed an \emph{append-only} authenticated
skip list (AAOSL), eliminating the need for commutative hashing, and
achieving a simpler structure better suited to representing
tamper-evident logs such as blockchains.  The concrete AAOSL achieved
in \Cref{sec:concretemodel} by instantiating our abstract model
presented in \Cref{sec:refinedblocks} is essentially the same
as~\cite{Baker2003}, modulo our changes to eliminate the possibility
of a prover providing different authenticators for the same index
within an advancement proof.  The ``skipchains'' used by
Chaniac~\cite{chainiac2017} use the same ``hop structure'' in the case
of deterministic skipchains with $b = 2$; Chaniac also maintains
``forward'' links built when new elements are appended.

Authenticated data-structures have been studied in many different
incarnations.  Miller et al.~\cite{miller14gpads} presents
\emph{lambda-auth}, a language used to write a large variety of
authenticated data structures with ease.  They provide pen-and-paper
proofs that a prover can fool a verifier only if it finds a hash collision.
Brun and Traytel~\cite{BrunTraytel2019} formalized
\emph{lambda-auth} in Isabelle/HOL, which corrected some issues in the
original paper, strengthening the overall result and reinforcing
the importance of formal verification. In authenticated data structures produced using \emph{lambda-auth},
the hash of a value depends exclusively on its type structure.
In our case, the hash of different indexes might have a different number
of dependencies, even though each log entry carries only one recursive
argument: its tail.  It therefore does not seem possible to use \emph{lambda-auth} to achieve
encode the original AAOSL presentation~\cite{Baker2003}, nor our
formalization thereof.

  Using trees instead of lists to enable efficient queries has
also been done in the CCF framework \cite{Russinovich2019},
with a verified incremental Merkle tree implementation \cite{Protzenko2020}.
This enables logarithmic-time access to past transactions and enables
peers to store only important transactions locally.
However, the CCF framework does not support advancement proofs.
It may be possible to extend incremental
Merkle trees to support advancement proofs
and to prove a property analogous to \textsc{EVO-CR}.
However, the translation is not immediate.  An important difference between incremental Merkle trees and
deterministic skiplogs is in how hashes are computed. The root hash
of the Merkle tree depends directly only on its immediate children,
whereas skiplog hashes depend directly on previous values carefully
chosen to provide logarithmic-sized advancement proofs.

There have been a number of formal verification
efforts related to blockchain consensus~\cite{Pirlea2018,Garay2015,Gopinathan2019}.
Consensus is orthogonal to proving \textsc{EVO-CR}. The former states
a property such as there is a single chain of blocks that is consistent
with every honest
participant's view, and commits must extend this chain.
In contrast,
our work entails formal verification of AAOSLs, which can be used to
enable verifying claims about past entries in the
presence of partial information.  In other words, Evolutionary
Collision Resistance states that if two peers agree at log index
$j$, even without possessing the entire log, then it is not possible to convince them
of conflicting membership claims for indexes $i < j$. The \emph{agree
on common} (\textsc{AOC}) property can be
related to the universal agreement properties of P\^{i}rlea and Sergey
\cite{Pirlea2018} or the common prefix property of Garay et al. \cite{Garay2015,Gopinathan2019}.
If we consider a set of $n$ ``degenerate'' advancement
proofs, each of which includes only hops between adjacent indexes, starting at genesis and arriving
at indexes $k_0, \cdots, k_{n-1}$, respectively, \textsc{AOC} implies that they all agree
on all indexes $i \leq \mathrm{min}\, \{ k_0 , \cdots , k_{n-1} \} $.
In our context, \textsc{AOC} is more general because it applies even in the case
of partial information: advancement proofs
typically do not both visit all indexes.

The most important future work aspect lies in the treatment of hash collisions.
We proved our results modulo finding an arbitrary hash collision, similar
to Miller et al.~\cite{miller14gpads} or Brun and Traytel~\cite{BrunTraytel2019}.
Yet, a full fledged security proof should provide a polynomial time reduction from an authentication failure
to a hash collision.  We are grateful to Profs. Seny Kamara and Jos\'{e} Carlos Bacelar Almeida
for useful discussions on this direction.

One promising option to improve our development in
this direction is to make our proofs return proof-relevant information
about which of its branches evidenced a collision.
For example, in the \ensuremath{\HV{\textsc{AgreeOnCommon}}} lemma, we could replace the \ensuremath{\HSCon{HashBroke}} disjunct as illustrated below.

\begin{myhs}
\begin{hscode}\SaveRestoreHook
\column{B}{@{}>{\hspre}l<{\hspost}@{}}%
\column{3}{@{}>{\hspre}l<{\hspost}@{}}%
\column{13}{@{}>{\hspre}l<{\hspost}@{}}%
\column{14}{@{}>{\hspre}l<{\hspost}@{}}%
\column{E}{@{}>{\hspre}l<{\hspost}@{}}%
\>[B]{}\HV{\textsc{AgreeOnCommon}}{}\<[E]%
\\
\>[B]{}\hsindent{3}{}\<[3]%
\>[3]{}\HSCon{\mathbin{:}}\HS{\forall}\;{}\<[13]%
\>[13]{}\HSSpecial{\HSSym{\{\mskip1.5mu} }\HSVar{j}\;\HV{i_1}\;\HV{i_2}\HSSpecial{\HSSym{\mskip1.5mu\}}}\;\HSSpecial{\HSSym{\{\mskip1.5mu} }\HV{t_1}\;\HV{t_2}\HSCon{\mathbin{:}}\HSCon{View}\HSSpecial{\HSSym{\mskip1.5mu\}}}{}\<[E]%
\\
\>[B]{}\hsindent{3}{}\<[3]%
\>[3]{}\HSSym{\to} \HSSpecial{\HSSym{\{\mskip1.5mu} }\HV{a_1}\HSCon{\mathbin{:}}\HSCon{AdvPath}\;\HSVar{j}\;\HV{i_1}\HSSpecial{\HSSym{\mskip1.5mu\}}}\;\HSSpecial{\HSSym{\{\mskip1.5mu} }\HV{a_2}\HSCon{\mathbin{:}}\HSCon{AdvPath}\;\HSVar{j}\;\HV{i_2}\HSSpecial{\HSSym{\mskip1.5mu\}}}{}\<[E]%
\\
\>[B]{}\hsindent{3}{}\<[3]%
\>[3]{}\HSSym{\to} \HSVar{rebuild}\;\HV{a_1}\;\HV{t_1}\;\HSVar{j}\HSSym{\equiv} \HSVar{rebuild}\;\HV{a_2}\;\HV{t_2}\;\HSVar{j}{}\<[E]%
\\
\>[B]{}\hsindent{3}{}\<[3]%
\>[3]{}\HSSym{\to} \HSSpecial{\HSSym{\{\mskip1.5mu} }\HSVar{i}\HSCon{\mathbin{:}}\HT{\mathbb{N}}\HSSpecial{\HSSym{\mskip1.5mu\}}}\HSSym{\to} \HSVar{i}\;\HT{\epsilon_{\textsc{ap}}}\;\HV{a_1}\HSSym{\to} \HSVar{i}\;\HT{\epsilon_{\textsc{ap}}}\;\HV{a_2}{}\<[E]%
\\
\>[B]{}\hsindent{3}{}\<[3]%
\>[3]{}\HSSym{\to} \HSCon{Either}\;{}\<[14]%
\>[14]{}\HSSpecial{(}\HT{\hbox{\it AOC\guydash{}Collision}}\;\HV{t_1}\;\HV{t_2}\;\HV{a_1}\;\HV{a_2}\HSSpecial{)}\;{}\<[E]%
\\
\>[14]{}\HSSpecial{(}\HSVar{rebuild}\;\HV{a_1}\;\HV{t_1}\;\HSVar{i}\HSSym{\equiv} \HSVar{rebuild}\;\HV{a_2}\;\HV{t_2}\;\HSVar{i}\HSSpecial{)}{}\<[E]%
\ColumnHook
\end{hscode}\resethooks
\end{myhs}

  Here, a value of type \ensuremath{\HT{\hbox{\it AOC\guydash{}Collision}}\;\HV{t_1}\;\HV{t_2}\;\HV{a_1}\;\HV{a_2}} would
witness a hash collision from the data provided in its parameters \ensuremath{\HV{t_1}},\ensuremath{\HV{t_2}},\ensuremath{\HV{a_1}} and \ensuremath{\HV{a_2}}.
The definition of \ensuremath{\HT{\hbox{\it AOC\guydash{}Collision}}} is directly dependent on the
structure of \ensuremath{\HV{\textsc{AgreeOnCommon}}}, and hence cannot be reused for \ensuremath{\HV{\textsc{evo-cr}}}, for instance.
The \ensuremath{\HV{\textsc{evo-cr}}} would have its own \ensuremath{\HT{\hbox{\it EvoCR\guydash{}Collision}}\HSSym{...}} datatype, which would actually use
values of \ensuremath{\HT{\hbox{\it AOC\guydash{}Collision}}} because \ensuremath{\HV{\textsc{evo-cr}}} calls \ensuremath{\HV{\textsc{aoc}}} internally and needs to
forward potential collisions. Consequently, every Agda function that
returns \ensuremath{\HSCon{Either}\;\HSCon{HashBroke}} will have its own collision-tracking datatype,
making this a non-trivial effort. Nevertheless, with these \ensuremath{\HSCon{Collision}} datatypes at hand, we would explicitly
construct the hash collisions and readers could convince themselves
of the polynomial runtime constraint based on the structure of these \ensuremath{\HSCon{Collision}} datatypes.

  Although the approach described above would surely work, it is labor intensive
and not transferable. As a future research question, we wonder whether there might
be a more automatic way of constructively keeping track of hash collisions. In fact,
with a little creativity one could even imagine enforcing the polynomial runtime
constraint at the type level using appropriate type systems~\cite{Hofmann1999}.

\section{Concluding Remarks}
\label{sec:conc}

We have presented formal, machine-checked proofs
that Authenticated Append-Only Skip Lists (AAOSLs) indeed provide
the properties claimed by its original authors~\cite{Baker2003}.

Our formalization effort also uncovered some improvements to the
original AAOSL.  For example, providing authenticators
separately in \ensuremath{\HSCon{View}}s (or with partial maps, as would be done
in practice) precludes representing proofs
with inconsistent authenticators, eliminating the need to
check that, and also results in slightly smaller advancement proofs.
Our proofs also highlighted that the actual implementation of \ensuremath{\HV{\hbox{\it \calcauth}}}
is a red herring. Any definition satisfying
\ensuremath{\HV{\hbox{\it \calcauth\guydash{}inj\guydash{}1}}} and \ensuremath{\HV{\hbox{\it \calcauth\guydash{}inj\guydash{}2}}} can be used with confidence. This showcases two
important aspects of a formal development. On one hand, it uncovers
the foundations of the security argument. In this case, we need an
injective \ensuremath{\HV{\hbox{\it \calcauth}}} function and a definition of hops that do not cross each other.
On the other hand, it provides an interactive playground to study variations
of the construction. After we finished the development using the definition of \ensuremath{\HV{\hbox{\it \calcauth}}}
provided by Maniatis and Baker, it was trivial to change it to a simpler variant
with confidence that the development is still valid.

This work provides increased confidence that
AAOSLs can be used in practice to dramatically reduce the
overhead of verifying a recent log state, thus enabling
sustainable dynamic participation in authenticated logs
such as blockchains. Our development is available in open source~\cite{aaosl-agda}.

\bibliography{references}

\appendix

%

\end{document}